\documentclass[12pt, a4paper]{article}
\pdfoutput=1

\usepackage{amsmath}
\usepackage{amsfonts}
\usepackage{amssymb}
\usepackage{graphicx, rotating}
\usepackage{epsfig}
\usepackage{latexsym}
\usepackage{graphicx}
\usepackage{color}
\usepackage{amsmath,bm,amssymb}
\usepackage{cite}
\usepackage{slashed}
\usepackage{hyperref}
\hypersetup{colorlinks, citecolor=bluscuro, linkcolor=black, urlcolor=bluscuro}
\definecolor{rossos}{cmyk}{0,1,1,0.55}
\definecolor{bluscuro}{rgb}{0.15, 0.2, .85}
\definecolor{bluchiaro}{cmyk}{1,.3,0.,0.1}

\numberwithin{equation}{section}

\newcommand{\be}{\begin{equation}}
\newcommand{\ee}{\end{equation}}
\newcommand{\bea}{\begin{eqnarray}}
\newcommand{\eea}{\end{eqnarray}}

\def\simlt{\stackrel{<}{{}_\sim}}
\def\simgt{\stackrel{>}{{}_\sim}}

\def\pp{{\scriptscriptstyle +}}
\def\mm{{\scriptscriptstyle -}}

\newcommand{\arXiv}[2]{\href{http://arxiv.org/pdf/#1}{{\tt [#2/#1]}}}
\newcommand{\arXivold}[1]{\href{http://arxiv.org/pdf/hep-#1}{{\tt [#1]}}}

\def\bma#1{\mbox{\boldmath{$#1$}}}

\begin{document}
\allowdisplaybreaks
\begin{titlepage}
\begin{flushright}
\end{flushright}
\vspace{.3in}

\vspace{-1cm}
\begin{center}
{\Large\bf\color{black} 
Bubbles of Nothing: \\
\vspace*{0.5cm}
The Tunneling Potential Approach  
} \\
\vspace{1cm}{
{\large J.J. Blanco-Pillado$^{1,2,3}$},
{\large J.R.~Espinosa$^4$, J. Huertas$^4$} and
{\large K. Sousa$^5$}
} \\[7mm]
{\it 
${}^1$ Department of Theoretical Physics, University of the Basque Country UPV/EHU \\
${}^2$ EHU Quantum Center, University of the Basque Country, UPV/EHU\\
${}^3$ IKERBASQUE, Basque Foundation for Science, 48011, Bilbao, Spain.\\
${}^4$ Instituto de F\'{\i}sica Te\'orica, IFT-UAM/CSIC, \\ 
C/ Nicol\'as Cabrera 13-15, Campus de Cantoblanco, 28049, Madrid, Spain\\
${}^5$  
 University of Alcal\'a, Department of Physics and Mathematics, \\
 Pza. San Diego, s/n, 28801, Alcal\'a de Henares (Madrid), Spain.
}
\end{center}
\bigskip

\vspace{.4cm}

\begin{abstract}
Bubbles of nothing (BoNs) describe the decay of spacetimes with compact dimensions and are thus of fundamental importance for many higher dimensional theories proposed beyond the Standard Model. BoNs admit a 4-dimensional description in terms of a singular Coleman-de Luccia (CdL) instanton involving the size modulus field, stabilized by some potential $V(\phi)$. Using the so-called tunneling potential ($V_t$) approach, we study which types of BoNs are possible and for which potentials $V(\phi)$ can they be present. We identify four different types of BoN, characterized by different asymptotic behaviours at the BoN core and corresponding to different classes of higher dimensional theories, which we also classify. Combining numerous analytical and numerical examples, we study the interplay of BoN decays with other standard decay channels, identify the possible types of quenching of BoN decays  and show how BoNs for flux compactifications can also be described in 4 dimensions by a multifield $V_t$.
The use of the $V_t$ approach greatly aids our analyses and offers a very simple picture of BoNs which are treated in the same language as any other standard vacuum decays. 
\end{abstract}
\bigskip

\end{titlepage}

\section{Introduction}
Many high energy physics models beyond the Standard Model predict several vacua
 allowing dynamical transitions between them, via quantum mechanical tunneling in particular. In field theory the study of these
processes was initiated by Coleman and collaborators \cite{Coleman:1977py,Callan:1977pt} who showed how these transitions proceed by the quantum nucleation of bubbles of the new vacuum that expand rapidly transforming large regions of spacetime to the new vacuum. 
These works also described how one can use Euclidean solutions of the equations of motion
(the instanton solutions) to compute the probability for these transitions to occur. Later on, Coleman and de Luccia (CdL) investigated the effects of gravity in vacuum decay finding
that gravitational instantons could deviate significantly from their flat space counterparts \cite{CdL}. 

All these considerations can be relevant not only for the physics of the early universe, where some of these phase transitions could occur, but also for the late-time universe. Many cosmological probes indicate that our universe is currently dominated by an effective cosmological constant. This observation can be easily accommodated in a low energy theory with a potential whose local minimum gives this positive energy density today. However, this
minimum is not necessarily the only vacuum and it is potentially unstable to transitions to other vacua, suggesting that our universe might have a finite lifetime \cite{SMdecay}. 

On the other hand, many extensions of the Standard Model require that our
universe is higher dimensional, compactified in order to agree with observation. One possibility is that our universe is described by a $M^4 \times X^{d}$ spacetime, where $M^4$ is the 4-dimensional space where we live and $X^d$ an internal space with $d$ dimensions. This
setup leads to a low energy effective theory that includes the degrees of freedom that parametrize the geometrical properties of the internal manifold and, in order to make this theory compatible with observations, one needs to fix all these massless degrees of freedom, {\it e.g.} inducing an effective potential that pins down these fields to their expectation values. Regardless of the particular mechanism that induces this potential, it seems reasonable to assume that this theory would also lead to multiple vacua. 

These arguments suggest that higher dimensional models have many possible bubble transitions of the kind discussed above. A prototypical example is String Theory: the original formulation of the theory has $10 - 11$ dimensions which must be compactified down to four dimensions by some mechanism, {\it e.g.} using fluxes of higher dimensional p-form fields along the internal dimensions \cite{Denef:2007pq}. The backreaction on the metric of these fluxes  fixes many of the degrees of freedom of the internal geometry leading to a 
perturbatively stable vacuum. The integral of the fluxes over the appropriate cycles of
the internal manifold are quantized and, therefore, the possible transitions between the
different vacua necessarily involve the presence of charged objects. Thus, one can think of these transitions as a higher dimensional version of the
Schwinger process \cite{Brown:1987dd}. The nucleation of the charged bubble wall decreases the flux through
the internal manifold in the interior of the bubble. This new configuration
with a different set of fluxes settles to a slightly different internal geometry,
in other words, a new vacuum. Such flux-changing transitions have been explored
in detail in several papers in String Theory \cite{Bousso:2000xa} and in other higher dimensional field
theory models in \cite{Blanco-Pillado:2009lan}.

However, in a higher dimensional theory there are new vacuum decay channels. In a seminal paper \cite{BoN}, Witten showed that compact extra dimensions could lead to the formation of the so-called {\it bubble of nothing} (BoN), a new decay
process that has many similarities with the usual CdL instantons. There is, however, a crucial
difference: the bubble interior is not a new vacuum since an extra dimension pinches off over the sphere of the bubble. Ref.~\cite{BoN} discussed the BoN in the simplest model with extra dimensions, namely, the pure Kaluza-Klein model in $5d$, in which  the decay is mediated by the Euclidean version of the Schwarzschild solution (with properties in agreement with the usual gravitational instantons). In particular, its analytic continuation along a surface of vanishing extrinsic curvature gives the Lorentzian 
evolution of this configuration. The bubble starts from rest and expands with a constant acceleration, ``eating'' the parent spacetime as do the usual CdL instantons. Moreover, the BoN has a single negative mode \cite{BoN} as expected for Euclidean solutions describing an instability \cite{ColemanNegative}.

Beyond the simplest model with extra dimensions, how generic is this type of process? Early attempts to generalize this type
of instability to models of flux compactification, with fluxes over the
internal dimensions, were hindered by important obstacles \cite{Young:1984jv}. In particular, a shrinking internal cycle endowed with flux causes a divergent backreaction of the energy-momentum tensor associated to the flux. This difficulty was first bypassed by placing a flux source in the
instanton solution such that the flux would be absorbed by this charged object at the location where the
internal dimension disappears \cite{BS}. This configuration leads to a smooth solution
everywhere and allows us to generalize this type of instantons to other compactification models
with similar ingredients. For similar solutions in other type of models with sources slightly different in nature, see \cite{YBD}.

The BoN decay channel can be nicely interpreted in models of flux compactification where one can find instantons interpolating between vacua with different set of fluxes.
In these models one may ask what kind of solution would appear in the limit where the transition happens to the vacua with zero flux. In such case, the interpolating
 domain wall soaks up all the flux leaving behind a
configuration without flux. Without flux, nothing prevents spacetime collapse and,  therefore, the vacuum on the interior of the bubble is replaced
by {\it nothing}, i.e. the internal manifold pinches off. This interpretation of the
bubble of nothing in models of flux compactifications was first put forward in \cite{BS}
and later discussed in similar scenarios in \cite{Brown:2011gt}. For further generalizations of this type of BoNs with more general internal manifolds 
in different dimensions see, for example, \cite{Blanco-Pillado:2010vdp,Blanco-Pillado:2016xvf,Ooguri:2017njy,Dibitetto:2020csn}.

Besides their use to assess the stability of String Landscape vacua, BoNs are also relevant for the Swampland program \cite{Swamp,SwmpRvw}, a relatively recent initiative that aims to characterize which effective field theories can be consistently coupled to gravity. Although general in its purpose, the evidence for its conjectures comes primarily from String Theory. One of these proposals, the Cobordism Conjecture \cite{CobConj}, states that all consistent quantum gravity theories are cobordant between them, that is, there exists a domain wall that connects them. This implies that every consistent quantum gravity must admit a cobordism to nothing, so there must exist a configuration ending spacetime. BoNs can be viewed as this type of configuration, as the spacetime ends smoothly over the surface of the bubble. 

BoN instantons can also be analyzed using an effective $4d$ description in terms of a singular CdL bounce\footnote{That the BoN solution becomes singular due to dimensional reduction may sound strange at first but is common as dimensional reduction or its opposite, oxidation, can change the nature of the singularities in different dimensions. For a somewhat related situation see \cite{BRS,Garriga:1998ri}, or, in a different context, \cite{Gibbons:1994vm}.}, as first explored in \cite{DFG} for Witten's BoN. This bottom-up effective approach is particularly well-suited to study the impact of a nonzero potential for the modulus field that controls the size of the compactification, thus generalizing Witten's BoN. For a recent discussion, see \cite{DGL},  where some of the necessary conditions on the potential for the
existence of a BoN were obtained.

In this paper we follow this bottom-up approach but using the so-called
{\it Tunneling Potential Approach} introduced in \cite{E} and further investigated in \cite{Eg,EFH,EK,ESM,EGrav}. In this approach, instead of using an Euclidean CdL bounce, vacuum decay is described in terms of a tunneling potential function, $V_t(\phi)$, that can be directly compared with $V(\phi)$ and minimizes a simple action functional defined directly in field space. This $V_t$ formalism can be applied to study BoN solutions and has a number of appealing properties, that we use and explain in this work, and are the following. The BoN configuration is described in terms of a function $V_t(\phi)$ on the same footing as the potential $V(\phi)$, without needing to examine the field profile or the metric (although these can be obtained from $V_t$ if needed), and the different types of possible BoNs can be classified simply by studying the asymptotic properties of $V$ and $V_t$ and their interrelation. The $V_t$ approach is also useful to study the interplay of BoNs with other decay configurations like CdL's, Hawking-Moss instantons \cite{HM} or pseudo-bounces \cite{PS}, which can be described by different $V_t$ solutions, all on the same footing with BoNs. The BoN action is given by a simple universal expression (without additional boundary contributions nor delicate cancellations between instanton and false vacuum terms, as in the Euclidean approach). Finally, the $V_t$ formalism is well suited to find analytic examples of BoNs.

Using this technique, we efficiently explore possible BoNs, identifying four possible types with characteristic asymptotic behaviour as the bubble core is approached. The different types correspond to different possible higher dimensional origins (depending on the topology and dimensionality of the compact space as well as on the possible presence of defects or other UV objects). We use simple toy examples to study (both numerically and analytically) the action and structure of these BoNs contrasting them with other decay channels that might be present for a given modulus potential, $V(\phi)$. For a fixed $V(\phi)$ one typically finds continuous families of possible BoN decays but, once the parameters  of the higher dimensional theory are fixed, only a  discrete number of BoNs are relevant, with asymptotic properties being directly related to the sizes of the compact space and the nucleated BoN. 

We also identify and study two types of critical cases for which the BoN decay is quenched. In the first, the action becomes infinite (CdL mechanism) and the BoN transforms into an end-of-the-world brane, while, in the second, the action remains finite. We also show explicitly how a two-field $V_t$ can describe a BoN in a $5d$ flux compactification, with the BoN selecting a direction in the two-field space with the right asymptotic behaviour to allow for a smooth shrinking of the compact dimension (the $4d$ description of the $5d$ mechanism of flux being absorbed by a source at the BoN core).

The paper is organized as follows. In section~\ref{sec:Vt}, we review the $V_t$ formalism and how it can be applied to describe vacuum decay in QFT, including CdL gravitational corrections. In section~\ref{sec:WBoN}, we review Witten's BoN, first giving the original $5d$ solution, then explaining its $4d$ CdL reduction and finally showing how the $V_t$ formulation gives a very simple description of it. The $4d$ $V_t$ approach is extended in Section~\ref{sec:BotUp} to more general settings with nonzero potential for the modulus field, $\phi$. In this section we identify four possible asymptotic behaviours of $V(\phi)$ and $V_t(\phi)$ in the neighbourhood of the BoN core ($\phi\to\infty$) required to have BoN solutions.  In section~\ref{sec:BCs} we analyze the interplay between boundary conditions for $V_t$ near the false vacuum and their asymptotic behaviour at large field values, describing as well how BoN decays can also be quenched by gravity. In section \ref{sec:BoNvsOther} we study how the BoN solutions and their action compares with other possible decay channels (like regular Coleman-De Luccia bounces, Hawking-Moss instantons or pseudo-bounces). In section~\ref{sec:anex} we provide analytic examples of all the different types of BoN found.

A bottom-up effective theory approach, as the one presented in Section~\ref{sec:BotUp}, lacks input from the higher-dimensional theory, which is ultimately responsible for the values of free parameters in the effective $4d$ description. This gap is closed in Section~\ref{sec:topdown} which shows how different compactification geometries and dimensions lead to the different types of BoN identified in Section~\ref{sec:BotUp}. Section~\ref{sec:quench} examines the critical limit in which BoNs turn into end-of-the world branes. Flux compactifications being of particular interest, we examine in Section~\ref{sec:BoNFlux} a particular flux BoN solution proposed in the literature showing how it can be described in terms of a tunneling potential in a multifield context.  We provide a summary and outlook in Section~\ref{sec:concl}. 

Finally, we have relegated further details of our work to several appendices. Appendix~\ref{App:E} deals with zero-energy considerations for the BoN as seen in $4d$.
Appendix~\ref{App:ActionAgrees} shows that the simple action calculation in the $V_t$ formalism reproduces the Euclidean result for all the different types of BoN discussed in Section~\ref{sec:BotUp}. Appendix~\ref{App:Vexp} analyzes in some detail the possible $V_t$ solutions for an exponential potential $V=V_A e^{a\sqrt{6\kappa}\phi}$, while Appendix~\ref{App:Vtexp} discusses which potentials would admit BoN decays described by a $V_t$ that is a simple exponential, $V_t=-e^{a\sqrt{6\kappa}\phi}$. Appendix~\ref{App:Vconstant} derives $V_t$ for the special case of a constant potential, $V=V_\infty>0$. Appendix~\ref{App:morex} presents more families of pairs of $V,V_t$ analytic examples. And, finally, Appendix~\ref{app:dSdA} derives useful formulas to calculate how the vacuum decay action depends on any parameter entering $V_t$ (and not $V$). A short paper with some of the main points developed here can be found in \cite{short}.

\section{Review of the Tunneling Potential Approach\label{sec:Vt}}

In this section we summarize the main features of the tunneling potential formalism, proposed in \cite{E,Eg}, to describe semiclassical false vacuum decay including the effects of gravitation. For simplicity we restrict ourselves to $4d$ single field theories, and refer the reader to \cite{EK,EF} for the generalisations to an arbitrary number of dimensions, $d\ge3$, and fields.

The tunneling potential approach reformulates the calculation of the tunneling action for the decay of a false vacuum at $\phi_\pp$ of a potential $V(\phi)$ in the following variational form: find the (tunneling potential) function $V_t(\phi)$, that goes from $\phi_\pp$ to some $\phi_0$ on the basin of the true vacuum\footnote{We assume $\phi_\mm>\phi_\pp$, so that $\phi_\pp<\phi_0< \phi_\mm$.} at $\phi_-$, and minimizes the action functional \cite{Eg}
\be
 S[V_t]=\frac{6\pi^2}{\kappa^2}\int_{\phi_\pp}^{\phi_0}d\phi\ \frac{(D+V_t')^2}{V_t^2 D}\ .
\label{SVt}
\ee
In this expression primes denote field derivatives, and
\be
D^2\equiv V_t'{}^2+6\kappa (V-V_t)V_t\ ,
\label{D2}
\ee
where $\kappa=1/m_P^2$, with $m_P$ the reduced Planck mass (that is, $m_P= 1/\sqrt{8 \pi G_4}$, with $G_4$ Newton's constant in $4d$). The method reproduces the Euclidean bounce result \cite{CdL} and has several good properties discussed elsewhere \cite{E,Eg,ESM,EK,EGrav}. 

\begin{figure}[t!]
\begin{center}
\includegraphics[width=0.48\textwidth]{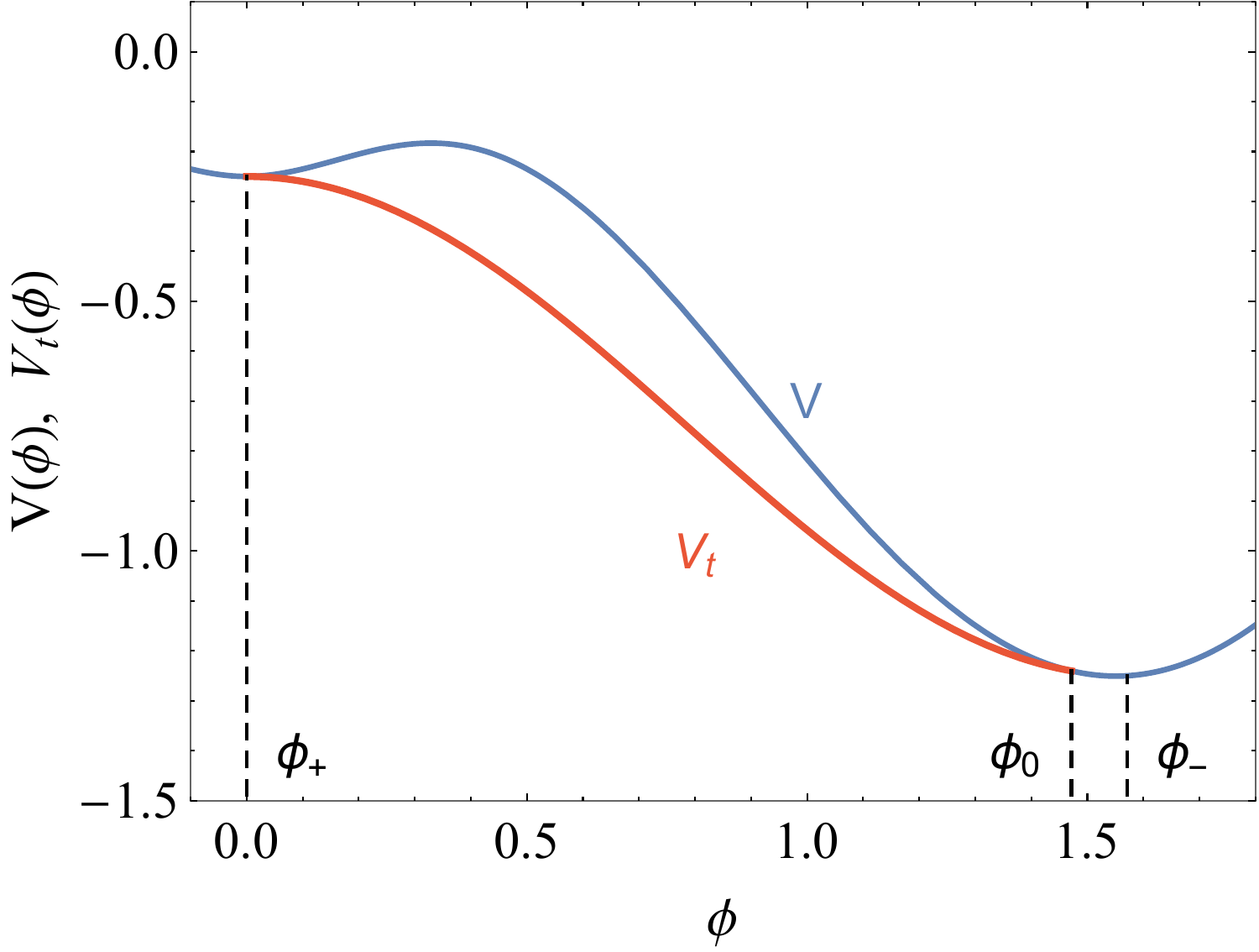}
\includegraphics[width=0.47\textwidth]{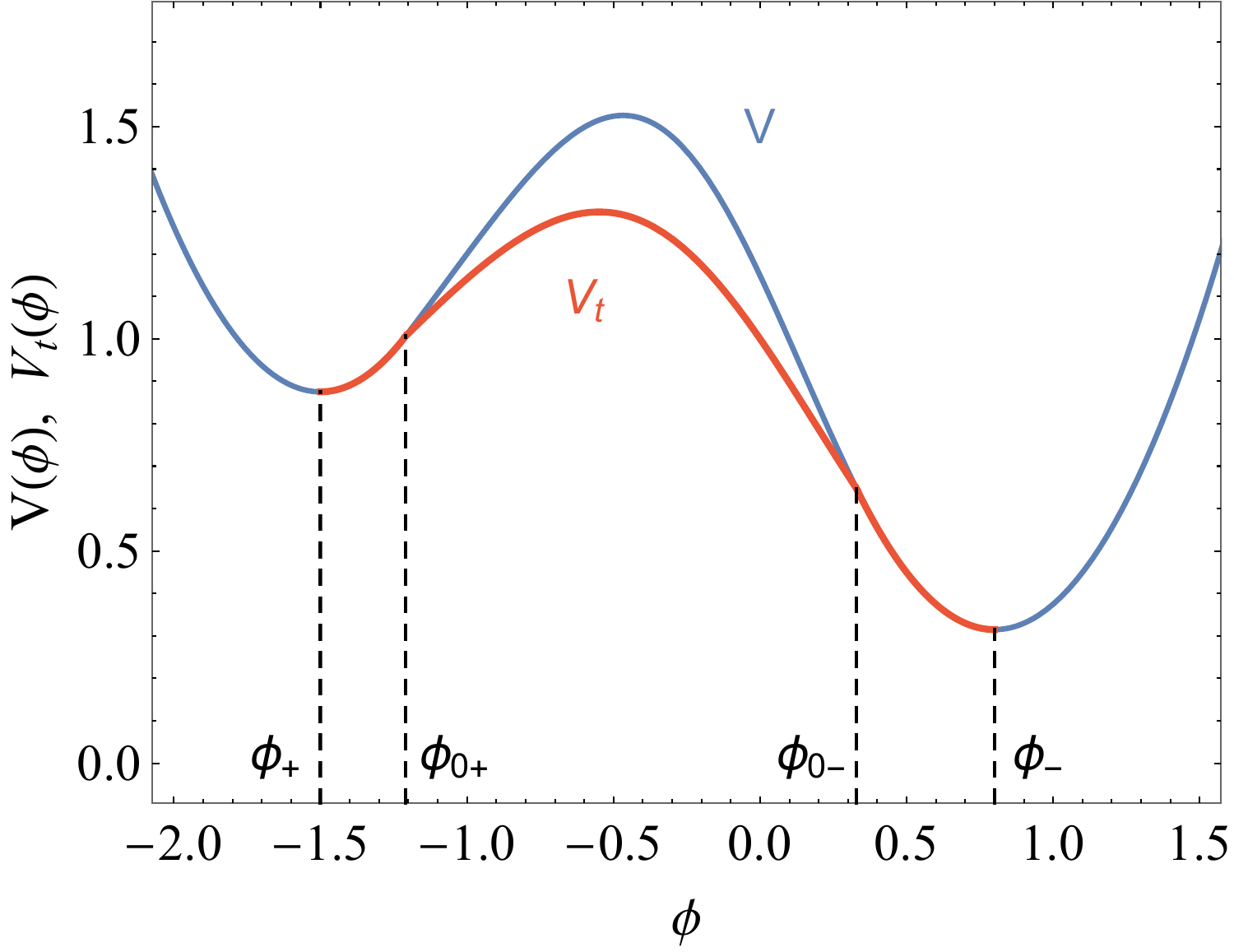}
\end{center}
\caption{Structure of the tunneling potential for the decay of a false AdS (or Minkowski) vacuum (left plot) or a dS vacuum (right plot).
\label{fig:Vttypes}
}
\end{figure}

The Euler-Lagrange equation, $\delta S[V_t]/\delta V_t=0$,  gives the ``equation of motion'' (EoM) for $V_t$:
\be
(4V_t'-3V')V_t' + 6(V-V_t)\left[V_t''+\kappa(3V-2V_t)\right]=0
\ ,
\label{EoMVt}
\ee
or, in terms of $D$,
\be
D'= \frac{\left(3V'-4\, V_t'  \right)}{6(V-V_t)}D\ .
\label{EoMVtD}
\ee
$V_t$ is qualitatively different depending on the potential value at the false vacuum  $V_+\equiv V(\phi_+)$:
\begin{itemize}

\item For $V_+\leq 0$ (decays of Minkowski or AdS false vacua), $V_t$ is monotonic with $V_t,V_t'\leq 0$, see Fig.~\ref{fig:Vttypes}, left plot, with boundary conditions
\be
V_t(\phi_+) = V(\phi_+), \quad V_t(\phi_0) = V(\phi_0), \quad
V_t'(\phi_+)=V'(\phi_+)=0,\quad V_t'(\phi_0)=\frac34 V'(\phi_0)\ ,
\label{MinkBCs}
\ee
where the field value $\phi_0$ is to be determined by the equations of motion and the previous boundary conditions.
As known from Coleman-De Luccia's work \cite{CdL}, for this type of false vacua, gravity can forbid decay (gravitational quenching, see discussion below). 

\item For $V_+>0$ (dS  vacua), $V_t$ is not monotonic and has the structure illustrated by Fig.~\ref{fig:Vttypes}, right plot.  In this case the field range covered by the bounce is $\phi \in (\phi_{0+},\phi_{0-})$, where $V_t<V$, and  the field values $\phi_{0\pm}\neq \phi_{\pm}$, are to be determined by the equation of motion and boundary conditions
\be
V_t(\phi_{0\pm}) = V(\phi_{0\pm}), \qquad V_t'(\phi_{0\pm})=\frac34 V'(\phi_{0\pm})\ .
\ee
The tunneling potential can be extended to the whole field range $(\phi_{+},\phi_{-})$ requiring that  away from the interval $(\phi_{0+},\phi_{0-})$ it satisfies $ V_t = V$.
The action for this decay splits in two contributions: a Hawking-Moss-like part from $\phi_\pp$ to $\phi_{0\pp}$ and a CdL-like part from $\phi_{0\pp}$ to $\phi_{0\mm}$ (the last part, from $\phi_{0\mm}$ to $\phi_\mm$ is zero)
\be
S=\frac{6\pi^2}{\kappa^2}\int_{\phi_\pp}^{\phi_\mm}\frac{(D+V_t')^2}{D V_t^2}=
\frac{24\pi^2}{\kappa^2}\left[\frac{1}{V(\phi_{\pp})}-\frac{1}{V(\phi_{0\pp})}\right]+\frac{6\pi^2}{\kappa^2}\int_{\phi_{0\pp}}^{\phi_\mm}\frac{(D+V_t')^2}{D V_t^2},
\label{SHMCdL}
\ee
As $V_\pp$ is increased  the range of the CdL interval shrinks to zero, $\phi_{0\pp}, \phi_{0\mm}\to\phi_B$ (the top of the barrier field value) there is no CdL decay, and the action tends to the Hawking-Moss one \cite{Eg}.
 
\end{itemize}

The $V_t$ formulation is ideally suited to study the gravitational quenching effect discussed above. To have a real tunneling action, $V_t$ should satisfy
\be
D^2 = V_t'{}^2+6\kappa (V-V_t)V_t>0\ ,
\label{quench}
\ee
(except at the false vacuum, point at which $D=0$)
and gravitational quenching occurs if this condition cannot be satisfied for any $V_t$ \cite{Eg}. For Minkowski or AdS vacua the second term in (\ref{quench}) is negative and, when gravitational effects are important (akin to a large $\kappa$), it might  be impossible to satisfy (\ref{quench}) for any $V_t$, in which case the potential is stabilized \cite{EGrav}.
The condition $D^2>0$ can be rewritten
as 
\be
V_t'< -\sqrt{6\kappa (V-V_t)(-V_t)}\ .
\label{newmono}
\ee
In other words, for AdS or Minkowski vacua, $V_t$ has to satisfy a condition stronger than mere monotonicity (which is recovered for $\kappa\rightarrow 0$). 

It is useful to introduce the function $\overline{V_t}$ (that we call the critical tunneling potential) as the solution to  $D\equiv 0$ with
\be
\overline{V_t}'= -\sqrt{6\kappa (V-\overline{V_t})(-\overline{V_t})}\ ,
\label{Vtc}
\ee
and boundary condition $\overline{V_t}(\phi_\pp)=V(\phi_\pp)\equiv V_\pp$.
Solutions of (\ref{Vtc}) with different values of $\overline{V_t}(\phi_\pp)$ span a family of non-intersecting integral curves for $D=0$. In order to have $D^2>0$, $V_t$ should have slope steeper than the $\overline{V_t}$ lines, from (\ref{newmono}), and cannot cross them from below. As a result, the  $V_t$ associated to the decay of $\phi_\pp$ [with $V_t(\phi_\pp)=V(\phi_\pp)$] must lie below the $\overline{V_t}$ line that leaves from the false vacuum.

Depending on the strength of gravitational effects, three cases are possible.
\begin{itemize}
\item {\it Subcritical case:} For weak gravity effects, $\overline{V_t}$ deviates somewhat from being horizontal and intersects $V$ at some $\phi_c<\phi_\mm$. This leaves room for $V_t<\overline{V_t}$ to intersect the $D=0$ integral lines from above and reach $V$ at some $\phi_0$ satisfying $D^2> 0$. Gravity makes the false vacuum more stable \cite{EGrav} but does not forbid its decay.  

\item {\it Supercritical case:} Strong gravity effects curve down $\overline{V_t}$ away from $V$ so that it never intersects it (except at $\phi_\pp$). As $V_t<\overline{V_t}$, there are no viable $V_t$'s with real $D$ and vacuum decay is forbidden by gravity. In such cases the impossibility of the decay can often be traced back to a positive energy theorem, which sets a lower bound to the  bubble-wall tension  (see e.g. \cite{Cvetic:1992st}).

\item
{\it Critical case:} $V_t\equiv \overline{V_t}$ solves  (\ref{EoMVtD}) with $D\equiv 0$ and has the right boundary condition $V_t'=V'=0$ at $\phi_0=\phi_\mm$. The tunneling action is infinite, and gravity forbids the decay of $\phi_\pp$ into $\phi_\mm$. $V_t$ describes a flat and static domain wall interpolating between false and true vacua \cite{Cvetic}. Solving $D\equiv 0$ for $V$ we see that any potential made critical by gravity (regarding vacuum decay) has the generic form
\be
V_c(\phi)=V_t-\frac{V_t'{}^2}{6\kappa V_t}\ ,
\label{Vc}
\ee
for some monotonic function $V_t(\phi)$.\footnote{Using $\mathcal{W}(\phi) \equiv \sqrt{-V_t(\phi)}$, one gets 
$V_c (\phi)  = \frac{2}{3\kappa}\left(\mathcal{W}'{}^ 2- \frac{3}{2} \kappa \mathcal{W}^2\right)$.
Scalar potentials with such structure can be found in supergravity models, and in the framework of fake-supergravity \cite{Freedman:2003ax}, where $\mathcal{W}$ can be readily identified  as the superpotential.}  
\end{itemize}

For later use we give now a dictionary between the Euclidean and tunneling potential formalisms to translate results between the two. In the Euclidean approach, false vacuum decay is described by an $O(4)$-symmetric bounce configuration $\phi(\xi)$, that extremizes the Euclidean action, and a metric function, $\rho(\xi)$, for the $O(4)$-symmetric Euclidean metric 
\be
ds^2= d\xi^2 +\rho(\xi)^2 d\Omega_3^2\ .
\ee
Here $\xi$ is a radial coordinate and $d\Omega_3^2$ is the line element on a unit three-sphere. 

The key relation between both formalisms is
\be
V_t (\phi)= V(\phi) -\frac12 \dot\phi^2\ ,
\label{Vtlink}
\ee
where $\dot x\equiv dx/d\xi$, and $\dot\phi$ is expressed in terms of the field via the bounce profile $\phi(\xi)$. Using the Euclidean EoM for $\phi$ we get
\be
\dot\phi = - \sqrt{2(V-V_t)}\ ,\quad \ddot\phi=V'-V_t'\ ,
\label{dphi}
\ee
where the minus sign for $\dot\phi$ follows from our convention $\phi_\pp<\phi_\mm$, and
\be
\rho=\frac{3\sqrt{2(V-V_t)}}{D}\ ,\quad 
\frac{\dot\rho}{\rho}=\frac{-V_t'}{3\sqrt{2(V-V_t)}}\ ,\quad
\frac{\ddot\rho}{\rho} = -\frac{\kappa}{3}\ (3V-2V_t)\ .
\label{rho}
\ee
Knowing $V_t$, both the field profile and the metric function can be derived from it using the previous formulas.

Finally, the $V_t$ approach is also quite convenient to deal with a class of decay modes  that are not extremals of the action and were called pseudo-bounces in \cite{PS} for that reason.
A pseudo-bounce solution would be an extremal of the action only if  $\phi_0$, the end point of the tunneling interval, is held fixed.\footnote{Thus, they are a special type of constrained instanton \cite{CI}.} These solutions have actions larger than the CdL one and are therefore generically subleading, but can become relevant when the CdL solution is ``pushed to infinity'' \cite{PS} (that is, the action has a runaway direction in field configuration space). In such case, vacuum decay is driven by non-CdL configurations and dominated by pseudo-bounces tracking the bottom of a sloping-valley in field configuration space. The tunneling potential method gets such pseudo-bounce solutions by solving (\ref{EoMVt}) with the modified boundary condition $V_t'(\phi_0)=0$ [while the CdL instanton satisfies $V_t'(\phi_0)=3V'(\phi_0)/4$].\footnote{This difference is connected with the fact that, in Euclidean formalism, pseudo-bounces have an inner core where the field takes a constant value $\phi_0$ out to some finite $\rho=\lim_{\phi\to\phi_0}3\sqrt{2(V-V_t)}/D\neq 0$, see \cite{PS}. }

\section{Witten's Bubble of Nothing\label{sec:WBoN}}

In order to show how the $V_t$ formalism can be used to study BoN solutions, we start with the simple BoN first discussed by Witten \cite{BoN}, for $5d$ Kaluza-Klein theory. After reviewing this solution in its $5d$ formulation, we derive its $4d$ CdL reduction, and then show how does the BoN solution look like in $V_t$ language.

\subsection{\texorpdfstring{$\bma{5d}$}{5d} Analysis}

Consider $5d$ KK spacetime, ($4d$ Minkowski $\times S^1$), which is unstable against semiclassical decay via the tunneling nucleation of a BoN, described by the Euclidean metric
\be
ds^2= \frac{dr^2}{1-\mathcal{R}^2/r^2}+r^2d\Omega_3^2 + R_{KK}^2 \left(1-\mathcal{R}^2/r^2\right)d\theta_5^2\ ,
\label{BoNmetric}
\ee
where $R_{KK}$ is the KK radius, $\mathcal{R}$ is the size of the bubble at the time of nucleation, $r\in [\mathcal{R},\infty)$, and $\theta_5\in [0,2 \pi)$ parametrises the KK circle. For $r\to\infty$ this metric approaches the KK vacuum $\mathbb{M}^4\times S^1$.  By continuing the Euclidean metric (\ref{BoNmetric}) to Minkowski space, Witten showed that this instanton solution describes the tunneling from a homogeneous KK spacetime to a spacetime in which the radius of the 5th dimension shrinks to zero as $r\to \mathcal{R}$.
Therefore, this spacetime has a ``hole'', or bubble of nothing, at $r=\mathcal{R}$ when the BoN is nucleated, which subsequently expands (with radius $\sqrt{\mathcal{R}^2+t^2}$, and $t\in [0,\infty$) ) and destroys the KK spacetime.  Nevertheless, provided we require the bubble nucleation and KK radii to be equal, this spacetime is regular and geodesically complete. Indeed, near $r=\mathcal{R}$, the metric above is smooth when the condition $\mathcal{R}=R_{KK}$ holds: writing $r=\mathcal{R}+\alpha^2/(2\mathcal{R})$  one gets $ds^2\simeq d\alpha^2+ \alpha^2 d\theta^2_5+R^2_{KK}d\Omega_3^2$. This ``gravitational bounce'' metric is an extremum of the $5d$ Euclidean action with a single negative eigenmode, as expected for a decay-mediating bounce.

The rate per unit volume for this decay process is $\Gamma/V\sim e^{-\Delta S_E}$, where $\Delta S_E$ is the difference between the Euclidean action of the bounce and the KK vacuum. The Euclidean action difference reads
\be
\Delta S_E=-\frac{1}{16\pi G_5}\int d^5x\sqrt{g}R_5-\frac{1}{8\pi G_5}\int
d^4x (K_4-K_{40})\sqrt{h}\ ,
\ee
where $G_d$ is Newton's constant in $d$ dimensions, with $G_5=2\pi R_{KK} G_4$ and $R_5$ is the $5d$ Ricci scalar. In the integral over the boundary ($r\to\infty$), $h$ is the determinant of the boundary induced metric, $K_4$ is the trace of the second fundamental form of the boundary, and $K_{40}$ represents the latter quantity when the boundary is embedded in vacuum.  For this BoN solution 
\be
R_5=0\ , \quad K_4=\frac{1}{r}\left(2\sqrt{1-R^2_{KK}/r^2}+\frac{1}{\sqrt{1-R^2_{KK}/r^2}}\right)\ ,\quad  K_{40}=\frac{3}{r}\ ,
\ee
and one gets the finite tunneling Euclidean action 
\be
\Delta S_E=(\pi m_P R_{KK})^2\ .
\label{SBoN5D}
\ee

 For later convenience we introduce here an alternative gauge to write the BoN line element \eqref{BoNmetric}, which is particularly useful to study instantons describing the decay of more general compactifications (see section \ref{sec:topdown}) 
\be
ds^2= d  \alpha^2 +  \mathcal{R}^2 B(\alpha)^2 d\Omega_3^2 + R_{KK}^2 C(\alpha)^2 d\theta_5^2.
\label{eq:WittenWgauge}
\ee 
Here the new radial coordinate takes values in the range $\alpha\in[0,\infty)$.
In this gauge the Witten bubble solution becomes
\be
  B(\alpha) = \sqrt{1 + \alpha^2/\mathcal{R}^2}\ ,     \qquad C(\alpha) = \frac{\alpha}{\sqrt{\alpha^2+\mathcal{R}^2}}\ .
\ee

\subsection{\texorpdfstring{$\bma{4d}$}{4d} Dimensional Reduction to a CdL Bounce\label{sec:4dReduction}}

To reduce the BoN solution \eqref{BoNmetric} to a $4d$ effective description, we integrate the 5th dimension $\theta_5$, and introduce the scalar modulus field $\phi$ with
\be
e^{-2\sqrt{2\kappa/3}\,\phi}\equiv 1-\frac{R^2_{KK}}{r^2}\ .
\ee
This maps the BoN instanton (\ref{BoNmetric})
into a field profile, $\phi(r)$, with $r\to R_{KK}$ (the BoN core) corresponding to $\phi\to\infty$ and $r\to\infty$ (the KK vacuum) to $\phi\to 0$.

Next we perform a Weyl rescaling of the $4d$ metric 
\be
g_{\mu\nu}\rightarrow g_{\mu\nu}\, e^{\sqrt{2\kappa/3}\,\phi}\ . 
\ee
The resulting tunneling Euclidean $4d$ action is
\bea
\Delta S_E &=& -\frac{1}{16\pi G_4}\int d^4 x \sqrt{g} \left(R_4-\sqrt{6\kappa}g^{\mu\nu}\nabla_{\mu\nu}\phi-\kappa g^{\mu\nu}\partial_\mu\phi\partial_\nu\phi\right)\nonumber\\
&-&\frac{1}{8\pi G_4}\int d^3x \sqrt{h} \left(K_3-K_{30}+\sqrt{\kappa g^{rr}/6}\,\partial_r\phi\right)\ ,
\eea
where we keep the total derivative term $g^{\mu\nu}\nabla_{\mu\nu}\phi$ as we have a boundary to care about.

The same reduction and Weyl rescaling transform the BoN metric into
\be
ds^2=\frac{dr^2}{\sqrt{1-R^2_{KK}/r^2}}+r^2\sqrt{1-R^2_{KK}/r^2}\,d\Omega_3^2\ ,
\label{4DBoN}
\ee
which can be written in the form of a Coleman-De Luccia (CdL) bounce metric
\be
ds^2=d\xi^2+\rho(\xi)^2d\Omega_3^2\ ,
\label{BoNCdL}
\ee
with the identifications
\be
\frac{d\xi}{dr}\equiv \frac{1}{(1-R^2_{KK}/r^2)^{1/4}}\ ,\quad
\rho(\xi)^2\equiv r^2\sqrt{1-R^2_{KK}/r^2}\ .
\ee
Now, $\xi\to 0$ corresponds to the BoN core (with $\phi\to\infty$ and $\rho=0$) and $\xi\to\infty$ to the KK vacuum (with $\phi=0$ and $\rho\to\infty$). We see that this CdL solution is not of the standard form describing vacuum decay as the field diverges at the bounce core\footnote{At $\xi\to\infty$ (or $r\to \infty$) we have
$\rho=r+{\cal O}(1/r)$, $\dot\rho=1+{\cal O}(1/r^4)$,
$\dot\phi=-\sqrt{3/(2\kappa)}R^2_{KK}/r^3+{\cal O}(1/r^5)$. At $\xi\to 0$ [or $r=R_{KK}(1+\epsilon)$] we have
$\rho  = (2\epsilon)^{1/4}R_{KK}+{\cal O}(\epsilon^{5/4})$,
$\dot\rho  =  (8\epsilon)^{-1/2}+{\cal O}(\epsilon^{1/2})$,
$\dot\phi  =  -\sqrt{3/(2\kappa)}/[R_{KK}(2\epsilon)^{3/4}]+{\cal O}(\epsilon^{1/4})$.\label{foot}} and so does the $4d$ curvature. Nevertheless, it inherits some good properties due to its $5d$ UV origin; in particular, its Euclidean action is finite and equal to (\ref{SBoN5D}). 

To show this, rewrite the $4d$ Euclidean action in terms of $\xi$, $\rho$ and $\phi$ as
\bea
\Delta S_E &=& 2\pi^2\int_0^\infty d\xi\, \rho^3\left[-\frac{R_4}{2\kappa}+\frac12 \dot\phi^2+\frac{1}{\sqrt{6\kappa}}\left(\ddot\phi+\frac{3\dot\rho}{\rho}\dot\phi\right)\right]\nonumber\\
&-&\frac{2\pi^2}{\kappa} \lim_{\xi\to\infty} \rho^2\left[3(\dot\rho-1) +\sqrt{\frac{\kappa}{6}}\rho\dot\phi\right]\ ,
\eea
where dots stand for $\xi$ derivatives. 
For the solution (\ref{BoNCdL}) we have
\be
R_4=\frac{6}{\rho^2}(1-\rho\ddot\rho-\dot\rho^2)\ ,\quad
K_3=3\frac{\dot\rho}{\rho}\ , \quad
K_{30}=\frac{3}{\rho}\ ,
\ee
leading to
\be
-\frac{1}{2 \kappa}R_4+\frac12 \dot\phi^2 =0\ ,\quad \ddot\phi+\frac{3\dot\rho}{\rho}\dot\phi=0\ ,
\ee
which shows that the bulk part of the action vanishes (with divergent quantities cancelling out). The boundary term (using the asymptotic behaviour in footnote \ref{foot}) is
\be
\Delta S_E=-\lim_{\xi\to\infty}2\pi^2\frac{\rho^2}{\kappa}\left[3(\dot\rho-1) +\sqrt{\frac{\kappa}{6}}\rho\dot\phi\right]= (\pi m_P R_{KK})^2 \ ,
\ee
which agrees with the $5d$ result (\ref{SBoN5D}).\footnote{This also agrees with the result of \cite{DGL} $\Delta S_E=-\pi^2m_P\sqrt{2/3}\rho^3\dot\phi|_{\xi=0}$ (which we rewrite for our definition of $\phi$ with opposite sign). To make contact with our expression above, simply note that $\rho^2(\dot\rho-1)|_{\xi\to\infty}=0$ implies that $\Delta S_E=-(2\pi^2/\sqrt{6\kappa})\rho^3\dot\phi|_{\xi\to\infty}=-(2\pi^2/\sqrt{6\kappa})\rho^3\dot\phi|_{\xi=0}$, where the last equality follows from $d(\rho^3\dot\phi)/d\xi=0$.}  

One can try to rewrite the $4d$ action in the standard CdL form by moving the boundary term to the bulk (as a total derivative term) but one should pay attention to the fact that the boundary term does not vanish at $\xi\to 0$. Indeed, using footnote \ref{foot}, we find
\be
-\lim_{\xi\to 0}2\pi^2\frac{\rho^2}{\kappa}\left[3(\dot\rho-1) +\sqrt{\frac{\kappa}{6}}\rho\dot\phi\right]= -2(\pi m_P R_{KK})^2 \ .
\ee
This rewriting leads to the bounce action
\be
\Delta S_E = 2\pi^2\int_0^{\xi_{max}} d\xi\ \rho^3\left\{\frac12\dot\phi^2+V-\frac{3}{\kappa\rho^2}(1-\dot\rho)^2-\frac{1}{\kappa}\delta(\xi)
\left[\frac{3(\dot\rho-1)}{\rho}+\sqrt{\frac{\kappa}{6}}\dot\phi\right]
\right\}\ .
\label{SEfinal}
\ee
For later use, we have added a potential term, (which is zero for Witten's BoN) and replaced the upper limit of the integral by $\xi_{max}$ (which is $\infty$ for decays from AdS or Minkowski, as for Witten's BoN, but is finite for decays from a dS false vacuum).

The equations of motion derived from this action are not affected by the 
delta function term and reproduce the CdL ones, which read [including a potential $V(\phi)$]:
\bea
&&\ddot \phi+\frac{3\dot\rho}{\rho}\dot\phi = V'\ ,\label{E1}\\
&& \dot\rho^2 =1+ \frac{\kappa}{3}\rho^2\left(\frac12\dot\phi^2-V\right)\ ,
\label{E2}
\eea
where dots (primes) stand for derivatives with respect to $\xi$ ($\phi$).
Using these equations of motion we can further massage the action and write it in the even simpler form
\be
\Delta S_E=-2\pi^2\int_0^{\xi_{max}} \rho^3 V d\xi -\left.\pi^2\sqrt{\frac{2}{3\kappa}}\rho^3\dot\phi\,\right|_{\xi=0}\ ,
\label{FinalSE}
\ee
where we have used $\rho^2(1-\dot\rho)|_{\xi=\xi_{max}}=0$ to simplify the final expression. The result (\ref{FinalSE}) takes the standard CdL form (for decays from a Minkowksi false vacuum), except for the additional term evaluated at $\xi=0$, which is a purely $5d$ input. For similar considerations regarding the energy of the nucleated tunneling bubble from the $4d$ point of view, see appendix~\ref{App:E}.

\subsection{Tunneling Potential Approach to Witten's BoN}

For the $5d$ Kaluza-Klein vacuum, moving from the $4d$ reduction to the tunneling potential approach is straightforward. Simply use the relation with the Euclidean 
CdL formalism in Eq.~(\ref{Vtlink}) and rewrite the result
as a function of the field. This leads to the simple expression
\be
V_t(\phi) = -\frac{6}{\kappa R^2_{KK}}\sinh^3(\sqrt{2\kappa/3}\,\phi)\ .
\label{WBoN}
\ee 
In a potential $V=0$ there is no proper CdL vacuum decay and this tunneling potential describes something different (a BoN).
In particular, the boundary conditions satisfied by $V_t$ are
\be
V_t(0)=0\ , \quad V_t(\phi\rightarrow\infty)\sim - e^{\sqrt{6\kappa}\phi}\ ,
\ee
so that $V_t$ diverges at $\phi\rightarrow \infty$. This is a generic property of the $V_t$'s that describe BoNs. 

Concerning the action calculation in the $V_t$ formalism, the standard formula assumes that there are no contributions from boundary terms. When one redoes the calculation paying attention to such terms, the end result turns out to be the same, and moreover, there is no need to add any boundary  term to the action
as done in the CdL formalism. Therefore one can use directly Eq.~(\ref{SVt}). 
Using
\be
D(\phi)=\frac{6}{R_{KK}^2}\sqrt{\frac{6}{\kappa}}\sinh^2(\sqrt{2\kappa/3}\phi)\ ,
\ee
the action density $s(\phi)$ takes the simple form
\be
s(\phi)=\frac{\pi^2R_{KK}^2}{2}\sqrt{\frac{3}{2\kappa}}\mathrm{sech}^4(\sqrt{\kappa/6}\,\phi)\ ,
\ee
is finite everywhere and integrates to the correct result:
\be
S[V_t]=\int_0^\infty s(\phi)d\phi = (\pi m_P R_{KK})^2\ ,
\label{SWBoN}
\ee
without having to include additional terms as in the Euclidean approach.
The agreement between the simple action $S[V_t]$ given by (\ref{SVt}) and the Euclidean action (\ref{FinalSE}) holds in general, not only for Witten's BoN. We give the proof of this remarkable fact in Appendix~\ref{App:ActionAgrees}.

\section{BoNs with Nonzero Potential. Bottom-up Analysis\label{sec:BotUp}}

In this section we consider how a nonzero scalar potential for the modulus field, $V(\phi)$, needed to stabilize the extra dimensions, can affect the existence and shape of the BoN. In the same spirit of \cite{DGL}, we derive the conditions that $V(\phi)$ must satisfy asymptotically to allow for BoN decays (using its interplay with the asymptotic behaviour of $V_t$). In this section we make no assumptions about the possible origin of $V(\phi)$, issue discussed in Section~\ref{sec:topdown}, and we simply identify different types of asymptotic behaviours of $V$ and $V_t$ compatible in principle with the existence of BoN solutions. The use of the tunneling potential for this purpose is quite convenient: instead of using the BoN profile and metric function, a single function $V_t(\phi)$, which is on the same footing as the potential, captures the key asymptotic behaviour in a simple way\footnote{The tunneling potential approach has been used for similar purposes  in the study of dynamical cobordisms and end-of-the-world branes in \cite{DynCo} with similar advantages.}. Moreover, the $V_t$ formalism can be used to easily generate analytic examples of potentials admitting BoN decays. We give a number of such analytic potentials in section \ref{sec:anex} to illustrate the different types of asymptotic behaviour that we find as well as their interplay. 

In what follows, we assume that the BoN vacuum decay happens towards $\phi=\infty$, the compactification limit\footnote{This convention is opposite to that in \cite{DGL}, which has $\phi=-\infty$ for that compactification limit. We also use a different normalization for the constant $a$ below, with ours being a factor $\sqrt{6}$ smaller.}, which corresponds to the core of the BoN, $\phi(\xi\to 0)$.

\subsection{General Asymptotics\label{genasy}}

The tunneling potential  $V_t$ describing a BoN decay is
a solution of the  differential equation (\ref{EoMVt}) [the Euler-Lagrange equation from the extremality of the action (\ref{SVt})]
\be
(4V_t'-3V')V_t' + 6(V-V_t)\left[V_t''+\kappa(3V-2V_t)\right]=0\ ,
\label{EoM}
\ee
with the same boundary conditions at the false vacuum $\phi_+$ (or $\phi_{0+}\neq \phi_+$ for the dS case) as for standard vacuum decay (see section \ref{sec:Vt}), but with unusual boundary conditions at $\phi\to\infty$  [$V_t(\phi\to\infty)\to-\infty$], as shown in the previous section. 

To determine which boundary conditions are compatible with  \eqref{EoM}  for $\phi \to \infty$, we have studied the asymptotic properties of $V_t$ (relative to those of $V$) using the equation of motion, as discussed below.\footnote{For a more detailed discussion of the freedom in choosing boundary conditions (both at the false vacuum and at $\phi \to \infty$) as needed to determine a solution of the differential equation \eqref{EoM}, see section~\ref{sec:BCs}.} We have classified the allowed boundary conditions in four different types depending on the behaviour of $\lim_{\phi\to\infty}V/|V_t|$, and we show numerical and analytic examples of these types of BoN in later sections. The different types are

\begin{itemize}

\item {\bf Type 0}: $V$ is subdominant
with respect to $V_t$ at $\phi\to\infty$, so that $\lim_{\phi\to\infty}V/|V_t|=0$. In this case, eq.~(\ref{EoM}) gives (see Appendix~\ref{App:Vexp} for more details)
\be
V_t(\phi\to\infty) \sim V_{tA} e^{\sqrt{6\kappa}\phi}\ ,
\label{asy0}
\ee 
with $V_{tA} <0$.
This holds whether $V$ is positive or negative at $\phi\to\infty$ and we must choose the negative sign for $V_t$ as $V_t\leq V$. As $V$ is irrelevant in the limit $\phi\to\infty$, this type of BoN  behaves as Witten's BoN. Indeed, Witten's $V_t$ in (\ref{WBoN}) conforms to (\ref{asy0}).

\item {\bf Types ${\bma +}$ and ${\bma -}$}: We can assume instead that $V$ and $V_t$ are of comparable size at $\phi\to\infty$, so that $\lim_{\phi\to\infty}V/|V_t|$ is a constant. We call such type $+$ or $-$ according to the sign of $\lim_{\phi\to\infty}V/|V_t|$. Writing\footnote{Arguments similar to those in the text show that other simple asymptotic behaviours, like $V,V_t\sim  e^{a\kappa\phi^2}$, etc. are not possible. However, $V,V_t\sim  \phi^n e^{a\sqrt{\kappa}\phi}$ do in principle occur in some special instances.} $V(\phi\to\infty)\sim V_A e^{a \sqrt{6\kappa}\phi}$ and $V_t(\phi\to\infty)\sim V_{tA} e^{a \sqrt{6\kappa}\phi}$, with $a>0$ and $V_{tA}<0$, eq.~(\ref{EoM}) gives the condition
\be
[V_A+(a^2-1)V_{tA}](3V_A-2V_{tA})=0\ .
\label{asymptcond}
\ee
To satisfy (\ref{asymptcond}), the first possibility is that $V_{tA}=V_A/(1-a^2)$.  If $V_A<0$ (type $-$)
then the condition $V_t<V$ implies $a<1$, while $V_A>0$ (type $+$) is allowed with $a>1$. 

\item {\bf Type ${\bma -^*}$:} The second possibility to satisfy (\ref{asymptcond}) is  $V_{tA}=3V_A/2$. As $V_t\leq V$ must hold, in this case there can be a BoN decay described by $V_t$ only if $V_A<0$ so that this case would be of type $-$, but we call it {\bf type ${\bma -^*}$} to distinguish it from the previous $-$ type. 
 In principle, the case with $a=1$ could be of this type. For this  type of BoN, the gravitational term $\kappa (3V-2V_t) $ in (\ref{EoM}) vanishes asymptotically for $\phi\to\infty$.  
 \end{itemize}
 The asymptotic behavior of the four different types of BoN is summarized in Table~\ref{table:types}. Notice, in particular, that the properties of type 0 solutions can be obtained from types $\pm$ as a limiting case with $V_A\to 0$ (subleading $V$) and $a\to 1$ (for $V_t$).   
 

\begin{table}
\begin{center}
\begin{tabular}{cccccc}
 Type & $V_t(\phi\to\infty)$ &  Constraints & $\beta$ & $D(\phi\to\infty)$ & UV
\\
\hline
0  & $V_{tA}e^{\sqrt{6\kappa}\phi}$ & $V_{tA}<0$ , $a<1$ & $1/3$ & $e^{\sqrt{8\kappa/3}\phi}$ & $S^1$ 
\\
$-$  & $ V_{A}/(1-a^2)e^{a\sqrt{6\kappa}\phi}$ & $V_A<0$\ ,\, $1/\sqrt{3}<a< 1$ & $1/(3a^2) $ & $e^{\sqrt{\kappa/6}(3a+1/a)\phi}$ & $S^d$
\\
$+$ & $ V_{A}/(1-a^2)e^{a\sqrt{6\kappa}\phi}$ & $V_A>0$\ ,\, $a> 1$ & $1/(3a^2) $ & $e^{\sqrt{\kappa/6}(3a+1/a)\phi}$ & Sing.
\\
\, $-^*$ &   $
(3V_{A}/2)e^{a\sqrt{6\kappa}\phi}$ & $V_A<0$\ , $a>1/\sqrt{3} $ & $1$ & $e^{a\sqrt{6\kappa}\phi}$ & {\rm Sing.}
\end{tabular}
\caption{\it Taking $V(\phi\to\infty)=V_{A}e^{a\sqrt{6\kappa}\phi}$ we show, for the four different types discussed in the text: the asymptotic behaviours at $\phi\to\infty$ of the tunneling potential, $V_t(\phi)$,  and the quantity $D(\phi)$; several constraints on their parameters; the exponent $\beta$ entering $\rho(\xi\to 0)\sim \xi^\beta$; and their possible higher dimensional origin (see section \ref{sec:topdown}), indicating the geometry of the compact space or the need for a UV object to avoid a singularity (label ``Sing.'').\label{table:types}}
\end{center}
\end{table}

From the  asymptotic behaviour  of $V$ and $V_t$ at $\phi\to\infty$, and using
the relations \eqref{Vtlink} and  \eqref{rho} (see Section~\ref{sec:Vt}),  we can derive the asymptotic behaviour of the Euclidean functions describing the BoN [the field profile $\phi(\xi)$ and the metric function $\rho(\xi)$] at $\xi\to 0$ .
In particular, the metric function $\rho(\xi)$ can be obtained using both the first and second formulae in \eqref{rho}, 
but the second one is more convenient in general, as $D$ vanishes (asymptotically) at leading order in some cases and should be computed with some care, see below.

For type $0$ cases, with subdominant $V$ at $\phi\to \infty$, we immediately get 
\be
\phi(\xi\to 0) \simeq -\sqrt{\frac{2}{3\kappa}}\log \left(\xi\sqrt{-3\kappa V_{tA}}\right)\ ,\quad
\rho(\xi\to 0)\simeq c_\rho \xi^{1/3}\ ,
\label{phirhotype0}
\ee
which shows that the instanton is singular in four dimensions, with the leading behaviour near the singularity determined by the parameter $V_{tA}$.  Although the constant $V_{tA}$ is undetermined in the $4d$ theory, it is fixed by the higher-dimensional theory, for example, as discussed in section \ref{sec:topdown}, by requiring the higher dimensional space-time to be regular.\footnote{Indeed, this is the behaviour of Witten's BON, for which $
\phi(\xi\to 0) \simeq -\sqrt{2/(3\kappa)}\log [3 \xi/(2 R_{KK})]$ and $
\rho(\xi\to 0)\simeq (3 R_{KK}^2/2)^{1/3} \xi^{1/3}$.
}
The $\rho$  prefactor, $c_\rho$, although it is also undetermined in our derivation,  given a particular value of $V_{tA}$, it could be computed from the subleading terms in $V_t$ or $V$ near $\phi\to \infty$ ($\xi \to 0$). The results found in (\ref{phirhotype0}) agree with those obtained in \cite{DGL} for a $5d$ theory with the extra dimension compactified in a circle. In that case, provided we impose regularity of the BoN space-time, the prefactor $c_\rho$ can be shown to encode another purely $5d$ quantity: the radius of the nucleated BoN \cite{DGL}. In order to get such a relation in our $V_t$ approach one also needs a top-down approach starting with the extra-dimensional theory, see section~\ref{sec:topdown}.

For the rest of cases, with $V(\phi\to \infty)\sim V_A e^{a\sqrt{6\kappa}\phi}$ and $V_t(\phi\to \infty)\sim V_{tA} e^{a\sqrt{6\kappa}\phi}$ we get
\be
\phi(\xi\to 0) \simeq -\frac{1}{a}\sqrt{\frac{2}{3\kappa}}\log \left[\xi a\sqrt{3\kappa (V_A-V_{tA})}\right]\ ,\quad
\rho(\xi\to 0)\simeq c_\rho\xi^\beta\ ,
\label{phirhotypepm}
\ee
with 
\be
\beta \equiv \frac{-V_{tA}}{3(V_A-V_{tA})}=\left\{
\begin{array}{ccl}
\frac{1}{3a^2}\ , & \mathrm{for\ types}\ +,-: & V_{tA}=V_A/(1-a^2)\ ,
\\
1\ , & \mathrm{for\ type}\; -^*: & V_{tA}=3V_A/2 \ .
\end{array} \right.
\label{betapm}
\ee
As in the previous case the bounce is singular, but here the asymptotic parameter $V_{tA}$ (and the prefactor of $\rho$, which depends on $V_{tA}$) is determined in the $4d$ theory, namely, by the asymptotic behaviour $V(\phi \to\infty)$. Therefore, in these cases we can use information about the higher-dimensional theory to \emph{constrain} the limiting behaviour  of the scalar potential $V$. In particular, imposing the BoN spacetime to be regular we find that the result for type $-$ agrees with that found in \cite{DGL} for a $4+d$ theory with $d>1$ dimensions compactified in a sphere. In that case, the regularity condition for the BoN instanton also determines $V_{tA}$, and it leads to a relation  between the prefactor $c_\rho$ and the bubble nucleation radius. Table~\ref{table:types} also compiles the previous information for the four different types of asymptotics we consider.

We can also check under what conditions the BoN action, given by the integral (\ref{SVt}), converges at $\phi\to\infty$ (and this can lead to additional constraints on the parameters).  For that purpose we need the asymptotic behaviour of $D(\phi\to\infty)$, which is generically subleading compared to $V_t$ and $V_t'$. To get the asymptotics of $D$ using $D^2=V_t'{}^2+6\kappa(V-V_t)V_t$ requires to know  all subleading terms in $V_t$ (and $V$) up to ${\cal O}(D)$.  This complication can be circumvented quite simply by resorting to the EoM for $V_t$ as a differential equation for $D$, as given in (\ref{EoMVtD}), using which the asymptotics of $D$ can be obtained just knowing the leading terms of $V$ and $V_t$. We  find the following:
\begin{itemize}
\item For {\bf type $0$} cases, $D$ vanishes at leading order ($e^{\sqrt{6\kappa}\phi}$) and, using (\ref{EoMVtD}), we get $D\sim e^{\sqrt{8\kappa/3}\phi}$. Plugging this into the action density $s(\phi)$ we find $s(\phi\to\infty)\sim e^{-\sqrt{8\kappa/3}\phi}$, whose integral is always convergent.

\item For {\bf type $+$ and $-$} cases, we again find that $D$ vanishes at leading order
and, calculated via (\ref{EoMVtD}) at subleading order, it is $D\sim e^{(1/a+3a)\sqrt{\kappa/6}\phi}$. For this result to be really subleading one needs $a>1/\sqrt{3}$, assuming which we get $s(\phi)\sim e^{-(1/a+3a)\sqrt{\kappa
/6}\phi}$, and thus a convergent action.

\item Finally, for {\bf type $-^*$} cases, we get $D\sim -3V_A\sqrt{\kappa(3a^2-1)/2}\, e^{a\sqrt{6\kappa}\phi}$ (so that $a>1/\sqrt{3}$ is required).
Plugging this $D$ asymptotics into the action density $s(\phi)$ we find $s(\phi\to\infty)\sim e^{-a\sqrt{6\kappa}\phi}$, whose integral is again convergent. In this case, $D$ is not subleading
and this allows to calculate the $\rho$ prefactor as $\rho\sim a\xi/\sqrt{a^2-1/3}$. 
\end{itemize}

Section~\ref{sec:BoNvsOther} presents a numerical analysis of decay channels for various potentials and clarifies the status and interplay of the different types of BoN solutions we have discussed while section \ref{sec:anex} adds several analytic examples of the different types of BoN. To complete the previous discussion, the asymptotic behaviours of $V_t(\phi)$ and $D(\phi)$ are studied, for a simple exponential potential $V(\phi)=V_A e^{a\sqrt{6\kappa}\phi}$
in Appendix~\ref{App:Vexp} and, for a simple exponential tunneling potential $V_t(\phi)=-e^{a\sqrt{6\kappa}\phi}$ in Appendix~\ref{App:Vtexp}, illustrating the four types of behaviour discussed above, with subleading terms explicitly obtained.

As we discuss in section \ref{sec:topdown}, both type 0 and type $-$ solutions can be uplifted to regular BoN solutions of a higher dimensional theory, provided certain restrictions are imposed on the parameters. Regarding type $+$ solutions, the asymptotic form of the potential is consistent with the presence of a higher dimensional flux on the internal dimension. Since for the BoN to be regular this flux has to be neutralised by some charged object \cite{BS}, we do not expect type $+$ solutions to represent the complete BoN geometry, as the potential would need to change near the BoN core.\footnote{See \cite{Draper:2023ulp} for a recent discussion on this subject using  a $4d$ perspective.} We discuss a concrete example of this behaviour in section~\ref{sec:BoNFlux}. Finally, we have not found a higher dimensional interpretation for type $-^*$ solutions.
Nevertheless, we do not discard type $-^*$ and $+$ BoN solutions as they might be relevant to describe bubbles of nothing with defects or compact geometries more complicated than spheres (like Calabi-Yau orientifolds) shrinking to zero through a defect \cite{Hebecker}.

\section{Interplay Between Boundary Conditions. Quenching of BoN Decay\label{sec:BCs}}

In this section we first clarify how $V_t$ solutions are determined by boundary conditions, both at low field values and at $\phi\to\infty$, and then discuss how the $\phi\to\infty$ asymptotics governs the possible gravitational quenching of BoN decays.

As $V_t$ is a solution of the second order differential equation (\ref{EoMVt}), it depends on two integration constants, typically obtained by fixing $V_t$ and $V_t'$ at some field value. For dS vacua, this is precisely the situation if we solve for $V_t$ starting at some initial value $\phi_i\neq\phi_\pp$ with $V_t(\phi_i)=V(\phi_i)$ and $V_t'(\phi_i)=3V'(\phi_i)/4$. (That is, $\phi_i$ is the starting point of the CdL range of the $V_t$ solution.) For Minkowski or AdS  false vacua, we start
with $V_t(\phi_\pp)=V(\phi_\pp)$ but $V_t'(\phi_\pp)=0$ does not fix completely the solution as $\phi_\pp$ is an accumulation point of an infinite family of solutions, and one needs to fix an additional constant to select a particular solution, see below. For the numerical exploration of vacuum decay solutions in the rest of the paper we solve the EoM for $V_t$ starting at low field values, using the low-field expansions derived below (including higher orders) for the low field boundary condition. We find this procedure to be more convenient than the opposite approach used in \cite{DGL} which starts at large field values and relies on the overshoot/undershoot method and interval splitting to find the solutions (both methods are, of course, complementary). Using our approach, as we show below, all starting boundary conditions correspond to a solution, be it a BoN, a CdL or a pseudo-bounce.\footnote{In other words, we could say that our solutions never undershoot or overshoot but are always on target.} We find out the type of solutions we get by looking at their asymptotic behaviour at large $\phi$, which we discuss for BoNs next.

Consider then a  BoN $V_t$ as $\phi\to\infty$. For the boundary value problem to be well defined (i.e.,  for the solution to be unique), it does no suffice to require $V_t$ to be divergent for large values of $\phi$, it is also necessary to specify the asymptotic behaviour of the tunneling potential in this limit.  For type 0 BoNs, the two integration constants in that field regime can be conveniently chosen to be $V_{tA}$ and $D_\infty$, the prefactors of the leading exponentials in $V_t$ and $D$, respectively. 
There is an interesting interplay between the boundary conditions satisfied by $V_t(\phi)$ at both ends of the field interval in which it is defined. 
In order to illustrate this, let us analyze this interplay for the simple toy potential 
\be
V(\phi)= V_\pp+\frac12 m^2\phi^2\ ,
\label{type0V}
\ee
(which is of type $0$), for false vacua of different kinds (Minkowski, dS or AdS). We comment on other types of BoN later on.

\begin{figure}[t!]
\begin{center}
\includegraphics[width=0.48\textwidth]{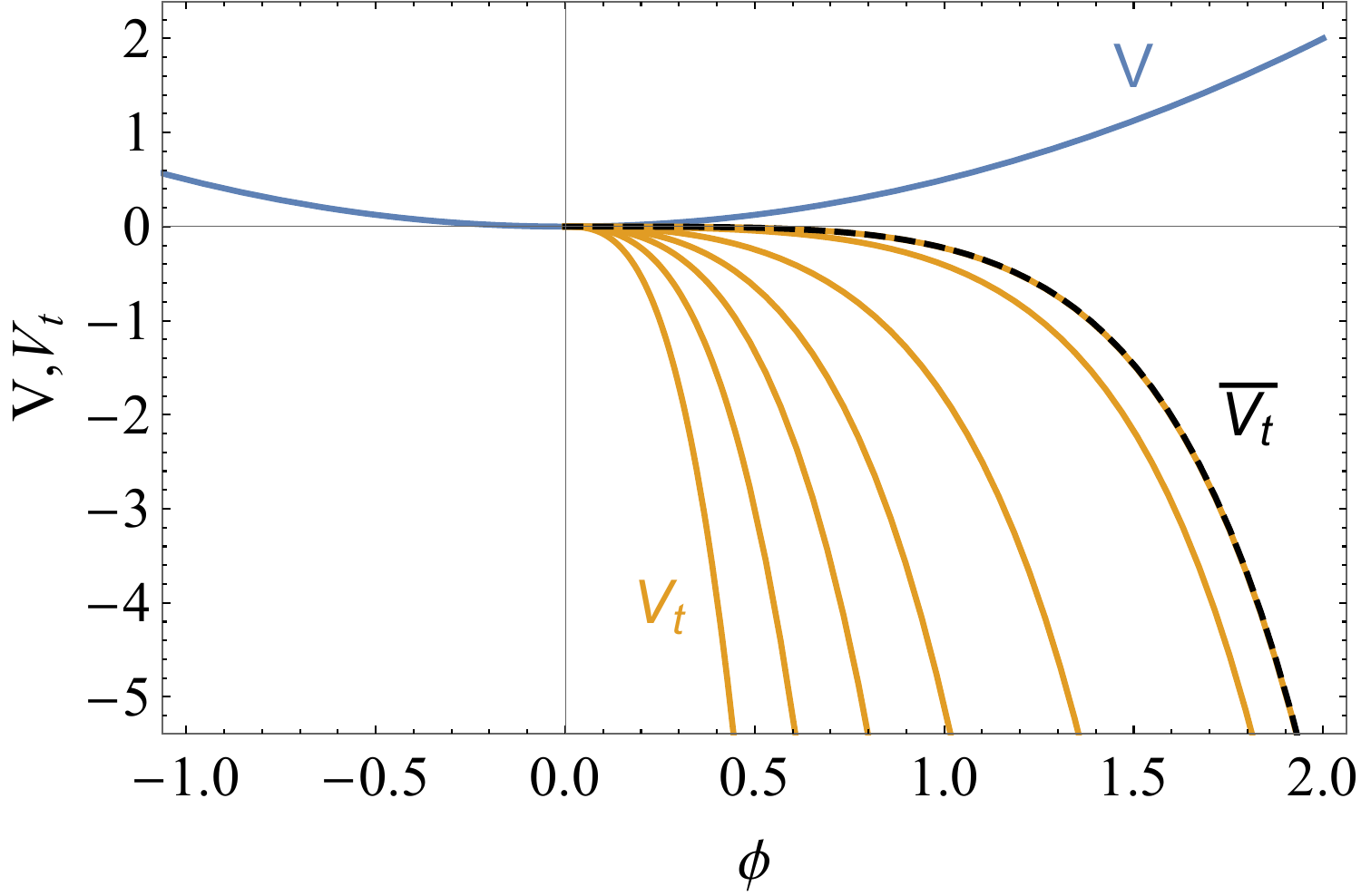}
\includegraphics[width=0.51\textwidth]{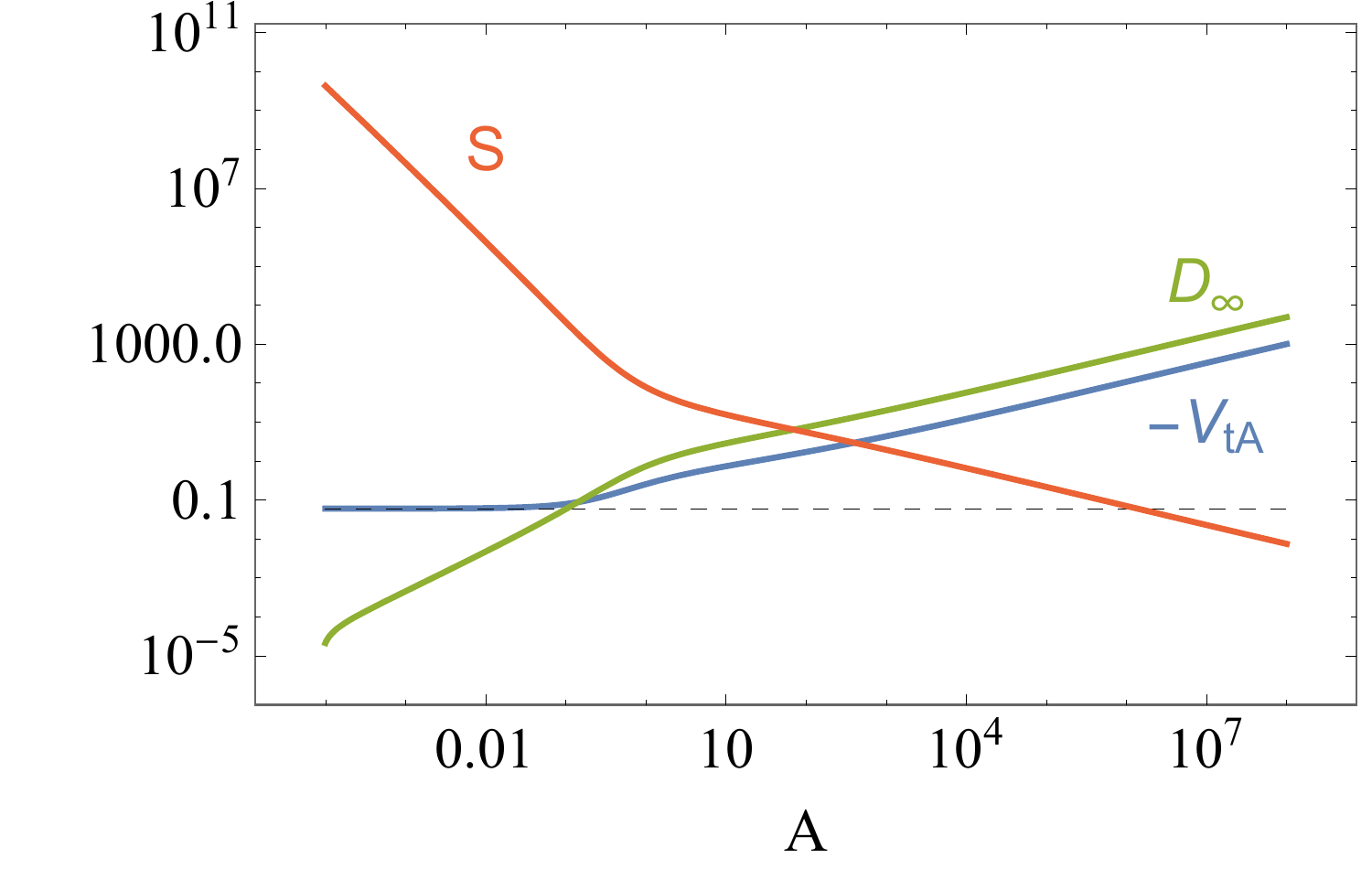}
\end{center}
\caption{Type 0 BoN decay for a \emph{Minkowski false vacuum}. Left: Potential (\ref{type0V}) with $V_\pp=0$, $\kappa=1$, $m^2=1$, and tunneling potentials $V_t(A;\phi)$ (bounded by $\overline{V_t}$, dashed line) for $A=\{0.001,0.1,1,10,100,10^3,10^4\}$. Right: Tunneling action $S$ and prefactors $V_{tA}, D_{\infty}$ that control the asymptotic behaviors $V_t(\phi\to\infty)\sim V_{tA}e^{\sqrt{6\kappa}\phi}$ and  $D(\phi\to\infty)\sim D_\infty e^{\sqrt{8\kappa/3}\phi}$. The lower bound on $-V_{tA}$ is indicated by the black dashed line.
\label{fig:BC0Mink}
}
\end{figure}

\subsection{Minkowski False Vacuum \label{sect:MFV}}

For a Minkowski vacuum ($V_\pp=0$), the low-field expansion of $V_t(\phi)$ is
\be
V_t(A;\phi) = -\frac{m^2\phi^2}{W}-\frac{3\kappa m^2\phi^4}{16}\left(1+\frac{5}{2W}\right)-\frac{9\kappa^2m^2\phi^6}{256}\left(1+\frac{8}{3W}\right)+{\cal O}\left(\phi^8,\frac{1}{W^2}\right)\ ,
\label{VtexpMink}
\ee
where 
\be
W\equiv W\left(A^{-1/3} e^{1/A}\phi^{-2/3}\right)\ ,
\label{eq:prodLog}
\ee
with $W(x)$ the product-log function [which satisfies $W(x)e^{W(x)}=x$ and has the large $x$
expansion $W(x)=\log x+ (1-\log x)\log(\log x)/(\log x)+...$] and $A>0$ a free parameter.\footnote{The particular functional dependence $A^{-1/3} e^{1/A}$ we choose is tailored to cover different regimes of the solutions within a sensible range of $A$ values. The particular values of $A$ are not significant as they are just labels without physical meaning. Moreover, the numerical value associated to a particular solution can be sensitive to the precision of the low-field expansion used to solve numerically the EoM for $V_t$.}
We therefore find an infinite family of $V_t$ solutions parametrized by $A$, $V_t(A;\phi)$, describing possible decay channels of the vacuum at $\phi_\pp=0$. Figure~\ref{fig:BC0Mink}, left plot, shows $V_t(A;\phi)$ for $A=\{0.001,0.1,1,10,100,10^3,10^4\}$. For $A\to 0$ we reach the critical tunneling potential $\overline{V_t}(\phi)$ (black dashed line in the figure) given by the $W\to\infty$ limit of $V_t$ in (\ref{VtexpMink})
\be
\overline{V_t}(\phi)=V_t(\infty ;\phi) = -\frac{3\kappa m^2\phi^4}{16}-\frac{9\kappa^2m^2\phi^6}{256}+{\cal O}\left(\phi^8\right)\ ,
\ee
which gives $D=0$ and represents an upper limit on allowed $V_t$'s (that should have $D^2>0$).

It can be checked numerically that the asymptotic behaviour of the different $V_t$ solutions follows the type 0 expectation, $V_t\sim V_{tA} e^{\sqrt{6\kappa}\phi}$ and $D\sim D_\infty e^{\sqrt{8\kappa/3}\phi}$ with different $V_{tA},D_\infty$ for different values of $A$. The functions $V_{tA}(A)$ and $D_\infty(A)$ are model-dependent functions, different for different $V(\phi)$, and are given for our toy example in the right plot of figure~\ref{fig:BC0Mink}. It is interesting that $-V_{tA}$ is bounded below, as shown by the dashed line, which corresponds to the $V_{tA}$ prefactor of $\overline{V_t}\sim \overline{V}_{tA}e^{\sqrt{6\kappa}\phi}$. The physical implications of this bound are discussed in subsection~\ref{sect:dynquench}.

\begin{figure}[t!]
\begin{center}
\includegraphics[width=0.48\textwidth]{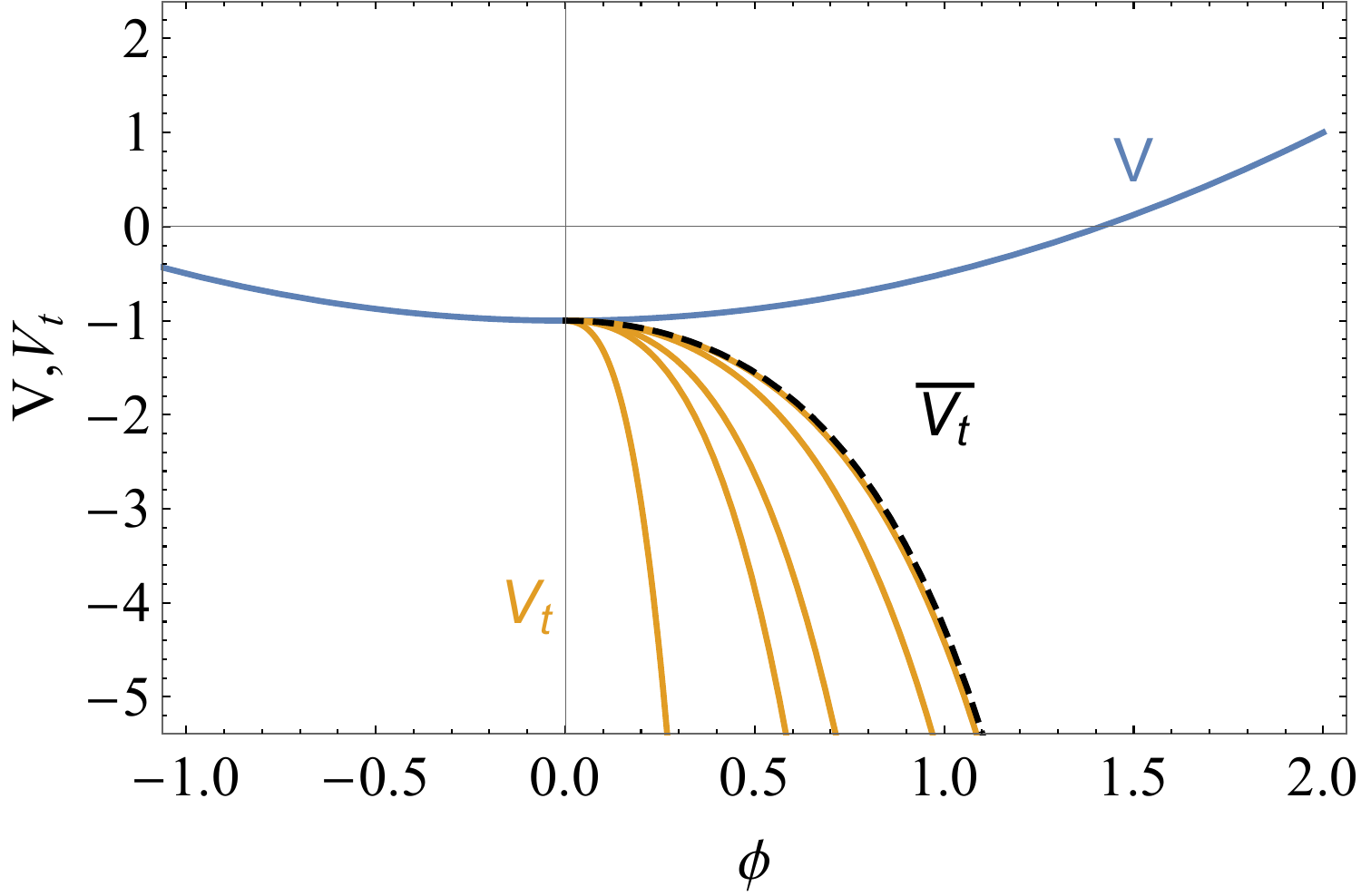}
\includegraphics[width=0.49\textwidth]{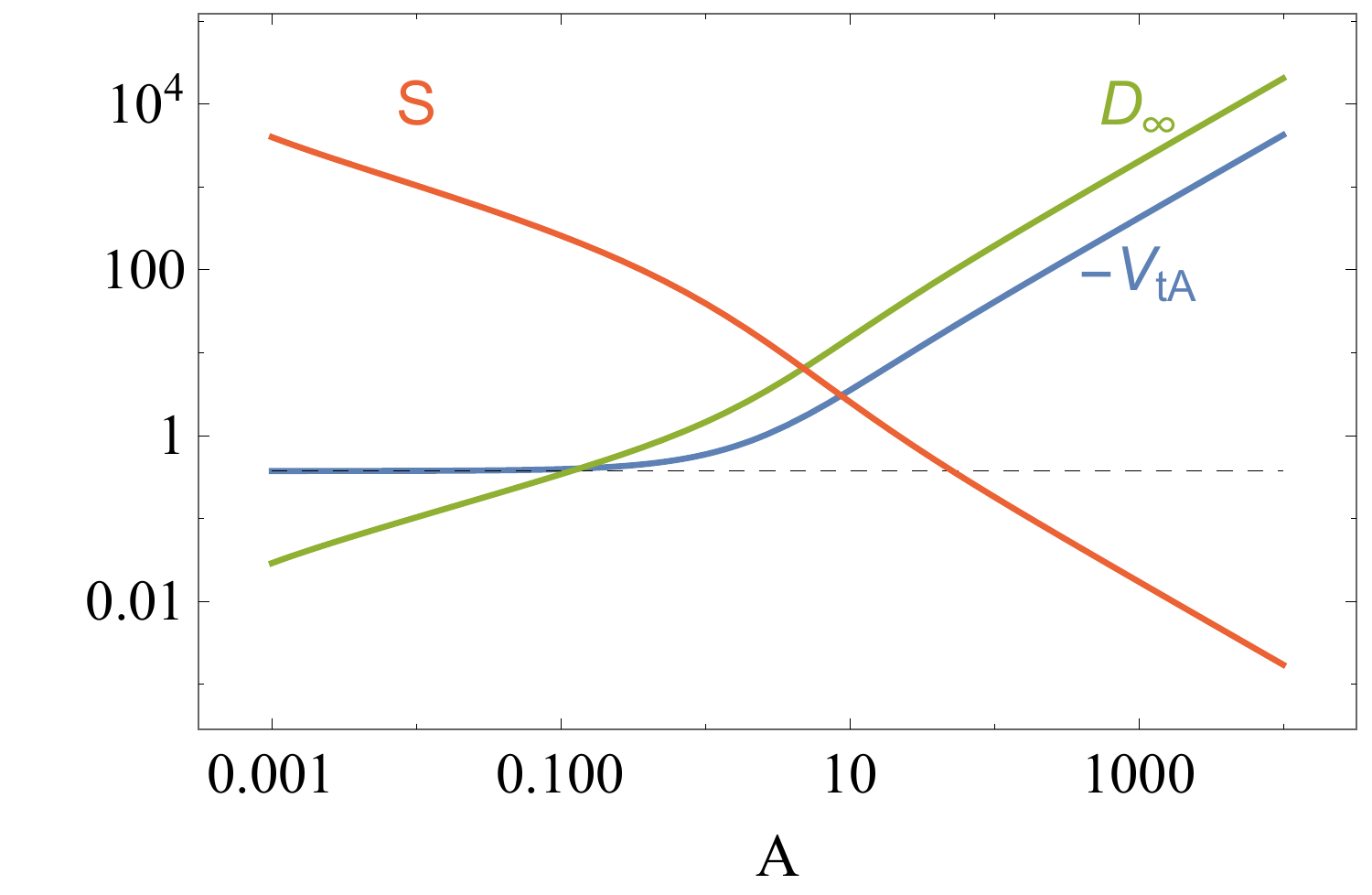}
\end{center}
\caption{Type 0 BoN decay for an \emph{AdS false vacuum}. Left: Potential (\ref{type0V}) with $V_\pp=-1$, $\kappa=1$, $m^2=1$ and different tunneling potentials $V_t(A;\phi)$ (bounded by $\overline{V_t}$,dashed line) for $A=\{0.1,1,5,10,100\}$. Right: Tunneling action $S$ and prefactors $V_{tA}, D_{\infty}$ that control the asymptotic behaviors $V_t(\phi\to\infty)\sim V_{tA}e^{\sqrt{6\kappa}\phi}$ and  $D(\phi\to\infty)\sim D_\infty e^{\sqrt{8\kappa/3}\phi}$.  The lower bound on $-V_{tA}$ is indicated by the black dashed line.
\label{fig:BC0AdS}
}
\end{figure}

\subsection{AdS False Vacuum}

Consider next the AdS vacuum case, with $V_\pp<0$. For such case, the low-field expansion of $V_t$ reads
\be
V_t(A;\phi) = V_\pp +\frac12 m_t^2\phi^2 -A \phi^{2+\alpha}+B_4\phi^4+...
\label{VtexpAdS}
\ee
with
\be
m_t^2=\frac32 \kappa V_\pp\left(1+\sqrt{1-\frac{4m^2}{3\kappa V_\pp}}\right)\ ,\quad
\alpha=\frac{2\kappa V_\pp}{m_t^2}\ ,\quad
B_4=\frac{3\kappa(2m_t^4-5m^2 m_t^2+3m^4)}{5m_t^2-12m^2+3V_\pp \kappa}\ ,
\ee
and $A$ a free parameter, which labels a family of $V_t$ solutions. The critical tunneling potential $\overline{V_t}(\phi)$
corresponds to the case $A=0$ and, to have $D^2>0$, one needs $A>0$.
Figure~\ref{fig:BC0AdS}, left plot, shows different $V_t$'s for $A=\{0.1,1,5,10,100\}$ taking $V_\pp=-1$, $m=1$ and $\kappa=1$. For $A\to 0$ we reach the critical tunneling potential $\overline{V_t}(\phi)=V_t(0;\phi)$ (black dashed line). 

As in the previous case, the asymptotic behaviour for $\phi\to\infty$ of the different $V_t$ solutions is of type 0, with $V_t\sim V_{tA} e^{\sqrt{6\kappa}\phi}$ and $D\sim D_\infty e^{\sqrt{8\kappa/3}\phi}$, with different $V_{tA}$ and $D_\infty$ for different values of $A$. The functions $V_{tA}(A)$ and $D_\infty(A)$ are given in the right plot of figure~\ref{fig:BC0AdS}. We again find that $-V_{tA}$ is bounded below, as shown by the dashed line, which corresponds to the $V_{tA}$ prefactor of $\overline{V_t}$. Compared to the previous Minkowski case the bound on $V_{tA}$ is stronger.

\begin{figure}[t!]
\begin{center}
\includegraphics[width=0.48\textwidth]{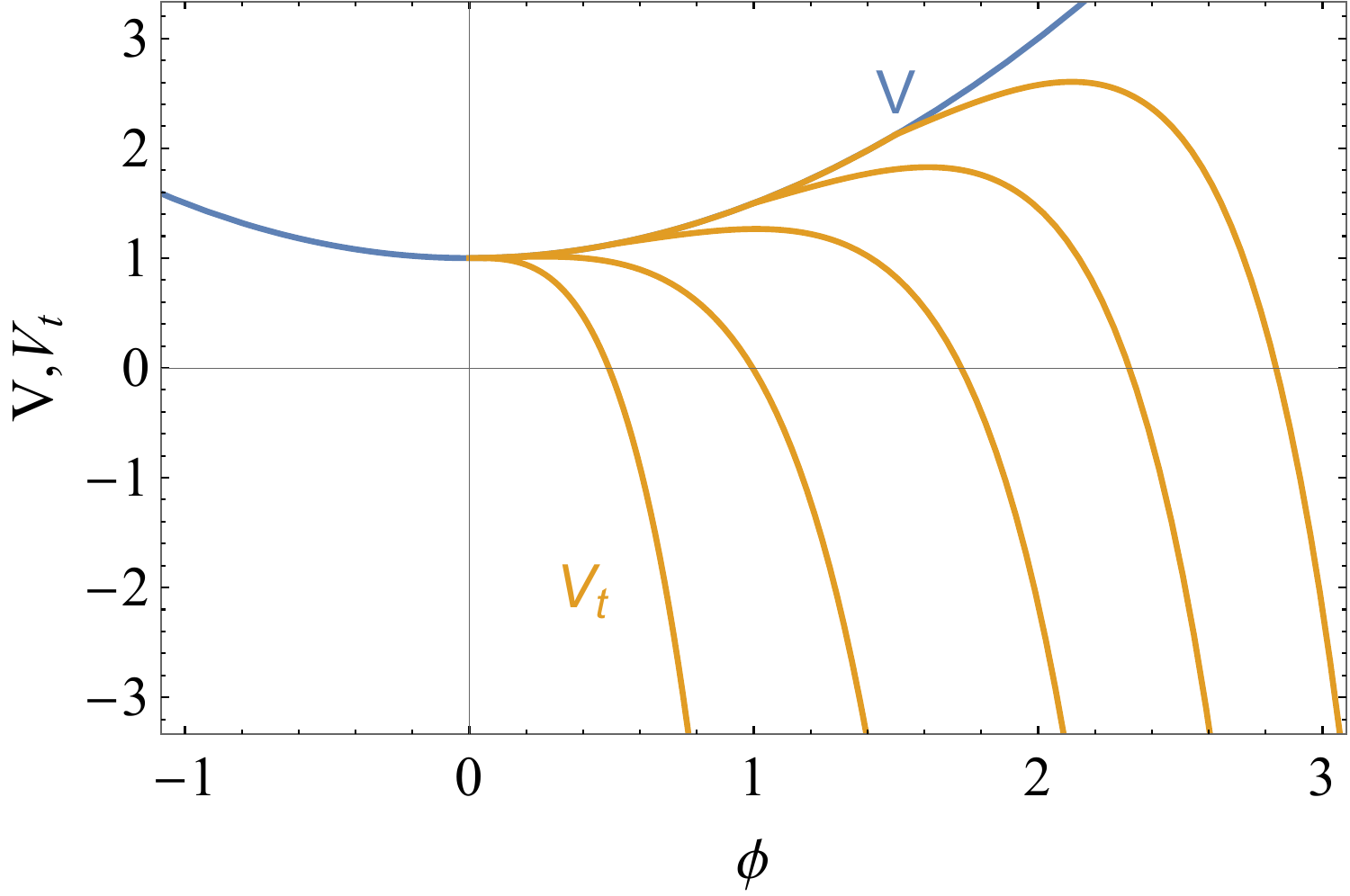}
\includegraphics[width=0.49\textwidth]{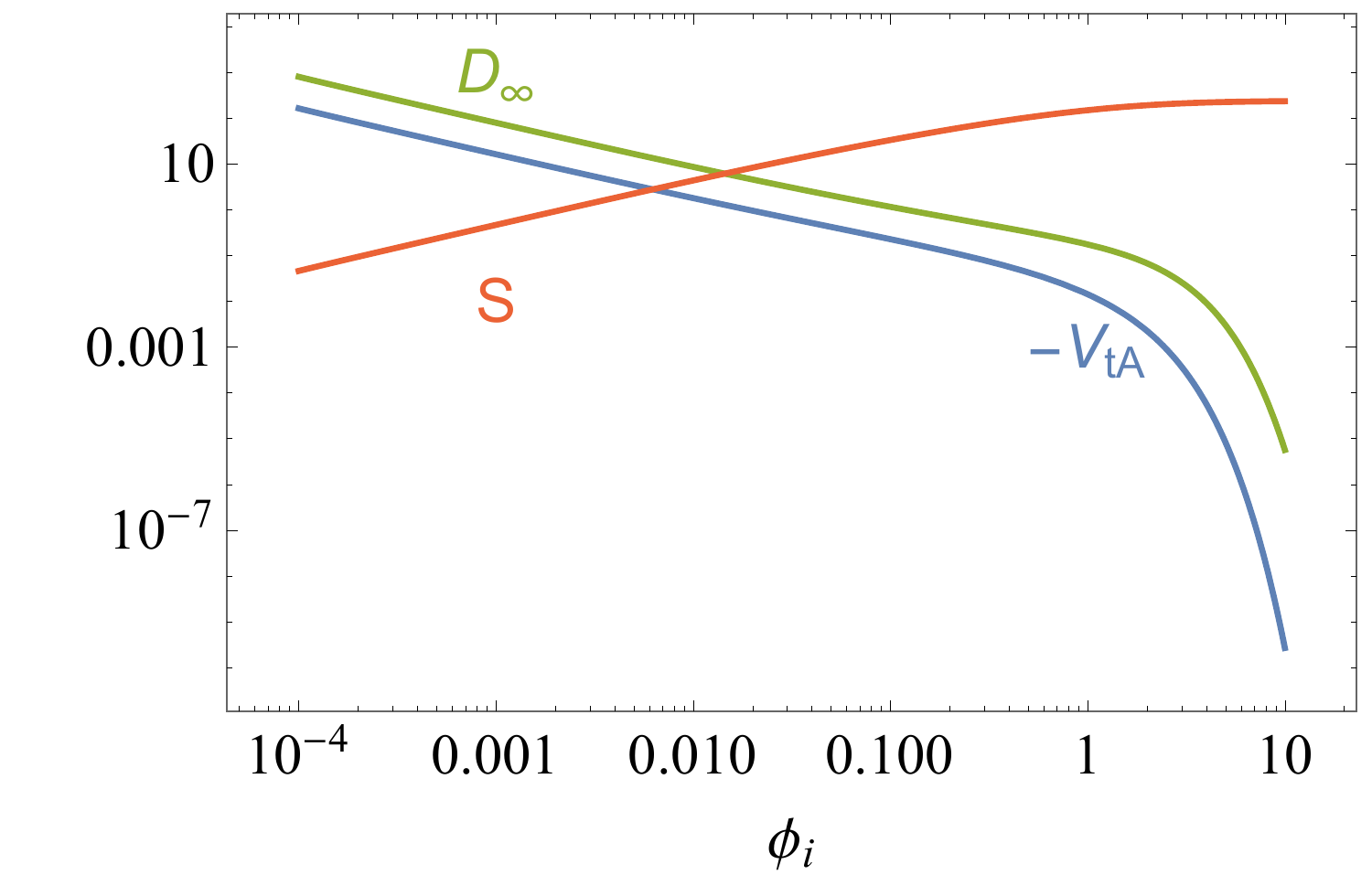}
\end{center}
\caption{Type 0 BoN decay for a \emph{dS false vacuum}. Left: Potential (\ref{type0V}) with $V_\pp=1$, $\kappa=1$, $m^2=1$ and different tunneling potentials $V_t(\phi_i;\phi)$ for $\phi_i=\{0.1,0.5,1,1.5\}$. Right: Tunneling action $S$ and prefactors $V_{tA}, D_{\infty}$ that control the asymptotic behaviors $V_t(\phi\to\infty)\sim V_{tA}e^{\sqrt{6\kappa}\phi}$ and  $D(\phi\to\infty)\sim D_\infty e^{\sqrt{8\kappa/3}\phi}$. 
\label{fig:BC0dS}
}
\end{figure}

\subsection{dS False Vacuum}

Consider finally the dS case, $V_\pp>0$. As happens for regular CdL dS decays, the CdL instanton reaches at $\xi\to\infty$ a field value $\phi_i$ different from the false vacuum $\phi_\pp$. The expansion of $V_t$ near $\phi_i$ takes the form
\be
V_t(\phi_i+\delta\phi)=V(\phi_i)+\frac34 V'(\phi_i) \delta\phi +\frac12 m_t^2\delta\phi^2+... 
\label{Vfornum}
\ee
with
\be
m_t^2=-\frac13 \kappa V_\pp + \frac12 m^2\left(1-\frac13\kappa\phi_i^2\right)\ .
\ee
In this case one can consider $\phi_i$ as the free parameter for a family of solutions of the EoM for $V_t$, this time leaving from different points rather than from the false vacuum. For regular CdL decay, this family of solutions features a single solution corresponding to the proper CdL instanton while all the rest correspond to pseudo-bounces \cite{PS}, see section~\ref{sec:Vt}.

For the $V_t$'s of BoN decays, different values of $\phi_i$ lead to different asymptotic behaviours at $\phi\to\infty$, $V_t\sim V_{tA} e^{\sqrt{6\kappa}\phi}$ and $D\sim D_\infty e^{\sqrt{8\kappa/3}\phi}$, and thus to different $V_{tA}(\phi_i)$ and $D_\infty(\phi_i)$, exactly as happens for Minkowski or AdS vacua. Figure~\ref{fig:BC0dS}, left plot, shows tunneling potentials for $V_\pp=1$, $m=1$, $\kappa=1$ and different values of $\phi_i=\{0.1,1,0.5,1,1.5\}$. The functions $V_{tA}(\phi_i)$  and $D_\infty(\phi_i)$ as well as the tunneling action $S(\phi_i)$ are shown in the right plot of the same figure. Now there is no bound on $V_{tA}$.

\subsection{Gravitational Quenching of BoN Decay\label{sect:dynquench}}

As reviewed in Sect.~\ref{sec:Vt} for standard vacuum decay, it is well known that Minkowski and AdS false vacua become more stable against semi-classical  decay when gravitational effects are taken into account \cite{CdL}. Actually, provided the effect of gravity is sufficiently strong, there might be a dynamical obstruction (e.g. a positive energy theorem) which forbids the nucleation of the tunneling bubble. This is referred to as \emph{gravitational quenching}. When the dynamical obstruction is just marginally satisfied (the critical case in section \ref{sec:Vt}), the model at hand could admit instanton solutions but with an infinite tunneling action, and therefore the decay would  still be forbidden.

For BoN decays we have in principle the same behaviour, with possible dynamical obstructions and critical cases \cite{Blanco-Pillado:2016xvf,GMSV} that satisfy the condition $D=0$, and represent an infinite and static BoN, that is, and End-of-the-World (ETW) brane \cite{DynCo} (we treat this case in more detail in section \ref{sec:quench}).
However, BoN decay is more subtle as, a priori, there might be topological obstructions preventing it \cite{BoN}. Nevertheless, the recently proposed Cobordism Conjecture \cite{CobConj} states that any consistent theory of quantum gravity should admit a cobordism to nothing. In other words, according to this conjecture, in a consistent quantum gravity theory no topological obstruction against BoN decay can be present regardless of the compactification. In that case, the only protection of a given compactification against BoN formation must be dynamical in origin (see \cite{Blanco-Pillado:2016xvf,GMSV}). 

The toy examples with a a simple quadratic potential discussed in the previous subsections can illustrate how the dynamical constraint on BoN decay appears.
As we show in section~\ref{sec:topdown}, $V_{tA}$ and $D_\infty$ are ultimately determined by quantities in the higher dimensional theory. Generically 
\be
-V_{tA}=\frac{C_d}{\kappa R_{KK}^2}\ ,\quad \quad  D_\infty=\frac{C'_d}{{\cal R}R_{KK}\sqrt{\kappa}}\ ,
\label{VtDUV}
\ee
where $C_d, C'_d$ are (positive) constants that depend on the dimension $d$ of the compactified space,  $R_{KK}$ is the typical radius of that space, while ${\cal R}$ is the bubble nucleation radius of the BoN. Once $V_{tA}$ is fixed, only one member of the full family of allowed $V_t$'s is relevant for the vacuum decay in that theory, the one giving the correct $V_{tA}$ that matches with $R_{KK}$. For such $V_{tA}$, $D_\infty$ takes a particular value, which then  fixes the radius ${\cal R}$ realized in that particular decay. 

We also see that, when the function $-V_{tA}(A)$ is bounded below by some value $-V_{tA*}\le -V_{tA}(A)$ (as shown in figs. \ref{fig:BC0Mink} and \ref{fig:BC0AdS}, for Minkowski and AdS vacua respectively), for the BoN decay to be allowed, the KK radius must satisfy a dynamical constraint of the form
\be
R_{KK}^2 =  \frac{C_d}{\kappa (-V_{tA})} \le \frac{C_d}{\kappa (-V_{tA*})}.
\label{eq:dyConst1}
\ee
Therefore, if the higher dimensional theory gives a $-V_{tA}$ below the lower limit (corresponding to $R_{KK}$ bigger than some critical value), then the vacuum in that theory is stable against BoN decay. On the other hand, $D_\infty$ is not bounded and goes to zero as $A\to 0$, when $V_t\to \overline{V_t}$. This leads to a tunneling action $S$, also shown in figs. \ref{fig:BC0Mink} and \ref{fig:BC0AdS}, that grows without bound as $A\to 0$, in which limit we have ${\cal R}\to \infty$ (we discuss further such cases in section~\ref{sec:quench}). In other words: since we have $V_{tA*} = V_{tA}(0)$, we observe that when the dynamical constraint  \eqref{eq:dyConst1} is just saturated (\emph{critical case}), the BoN becomes  infinite and static, an ETW brane.

This dynamical constraint can be interpreted in the context of the Swampland program. In order to have no BoNs in the effective field theory, \eqref{eq:dyConst1} requires  $R_{KK}^2>1/(\kappa E_{EFT}^4)$, where we have identified $V_{tA*}\sim E_{EFT}^4$\footnote{$V_{tA*}$ is determined entirely by the $4d$ scalar potential, and we also assume there are no large energy hierarchies in the EFT regime of validity.}, but this condition might set us outside the regime of validity of the $4d$ EFT. Indeed,  for AdS vacuum decay the previous condition requires the KK radius to be at least of the same order (or larger) than the AdS scale, $R_{KK} \gtrsim L_{AdS}$, and thus, the use of the $4d$ EFT might not be justified due to the absence of scale separation. Thus, an EFT without BoNs seems to be in the Swampland. Conversely, in the regime where the EFT is consistent, BoNs are unavoidable. Note however that, while scale separation is required when the $4d$ EFT is obtained integrating out the physics above the KK scale, this condition is no longer necessary when the $4d$ theory is a consistent dimensional reduction of a higher dimensional theory (see e.g. section \ref{sec:BoNFlux}).

For the dS false vacua case there is no bound on $V_{tA}$, see fig. \ref{fig:BC0dS}, and therefore there is no dynamical constraint on BoN decay. Interestingly, while in this simple model the dynamical constraint is always connected to the CdL mechanism, this is not always the case: indeed, in the next section we show that in more complicated models BoN decay might be dynamically forbidden (even for dS false vacua), but without CdL suppression to enforce the constraint in the critical case.

\section{BoNs vs. Other Decay Channels\label{sec:BoNvsOther}}

{In the present section we study the interplay of BoN nucleation with other decay channels, such as standard CdL decay, HM bounces, and pseudo-bounces and compare as well their decay rates. To illustrate this interplay between the different decay channels we consider in this section more realistic toy potentials for $\phi$, of the form 
\be
V(\phi)=V_\pp+\frac12 m^2\phi^2-\lambda \phi^4+\lambda_6\phi^6\ ,
\label{Vcex}
\ee
which gives examples of type 0 BoNs. At the end of this section we also discuss a type $-$ case, after including appropriate exponential contributions to the potential. For the numerical work that follows we take $m=1$, $\lambda=17/4$, $\lambda_6=8/3$ and $\kappa=1$ such that the potential has a false vacuum at $\phi_\pp=0$, separated  from the true vacuum at $\phi_\mm=1$ by a shallow barrier that peaks at $\phi_B=0.25$. We then vary $V_\pp$ to consider in turn Minkowski, AdS or dS false vacua. 

Consistently with the findings of \cite{DGL} we observe that, while BoN decay becomes the dominant channel when the KK scale is well above the $4d$ EFT scale, there are regions of the parameter space where the standard vacuum decay channels have faster decay rates that BoN nucleation. In particular, as anticipated in the previous section, we present an example  of a dS vacuum which is dynamically protected against the BoN formation, but is still non-perturbatively unstable due to other decay channels. Contrary to the situation for Minkowski and AdS vacua in standard false vacuum decay, we find examples of critical  vacua (those marginally satisfying the corresponding dynamical constraint) with a finite BoN nucleation rate, and therefore  not protected by a CdL suppression mechanism.

\subsection{Minkowski False Vacuum}
Consider first the Minkowski case, with $V_\pp=0$, see figure~\ref{fig:PSCdLBoN}, upper left plot.
Solving first the EoM for the critical tunneling potential $\overline{V_t}$, eq.~(\ref{Vtc}), we find that $\overline{V_t}$ (black dashed line in the plot) is not curved down much by gravity and touches the potential beyond the barrier, signaling that the false vacuum is unstable against CdL decay.

Solving then the EoM for $V_t$, eq.~(\ref{EoMVt}), we find three different types of solutions, all of them lying below $\overline{V_t}$. At low $\phi$, the expansion of these solutions is as derived in (\ref{VtexpMink}) and the free parameter $A$ labels the family of solutions. For $A<A_{CdL}\simeq 2.82686$, we find pseudo-bounce solutions (green lines in figure~\ref{fig:PSCdLBoN}, with $A=\{1,1.7,2.5\}$), for which the tunneling proceeds to some fixed field value on the slope beyond the barrier. At $A=A_{CdL}$ we get the CdL instanton solution (red line), for which the tunneling action is stationary. For $A>A_{CdL}$ we obtain unbounded BoN solutions (orange lines, with $A=\{2.85,3,5,10,50\}$).

\begin{figure}[t!]
\begin{center}
\includegraphics[width=0.5\textwidth]{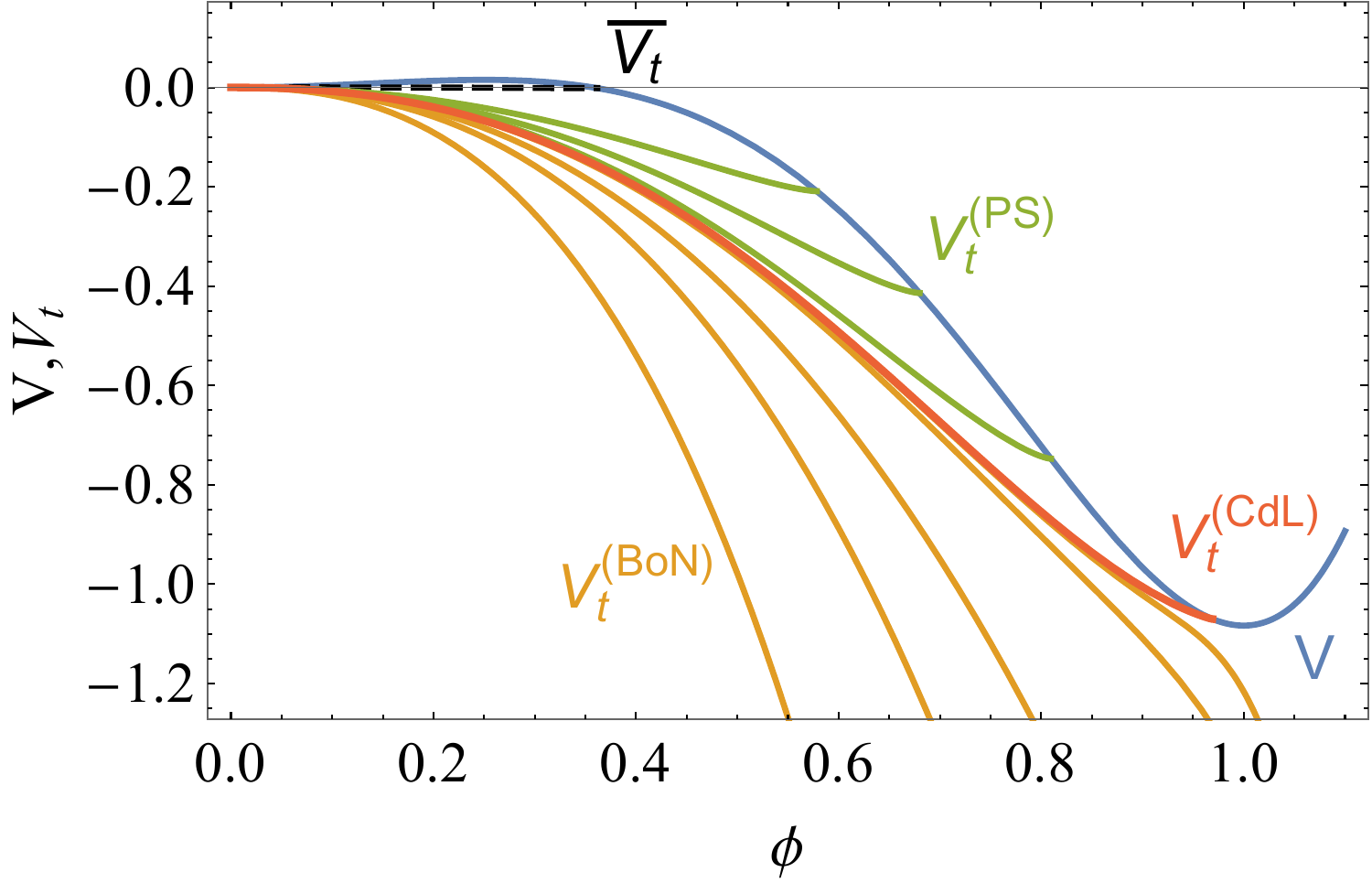}
\includegraphics[width=0.49\textwidth]{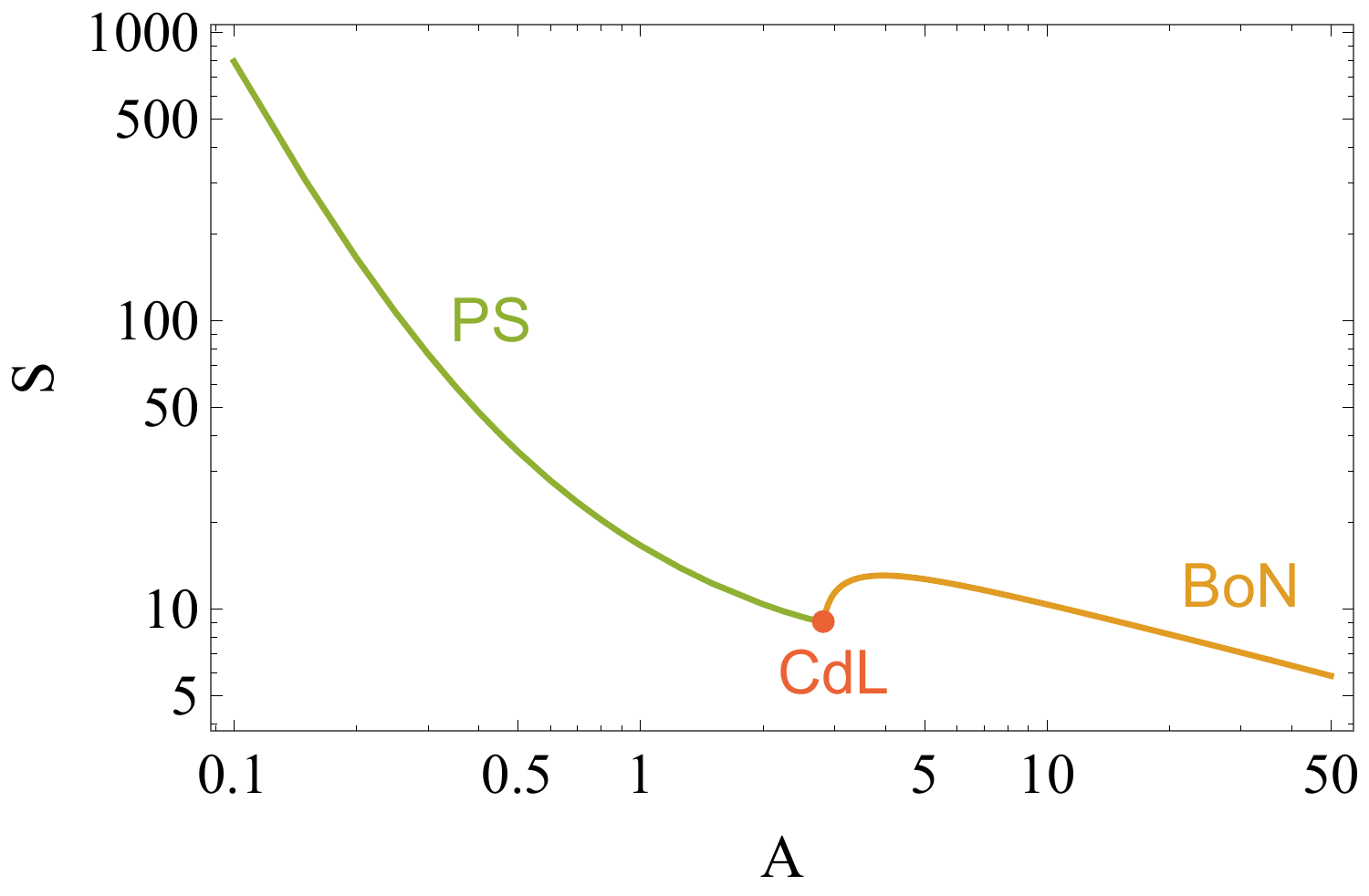}\\
\includegraphics[width=0.49\textwidth]{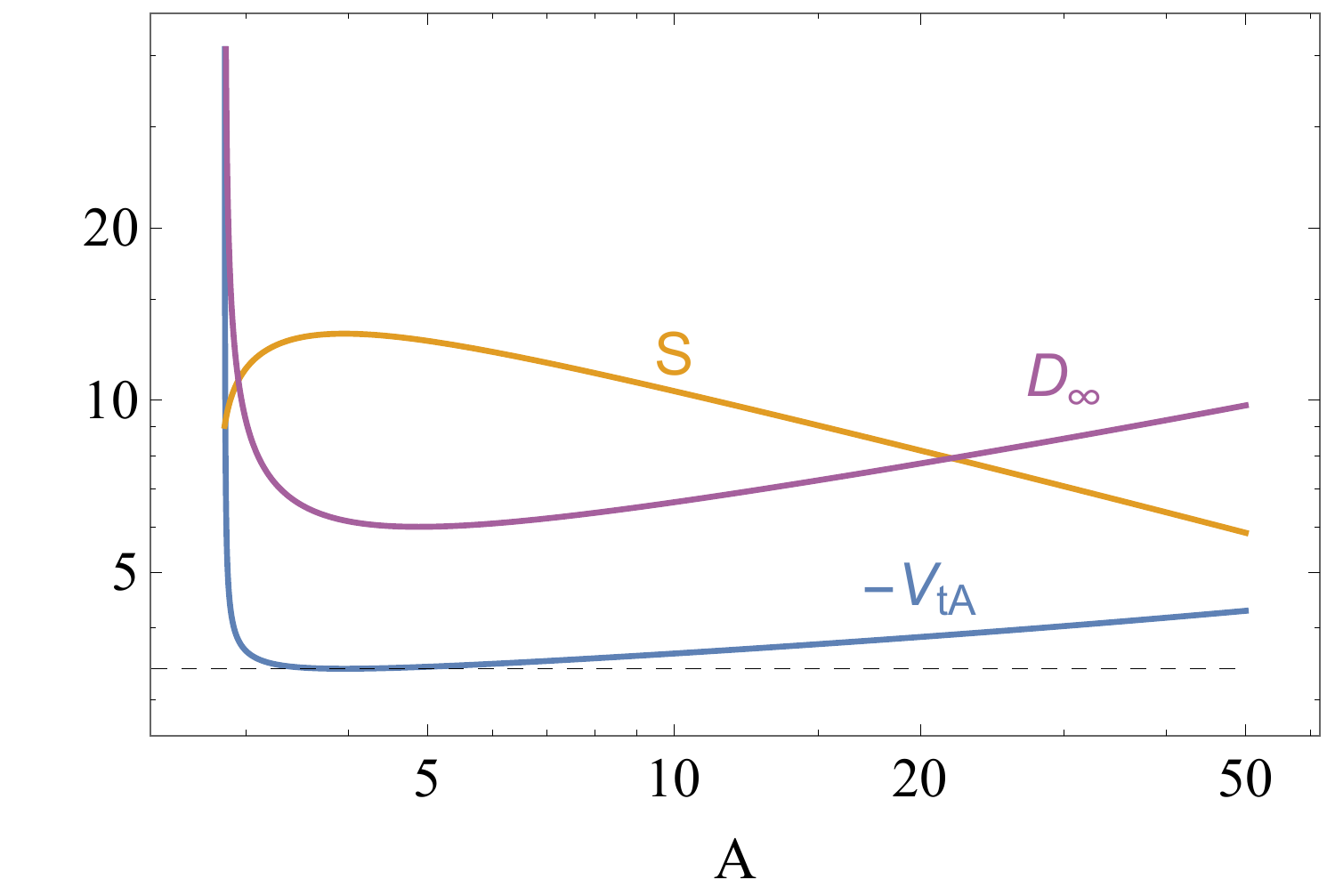}
\includegraphics[width=0.48\textwidth]{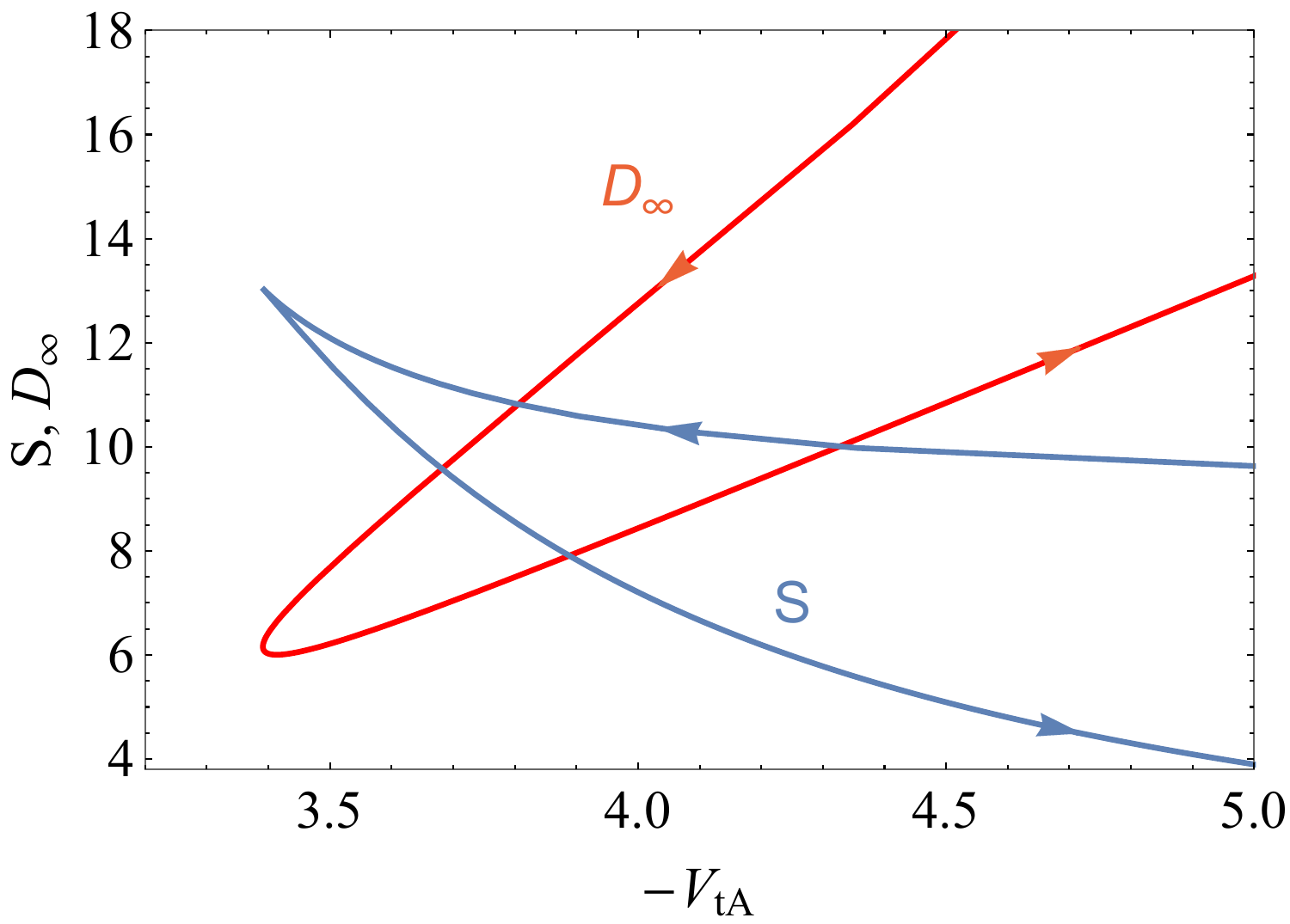}
\end{center}
\caption{
Upper left: Potential (\ref{Vcex}) with $V_\pp=0$ and tunneling potentials $V_t(A;\phi)$ of different types: $\overline{V_t}$ for $A=0$ (black dashed); pseudo-bounces for $0<A<A_{CdL}$ (green); CdL bounce for $A=A_{CdL}$ (red) and BoNs for $A>A_{CdL}$ (orange). 
Upper right: Tunneling action $S$ with labels/colors indicating different types. 
Lower left: For the BoN range of $A$, action $S$ and prefactors $V_{tA}, D_{\infty}$ that control the asymptotic $\phi\to\infty$ behaviors $V_t\sim V_{tA}e^{\sqrt{6\kappa}\phi}$ and  $D\sim D_\infty e^{\sqrt{8\kappa/3}\phi}$. The black dashed line shows the lower bound on $-V_{tA}$. 
Lower right: Action and $D_\infty$ as functions of $-V_{tA}$. 
\label{fig:PSCdLBoN}
}
\end{figure}

The same figure~\ref{fig:PSCdLBoN}, upper right plot, gives the tunneling action corresponding to the $V_t$ solutions just described. We see that the action of pseudo-bounces diverges at $A\to 0$ (or $V_t\to\overline{V_t}$) and is monotonically decreasing until
the CdL instanton is reached, at which point the action is stationary (as it corresponds to a true bounce). The slope of the pseudo-bounce action is given by (see appendix~\ref{app:dSdA})
\be
\frac{dS}{dA}=\frac{6\pi^2}{\kappa^2 V_e^2}(1-x_e)^2(2+x_e)\left.\frac{dV_t}{dA}\right|_{\phi_e}\ ,
\label{dSPSdA}
\ee
with $x_e\equiv\sqrt{1-\kappa V_e\rho_e^2/3}$, $V_e\equiv V(\phi_e)$ and $\rho_e=\rho(\xi=0)$, where $\phi_e$ is the end-point of the pseudo-bounce interval. We have confirmed numerically this expression.
Interestingly, the BoN action beyond the CdL point first increases, reaches a maximum and then monotonically decreases, eventually becoming smaller than $S_{CdL}$.\footnote{The limit $S\to 0$ (in this example and previous ones) corresponds to Witten-like solutions that drop almost vertically right after leaving the false vacuum, $V_t\simeq V_\pp -6/(\kappa R_{KK}^2)\sinh^3(\sqrt{2\kappa/3}\phi)\simeq V_\pp +V_{tA}e^{\sqrt{6\kappa}\phi}$ with $\-V_{tA}\sim 1/R_{KK}^2\to\infty$, but such small action indicates the breakdown of the semiclassical expansion. }
From this example we conclude that it is not always the case that the BoN decay channel dominates. The structure of the BoN action and solutions as $A\to A_{CdL}$ are discussed below.

The lower left plot of figure~\ref{fig:PSCdLBoN} gives $V_{tA}(A)$ and $D_\infty(A)$, as well as $S$, for BoN solutions.
The function $-V_{tA}(A)$ is bounded below by a minimum value $-V_{tA*}$ (black dashed line), implying the presence of a dynamical constraint which could prevent BoN decay. 
Indeed, we see that for a given higher-dimensional theory with a fixed value of $V_{tA}=V_{tA}^{(th)}$, determined by the dimension of the compact space and $R_{KK}$ as in (\ref{VtDUV}), we might have a vacuum that cannot decay via BoNs (if $V_{tA}^{(th)}>V_{tA*}$). When BoN nucleation is dynamically forbidden, the vacuum is still unstable to decay via the standard CdL channel.
Alternatively, when $V_{tA}^{(th)}<V_{tA*}$,  there are two possible BoN decay channels, corresponding to the two solutions of the equation $V_{tA}(A)=V_{tA}^{(th)}$. Among these two solutions, the one with lowest tunneling action is the one with highest $A$ and lowest $D_\infty$ [and thus highest ${\cal R}$, according to (\ref{VtDUV})]. This can be seen in the lower right plot of figure~\ref{fig:PSCdLBoN}, that shows the values of $S$ and $D_\infty$ for the two solutions, as a function of $-V_{tA}$ (the arrows indicate a growing $A$). We also see from the lower left plot of figure~\ref{fig:PSCdLBoN} that the value of $A=A_*\simeq 3.95$ where $V_{tA}$ reaches its maximum ($V_{tA*}$) corresponds precisely to the value at which $S$ is maximal. This can be understood from the simple relation
\be
\frac{dS}{dA}=-36\pi^2\sqrt{\frac{6}{\kappa}} \frac{V_{tA}}{D_\infty^3}\,\frac{dV_{tA}}{dA}\ ,
\label{dSBoNdA}
\ee
derived in Appendix \ref{app:dSdA}, that we have confirmed numerically. It is interesting that, in the critical case when the dynamical bound is saturated, $V_{tA}^{(th)}=V_{tA*}$, the action is finite so that there is no CdL suppression mechanism, as anticipated at the beginning of this section.

\begin{figure}[t!]
\begin{center}
\includegraphics[width=0.49\textwidth]{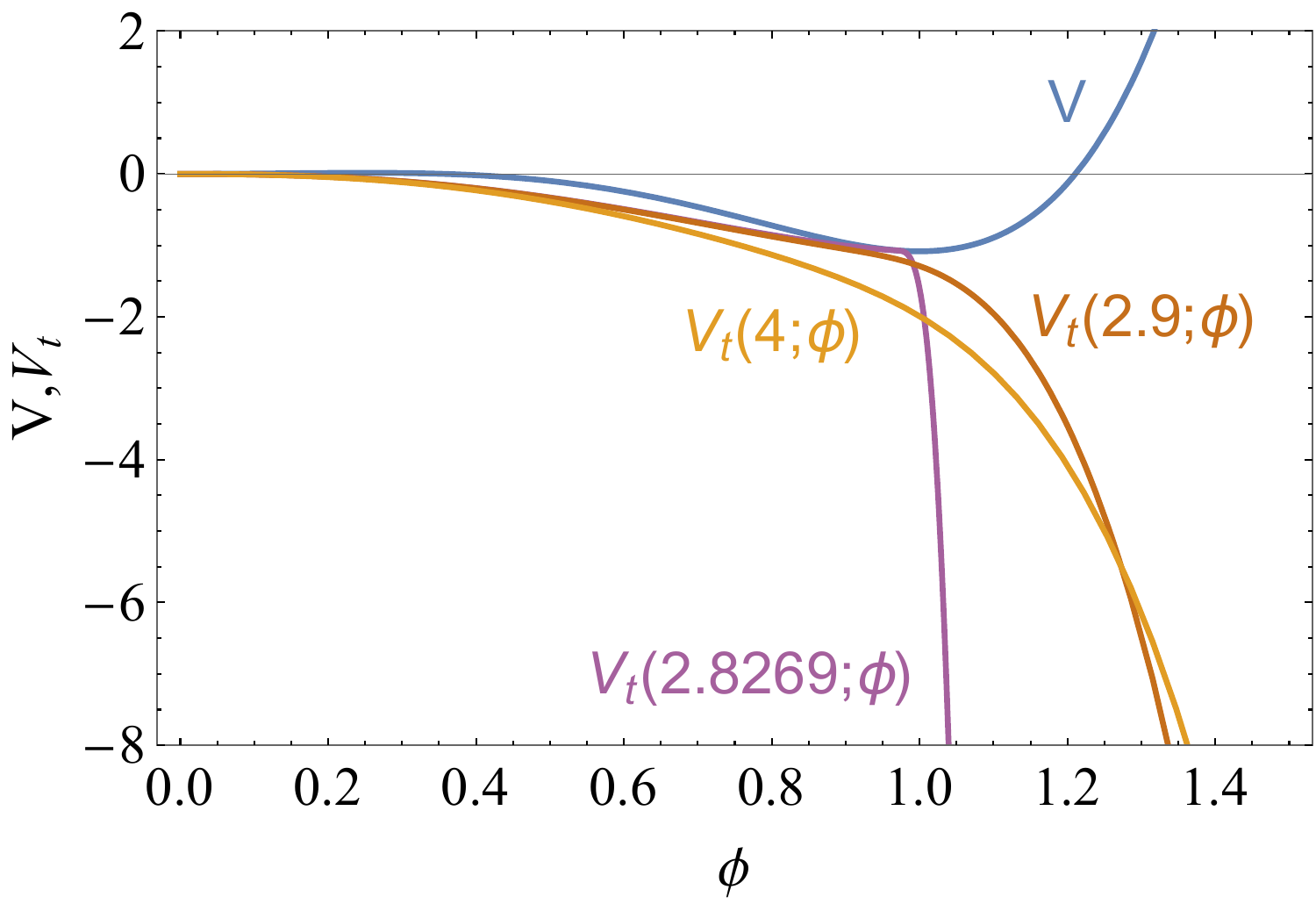}
\includegraphics[width=0.49\textwidth]{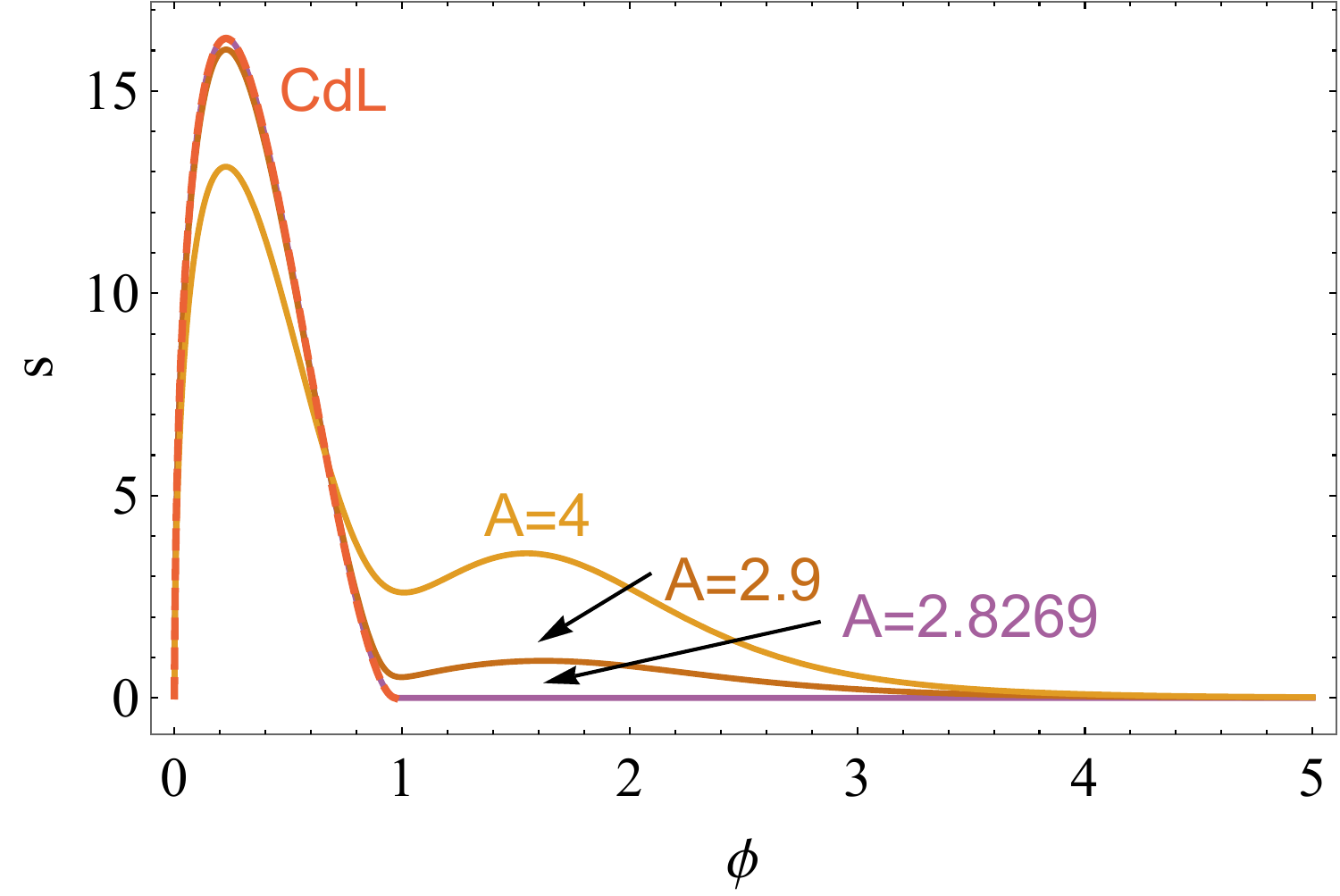}
\end{center}
\caption{
For the potential (\ref{Vcex}) with $V_\pp=0$, (left) example of crossing $V_t$'s for $A\simgt A_{CdL}$ and (right) action densities for $A\simgt A_{CdL}$ compared to the CdL case (red dashed line).
\label{fig:PSCdLBoN2}
}
\end{figure}

In order to understand the behaviour of the BoN action for $A\simgt A_{CdL}$, we must understand the non-monotonic behaviour of $V_{tA}(A)$ as both are related via (\ref{dSBoNdA}). Although close to the false vacuum we always have $V_t(A_1;\phi)<V_t(A_2;\phi)$ for $A_1>A_2$, this ordering can be reversed at high $\phi$ as different $V_t$ solutions can cross each other. This is what happens for $A\simgt A_{CdL}$, when $V_t$ solutions get very close to $V$ and get deflected down by it, causing $-dV_{tA}/dA<0$. An example of this crossing is shown in figure~\ref{fig:PSCdLBoN2}, left plot, which shows how the deflection is stronger the closer $A$ is to $A_{CdL}$. Solutions with higher values of $A$ do not suffer that deflection and once again one gets the normal situation with $-dV_{tA}/dA>0$. In other words, the presence of a minimum in the potential $V(\phi)$ distorts the shape of $V_{tA}(A)$  bending it upwards at low $A$ (the range of BoNs that probe the potential structure close to the minimum) as shown in figure~\ref{fig:PSCdLBoN}, while for large $A$ the BoNs are insensitive to the potential structure and $V_{tA}(A)$ looks like in the Minkowski example of section~\ref{sect:MFV}, see figure~\ref{fig:BC0Mink}.

As a result of the behaviour just discussed, we find that $S_{BoN}(A) \geq S_{CdL}$ (BoN decay has a rate lower than CdL tunneling) for a range of values of $A\simgt A_{CdL}$. That  translates into a range of values of the compactification radius $R_{KK}$, via the functional dependence $V_{tA}(A)$ and the relation \eqref{VtDUV}.  However, for larger values of $A$ (and large $-V_{tA}$) the BoN tunneling action becomes smaller than $S_{CdL}$ and equation \eqref{VtDUV} implies that, for  very small $R_{KK}$ (compared to the typical $4d$ EFT length scales), BoN decay always dominates. In general, this last regime is the most relevant, as it is precisely where the KK scale is well above the $4d$ EFT  energy scale, and thus, the region of parameter space where the EFT is well under control.    

We can also understand the continuity of $S(A)$ across $A_{CdL}$ in spite of the large difference between the CdL solution and BoN solutions with $A\simgt A_{CdL}$: the latter are very close to $V_t^{(CdL)}$ in the CdL field range and have a very large slope afterwards (see how $D_\infty$ and $-V_{tA}$ shoot up at $A\to A_{CdL}$ in the lower left plot of figure~\ref{fig:PSCdLBoN}). This large slope causes the action density to plummet 
exponentially. This effect is shown in the right plot of figure~\ref{fig:PSCdLBoN2},
which gives the action densities for several BoNs with $A\simgt A_{CdL}$ compared to the action for the CdL solution. Therefore, BoN solutions with $A\simgt A_{CdL}$ look like a mixed field configuration with a BoN-like core and a CdL outer part. They correspond to the 
hybrid CdL-BoN Euclidean solutions identified in \cite{DGL}. In fact, the two BoN solutions we find for a fixed $V_{tA}$ correspond to the two branches of solutions called in \cite{DGL} BoN false-vacuum branch and a BoN-CdL branch. As we have seen, the first branch is Witten-like, not very sensitive to the potential structure and reaches down to $S=0$, while the second branch is sensitive to the additional vacuum structure of the potential and the existence of a CdL solution, on which it ends. The two branches merge at the critical $V_{tA}$ as also found in \cite{DGL}.

\subsection{dS False Vacuum}
Consider next the case of a dS false vacuum, with $V_\pp>0$. The upper plots of figure~\ref{fig:PSCdLBoNdS} show the potential (\ref{Vfornum}), with $V_\pp=1$, and several tunneling potentials of different types obtained by solving numerically the EoM for $V_t$ starting at different values of $\phi_i$, which we use to parametrize the family of different solutions. From upper to lower lines (or upper to lower $\phi_i$), we first have the trivial solution (see section~\ref{sec:Vt}) corresponding to the Hawking-Moss instanton (purple line), corresponding to $\phi_i=\phi_B$, at the barrier top. This solution simply joins the false vacuum and the top of the barrier with $V_t^{(HM)}(\phi)=V(\phi)$. Next come the family of pseudo-bounces (green lines), with $\phi_{0\pp}<\phi_i<\phi_B$ 
and then the CdL instanton (red line), with $\phi_i=\phi_{0\pp}\simeq 0.044$. Finally, below the CdL instanton we find the family of unbounded-from-below BoNs (orange lines), which are of type 0 and correspond to $0<\phi_i<\phi_{0\pp}$.

The lower left plot in the same figure~\ref{fig:PSCdLBoNdS} shows the action associated to the different types of $V_t$ just described. We see that, as expected, pseudo-bounces connect the higher HM action to the lower CdL one. The action for BoNs is continuous across the CdL point, as already remarked in the previous subsection and for similar reasons. We also see that there is a region with $\phi_i\simlt \phi_{0\pp}$ for which the BoN decay is subdominant as its action is larger than the CdL one. The slope of the action for pseudo-bounces and BoNs agrees with expressions similar to (\ref{dSPSdA}) and (\ref{dSBoNdA})
with derivatives with respect to $p=\phi_i$ rather than $p=A$ (see appendix~\ref{app:dSdA}).

As already mentioned, the BoNs obtained are of type 0, with $V_t\sim V_{tA}e^{\sqrt{6\kappa}\phi}$ and $D\sim D_\infty e^{\sqrt{8\kappa/3}\phi}$ as $\phi\to\infty$. The lower right plot of figure~\ref{fig:PSCdLBoNdS} shows $V_{tA}(\phi_i)$ and $D_\infty(\phi_i)$ in the BoN range $\phi_i\in (0,\phi_{0\pp})$. As happened in the previous case, the maximum of the action (also plotted) occurs at the minimum of $-V_{tA}$. Similarly to what happened for the Minkowski case analyzed in the previous subsection, the plot also illustrates that, for a given UV theory fixing $V_{tA}$, two possible BoN decay channels are available (provided $-V_{tA}$ is above a minimum value), with different actions (and nucleation radius). 

\begin{figure}[t!]
\begin{center}
\includegraphics[width=0.5\textwidth]{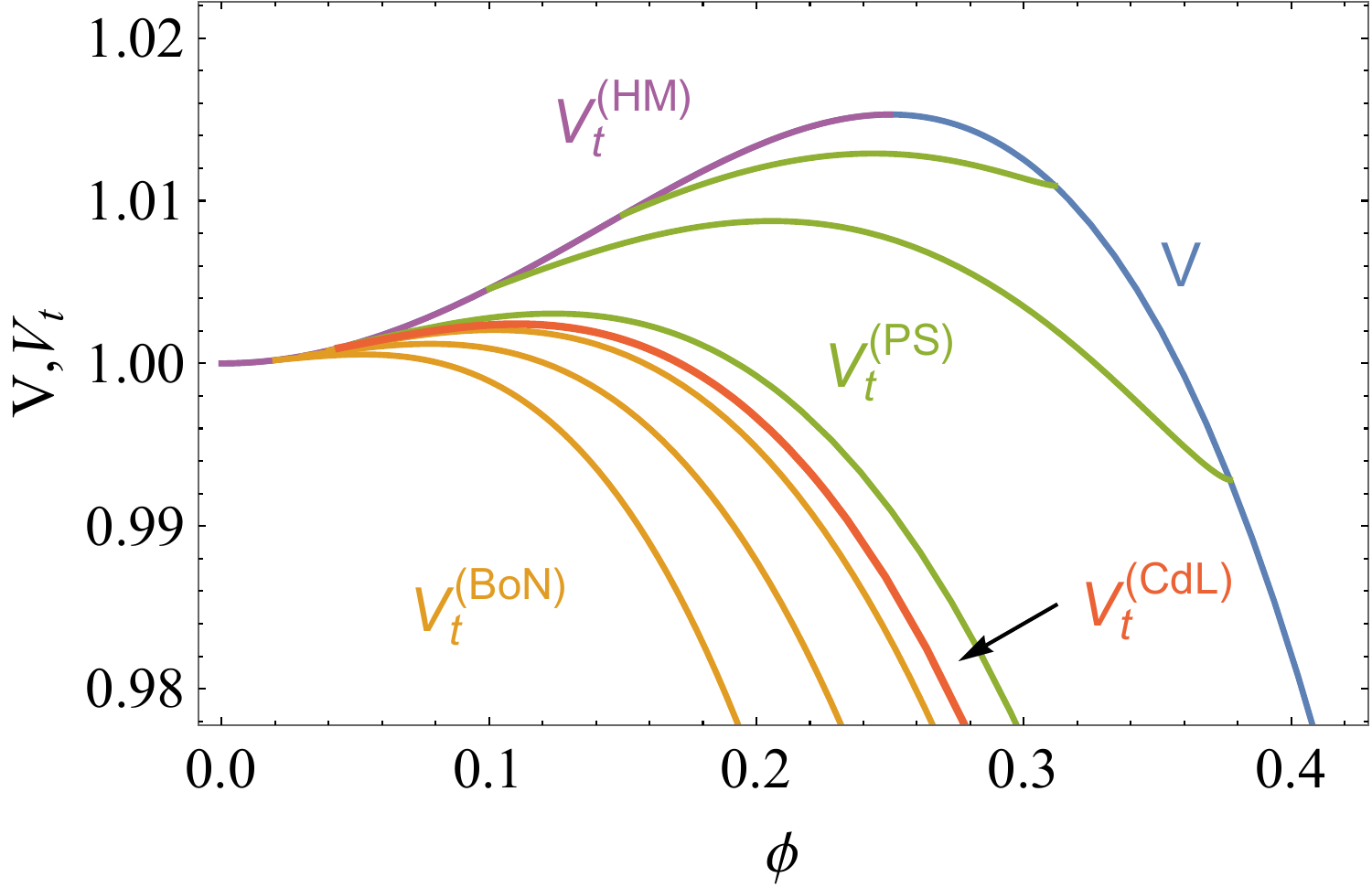}
\includegraphics[width=0.49\textwidth]{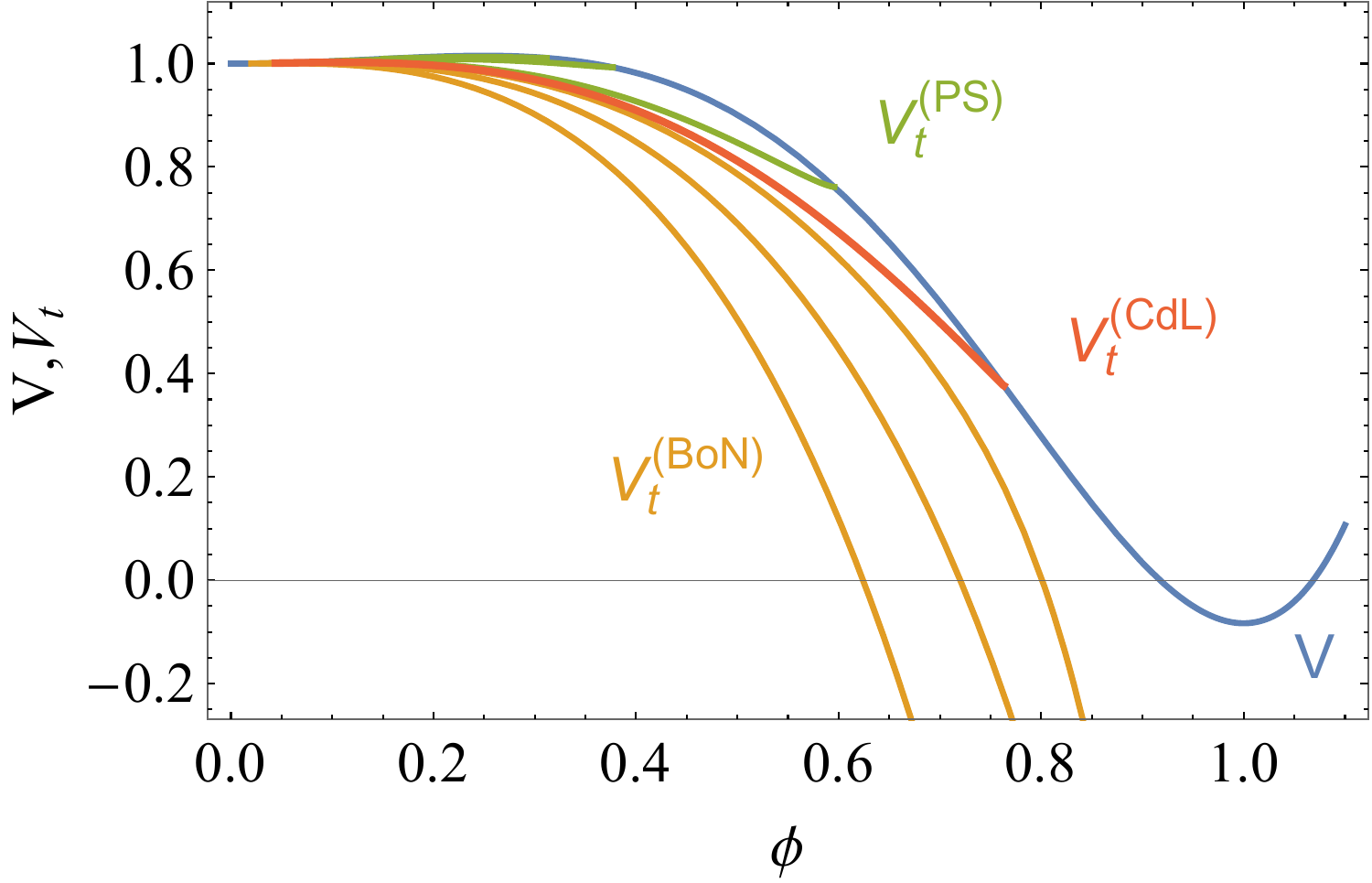}\\
\includegraphics[width=0.47\textwidth]{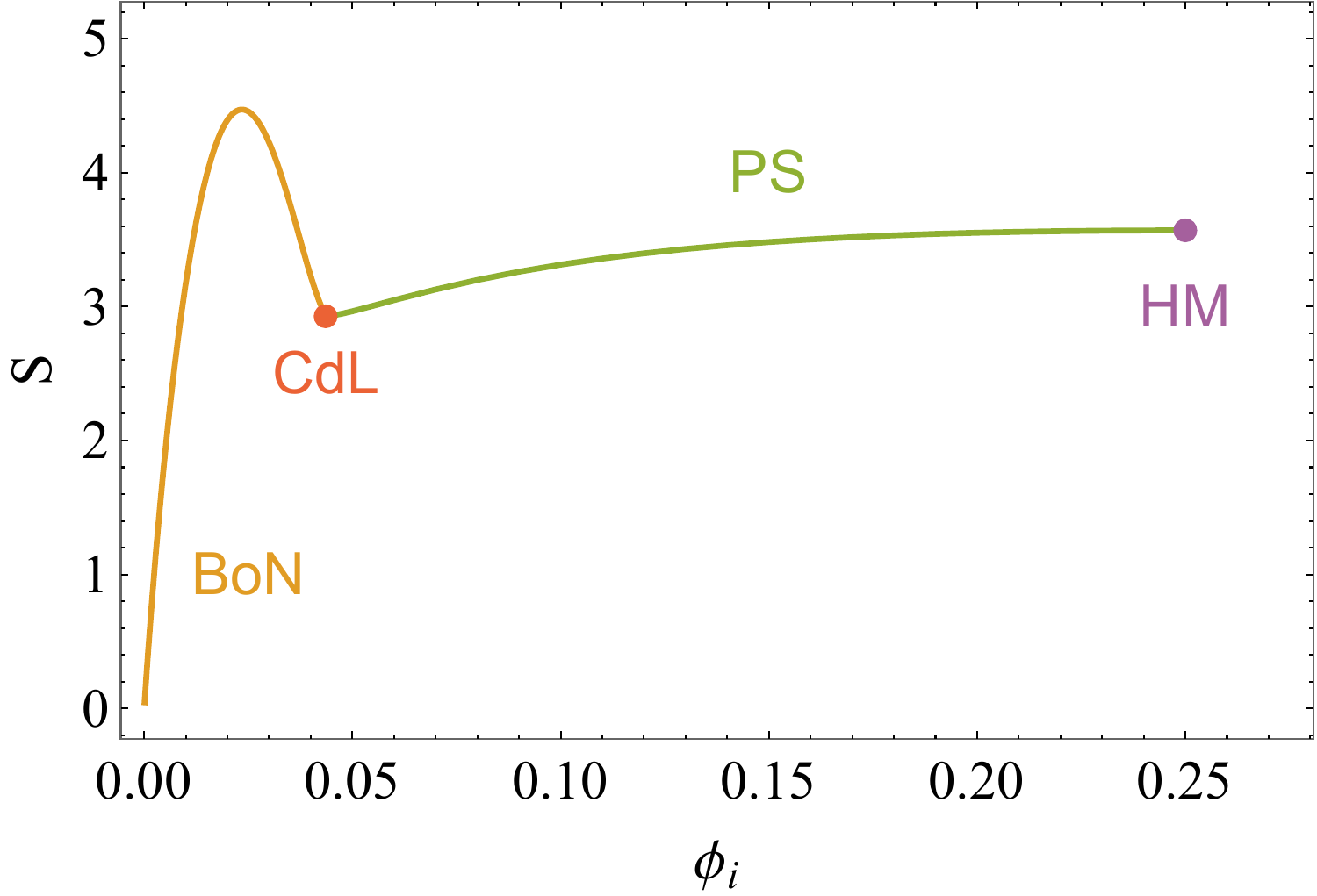}
\includegraphics[width=0.5\textwidth]{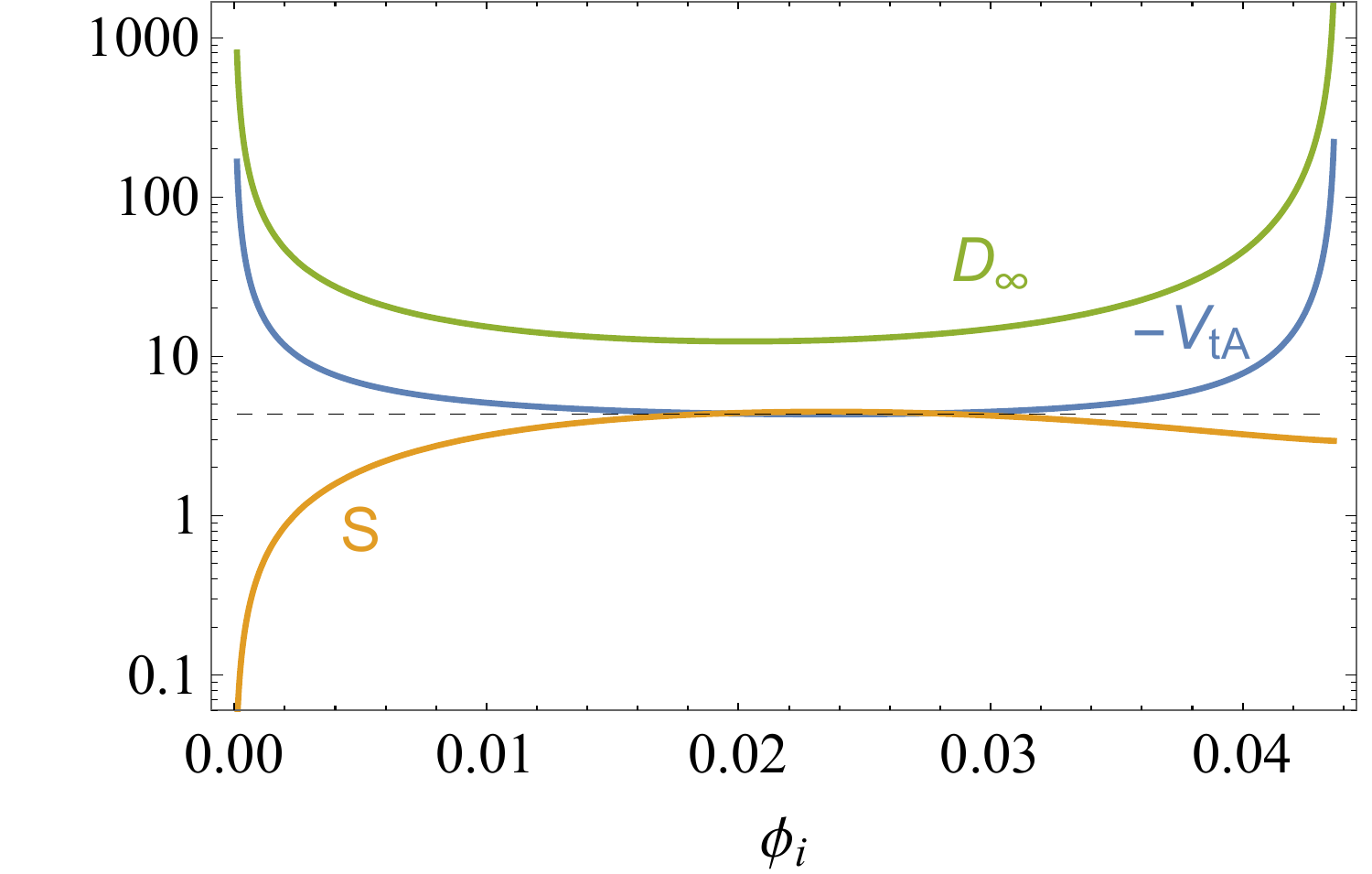}
\end{center}
\caption{
Upper left: Potential (\ref{Vcex}) with  $V_\pp=1$ and tunneling potentials $V_t(\phi_i;\phi)$ of different types: Hawking-Moss (purple); pseudo-bounces for $\phi_{0\pp}<\phi_i<\phi_B$ (green); CdL bounce for $\phi_i=\phi_{0\pp}$ (red) and BoNs for $\phi_i<\phi_{0\pp}$ (orange). 
Upper right: zoomed out version of the previous plot.
Lower right: Tunneling action $S$ with labels/colors indicating different types. 
Lower left: For the BoN range of $\phi_i$, action $S$ and prefactors $V_{tA}, D_{\infty}$ that control the asymptotic $\phi\to\infty$ behaviors $V_t\sim V_{tA}e^{\sqrt{6\kappa}\phi}$ and  $D\sim D_\infty e^{\sqrt{8\kappa/3}\phi}$. 
\label{fig:PSCdLBoNdS}
}
\end{figure}

On the other hand, when the  value of $-V_{tA}$ determined by the higher dimensional theory is below the lower bound [i.e. when $R_{KK}$ is above a certain value, see \eqref{VtDUV}], BoN decay is dynamically forbidden. In other words, the present example illustrates the possibility that BoN decay is obstructed by a dynamical constraint of the form \eqref{eq:dyConst1}, even when the false vacuum is dS. This result contrasts with the situation in standard false vacuum decay, where gravitational quenching only occurs for Minkowski and AdS vacua.

Notice that, as for the Minkowski case, even if BoN is dynamically forbidden, the false vacuum may still decay via the nucleation of CdL bubbles. If we require the KK and $4d$ EFT scales to be well separated, then $R_{KK}$ has to be small compared to the typical EFT length-scale, which in turn implies that $-V_{tA}$ is large due to the relation \eqref{VtDUV}. In this limit, where the EFT is well under control,  BoN is always dynamically allowed, and moreover, it is the fastest decay channel, as can be seen  in the two lower plots of  figure \ref{fig:PSCdLBoNdS}  (when $\phi_i\to 0$).  

As is well known, if $V_\pp$ is increased, the range of $\phi_i$ values for which there are pseudo-bounce solutions shrinks, with the CdL and HM solutions getting closer to each other. Eventually, the two solutions merge and only the HM solution remains. This case is illustrated for $V_\pp=2$ in figure~\ref{fig:BoNHMdS}, left plot. This plot also shows how the solutions cross each other for $\phi_i$ close to the top of the barrier. This, once again, explains the non-monotonic behaviour of $S_{BoN}$ shown on the same figure, right plot.  When $\phi_i$ is very close to the top of the barrier, we have again a strong deflection of the solutions away from the potential. This solutions have two well defined regions: a BoN core (for $\phi\simgt \phi_B$) and a HM tail. These are the hybrid BoN-HM solutions discussed in \cite{DGL}. As in the Minkowski case, the two BoN solutions for a fixed $V_{tA}$ correspond to the two branches of solutions called in \cite{DGL} BoN false-vacuum branch and BoN-HM branch. As before, the first branch is Witten-like, not very sensitive to the potential structure and reaches down to $S=0$, while the second branch is sensitive to the additional structure of the potential and the existence of a HM solution, on which it ends. The two branches merge at the critical $V_{tA}$ as also found in \cite{DGL}.

\begin{figure}[t!]
\begin{center}
\includegraphics[width=0.5\textwidth]{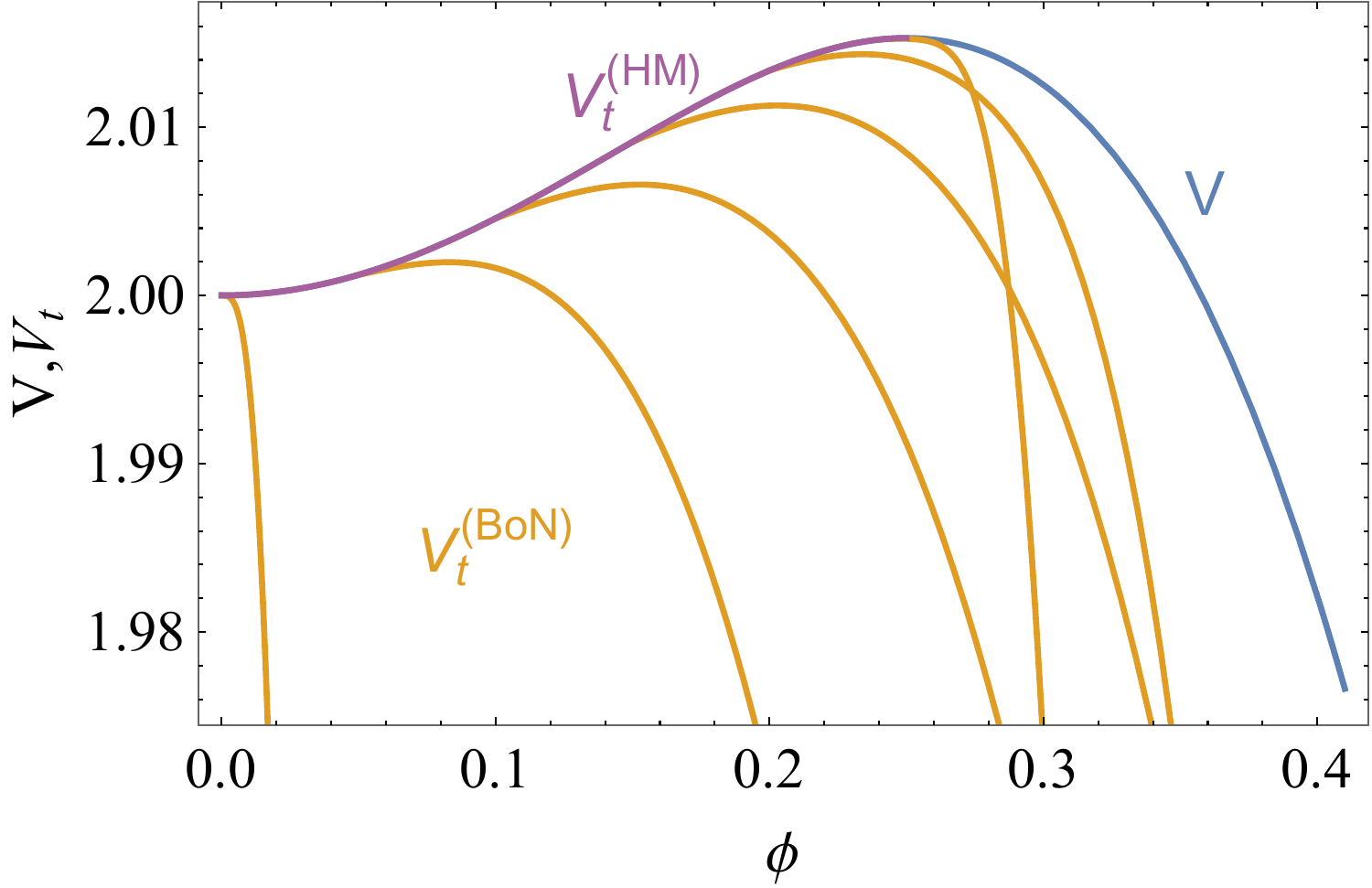}
\includegraphics[width=0.49\textwidth]{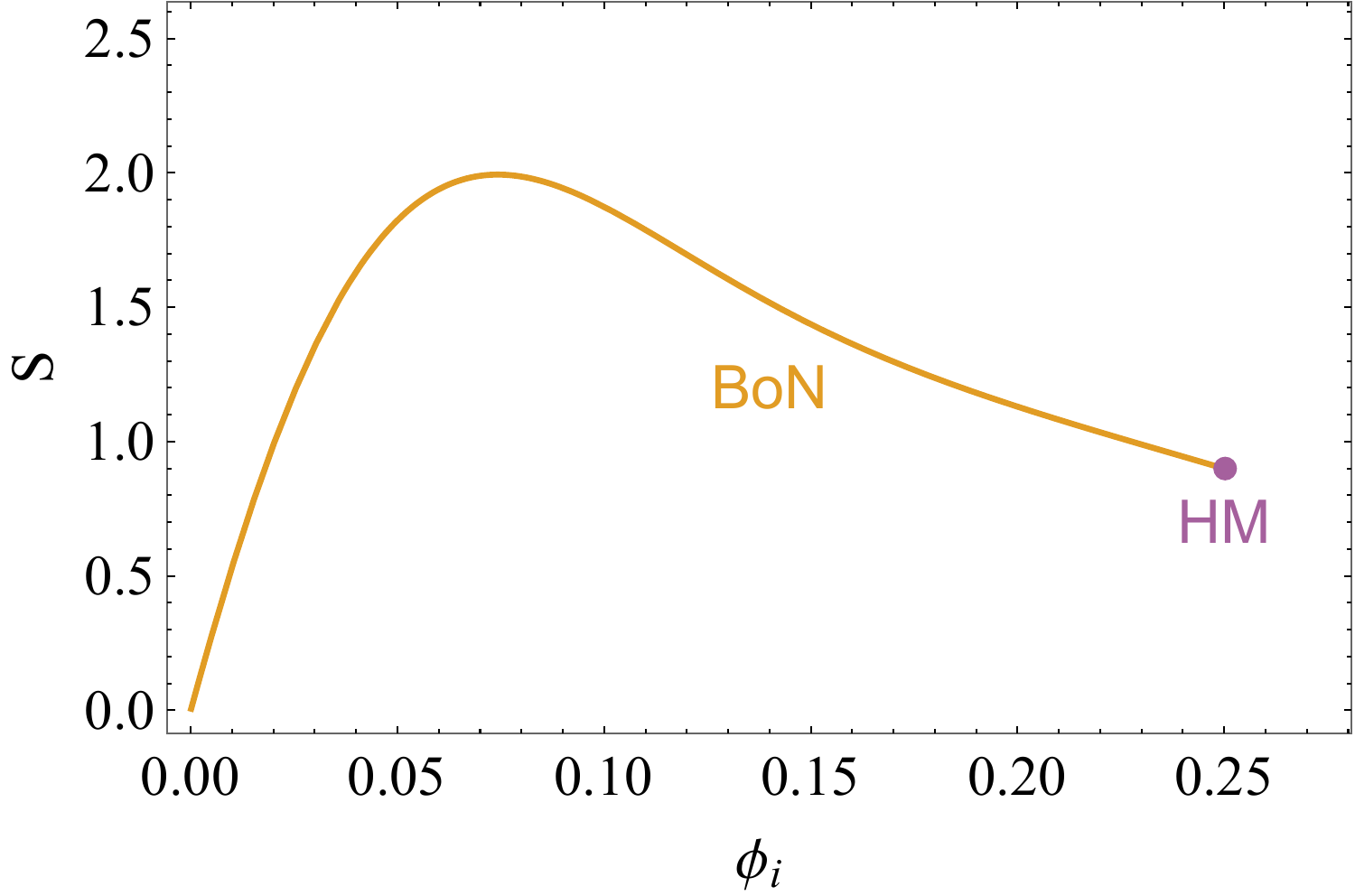}
\end{center}
\caption{
Left: Potential (\ref{Vcex}) with  $V_\pp=2$ and tunneling potentials $V_t(\phi_i;\phi)$ of different types: BoNs for $\phi_i<\phi_B$ (orange) and Hawking-Moss (purple). 
Right:Tunneling action $S$ with labels/colors indicating different types. 
\label{fig:BoNHMdS}
}
\end{figure}

\subsection{AdS False Vacuum}
Consider next the case of an AdS false vacuum, with $V_\pp<0$. Figure~\ref{fig:AdSBoN} shows the potential (\ref{Vfornum}), with $V_\pp=-1$, and different $V_t$ solutions, obtained numerically by solving the EoM for $V_t$, varying the $A$ parameter that appears in the low field expansion of $V_t$, (\ref{VtexpAdS}). We show as well the solution $\overline{V_t}$ of $D=0$ with $\overline{V_t}(0)=V_\pp$. For this particular value of $V_\pp$ we see that $\overline{V_t}$ does not intersect the potential beyond the barrier and this means that the false vacuum is stable against CdL decay (CdL solutions  reappear for small enough $V_\pp$). Thus we only have a family of (type 0) BoN solutions, as shown. The action corresponding to these solutions is shown, as a function of $A$, on the right plot of the same figure, with $S\to\infty$ as $A\to 0$ (or $V\to\overline{V_t}$). This time the action is a monotonic function and so must be $V_{tA}(A)$, implying that for a fixed value of $V_{tA}$ in the UV theory there is a single BoN decay channel. CdL suppression of BoN decay ($S \to \infty$, $\mathcal{R}\to \infty$) for a vacuum saturating \eqref{eq:dyConst1} can only occur when the standard  CdL decay is dynamically forbidden.  Indeed, for CdL decay to be allowed $\overline V_t$ must intersect the scalar potential $V$ at some field value, and therefore a tunneling potential satisfying $D=0$ ($V_t =\overline V_t$) does not have the right asymptotic behaviour at large $\phi$ to represent BoN nucleation.

\begin{figure}[t!]
\begin{center}
\includegraphics[width=0.5\textwidth]{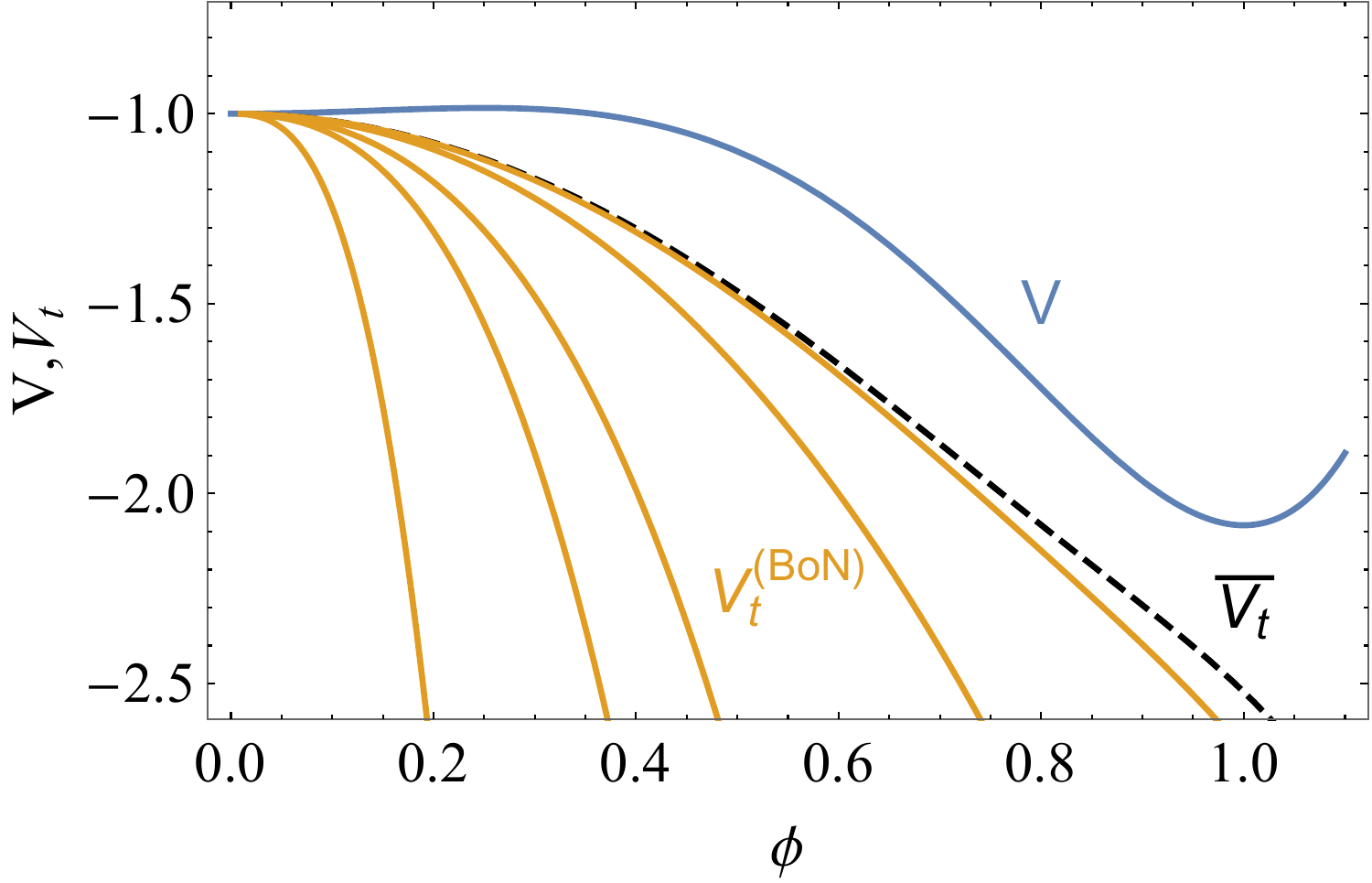}
\includegraphics[width=0.47\textwidth]{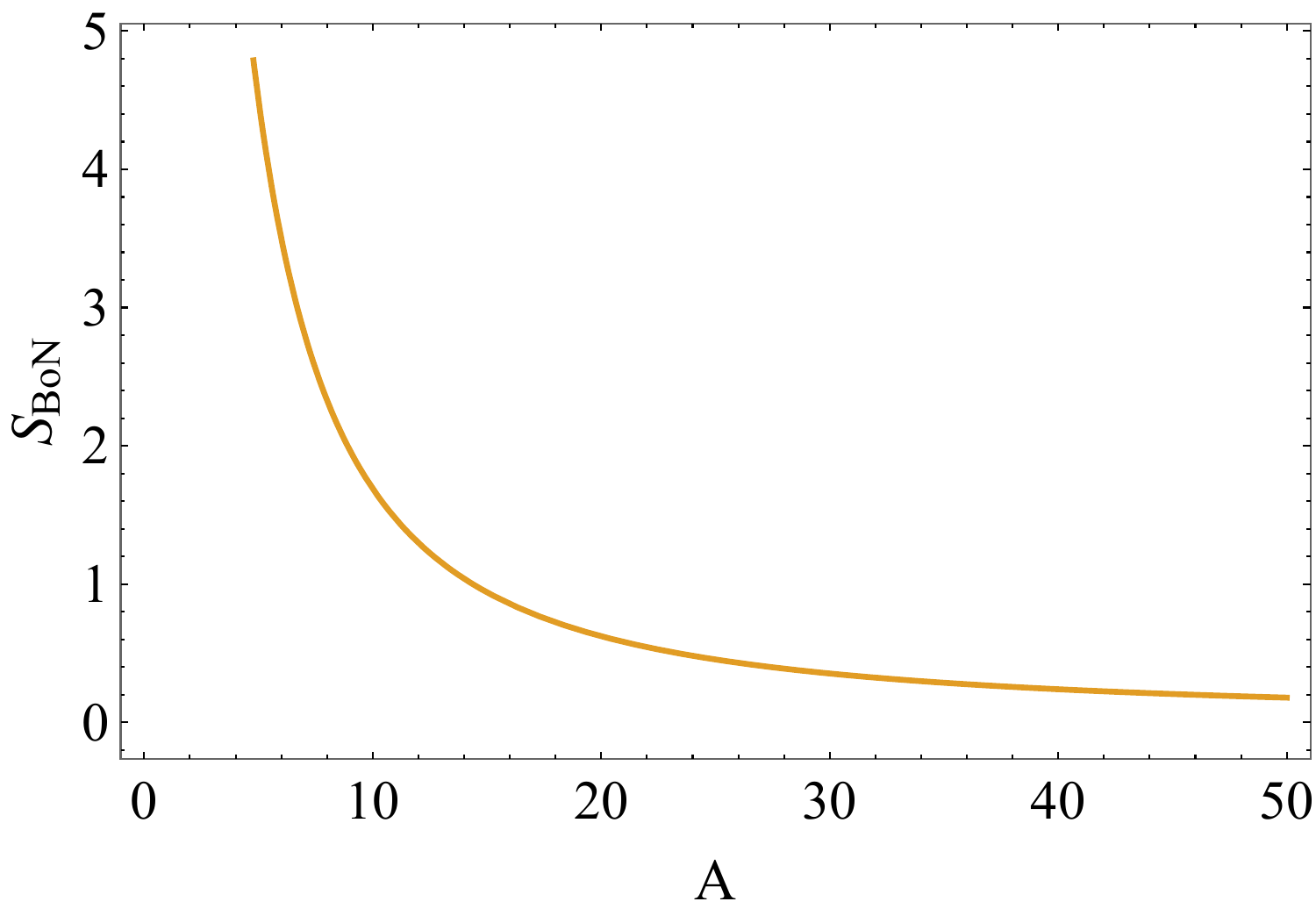}
\end{center}
\caption{
Left: Potential (\ref{Vcex}) with $V_\pp=-1$ and tunneling potentials $V_t(\phi_i;\phi)$ of different types: BoNs for $\phi_i<\phi_B$ (orange) and Hawking-Moss (purple). 
Right:Tunneling action $S$ with labels/colors indicating different types. 
\label{fig:AdSBoN}
}
\end{figure}

\begin{figure}[t!]
\begin{center}
\includegraphics[width=0.49\textwidth]{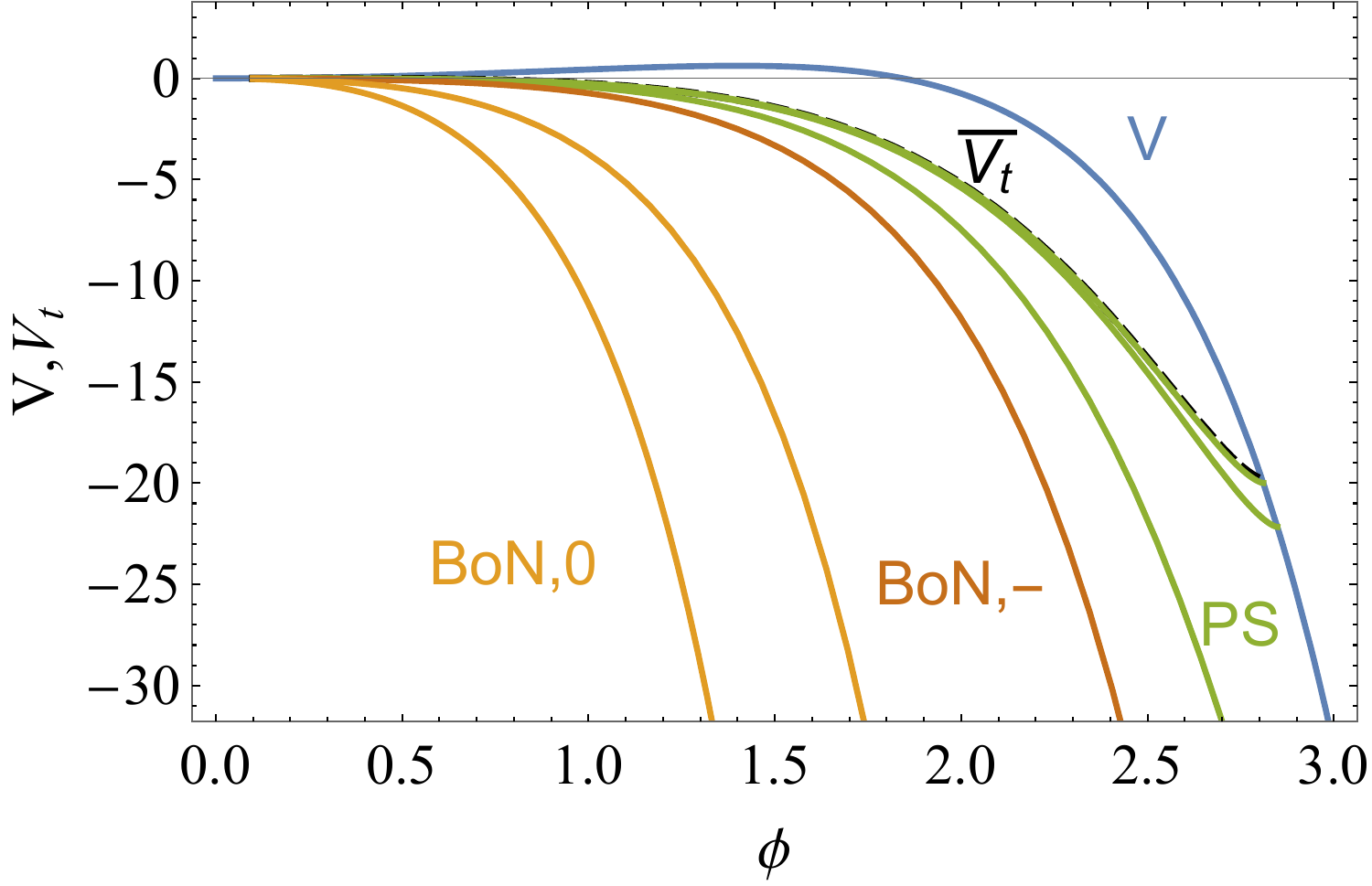}
\includegraphics[width=0.5\textwidth]{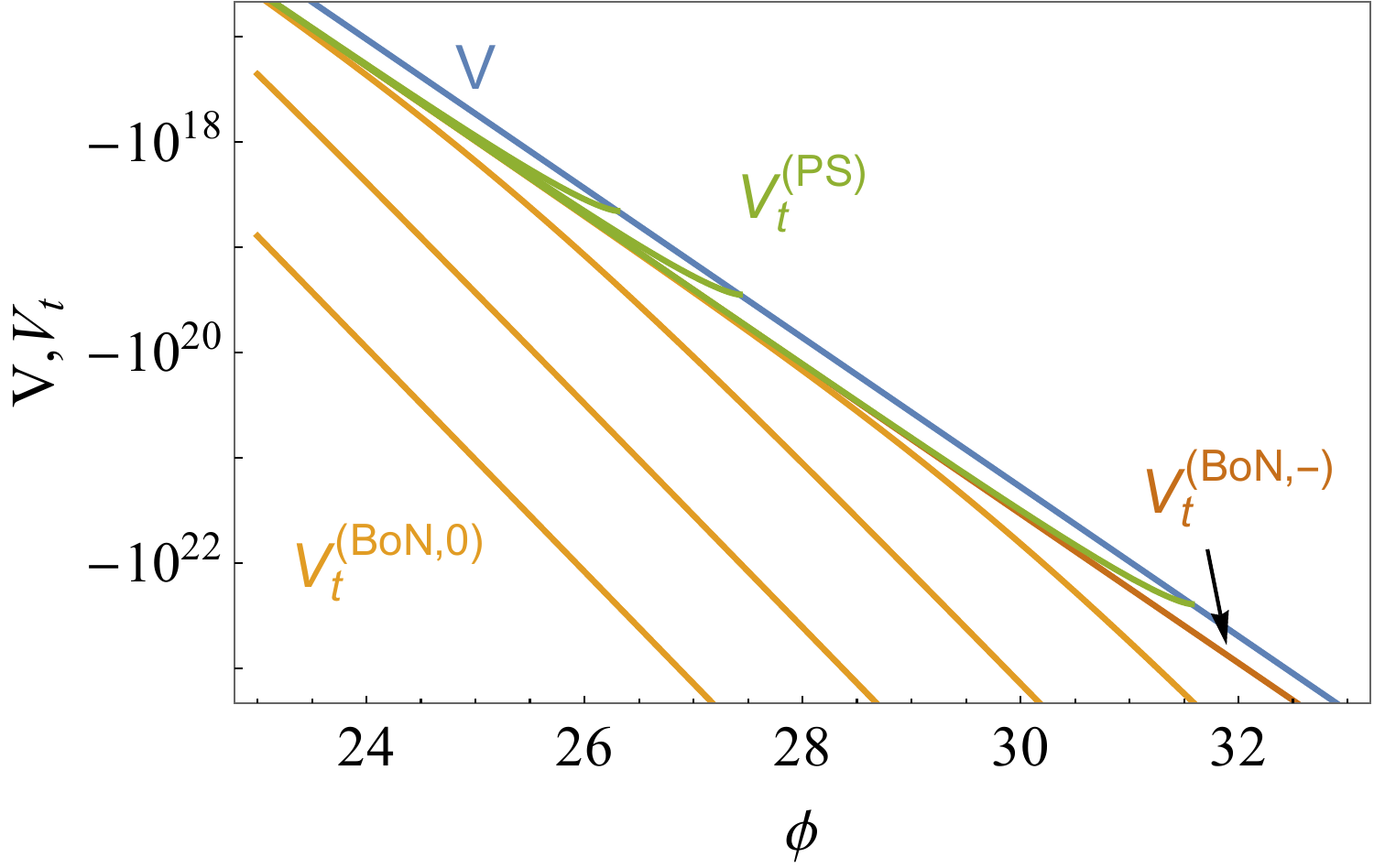}
\\
\includegraphics[width=0.49\textwidth]{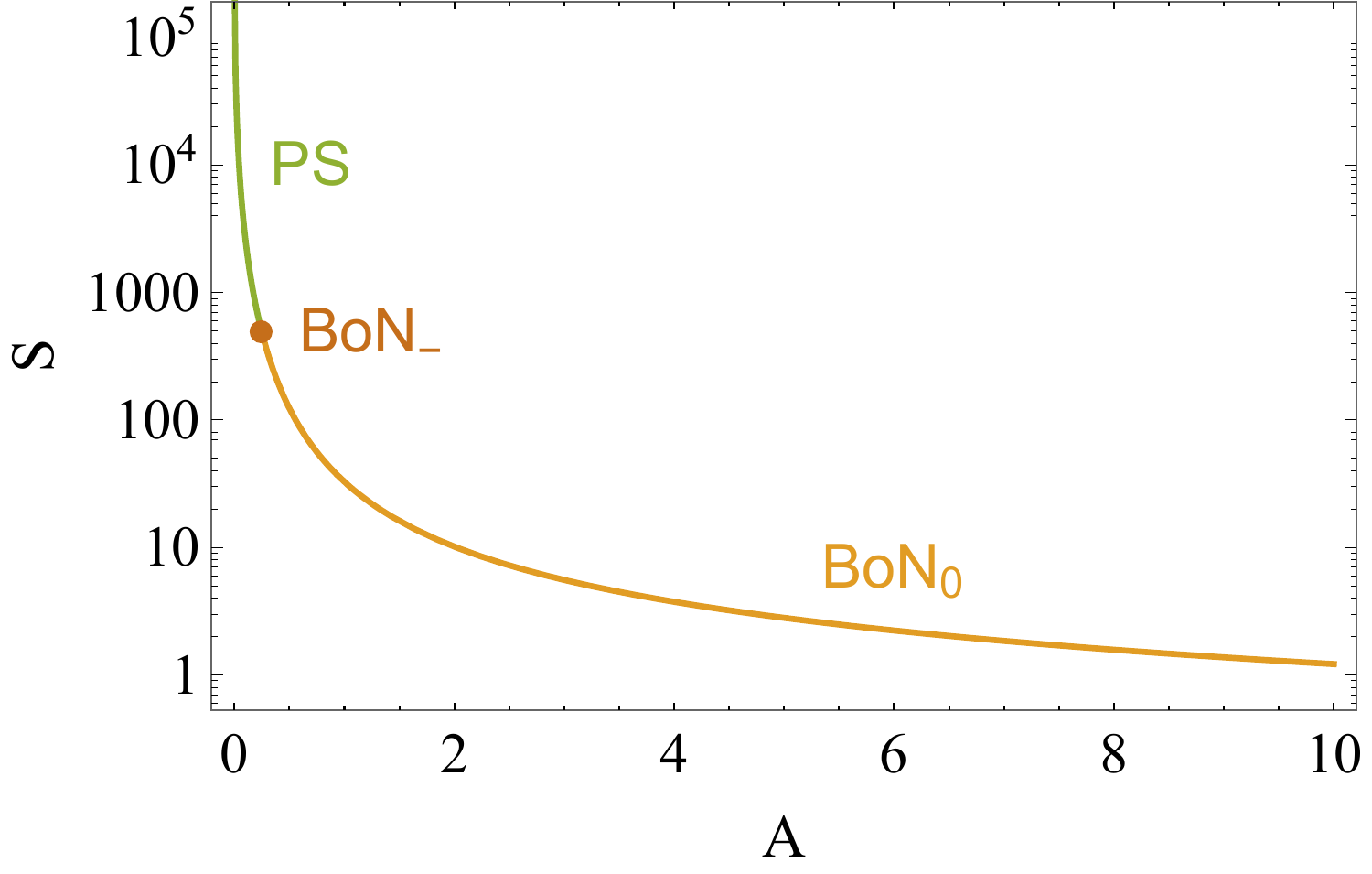}
\includegraphics[width=0.5\textwidth]{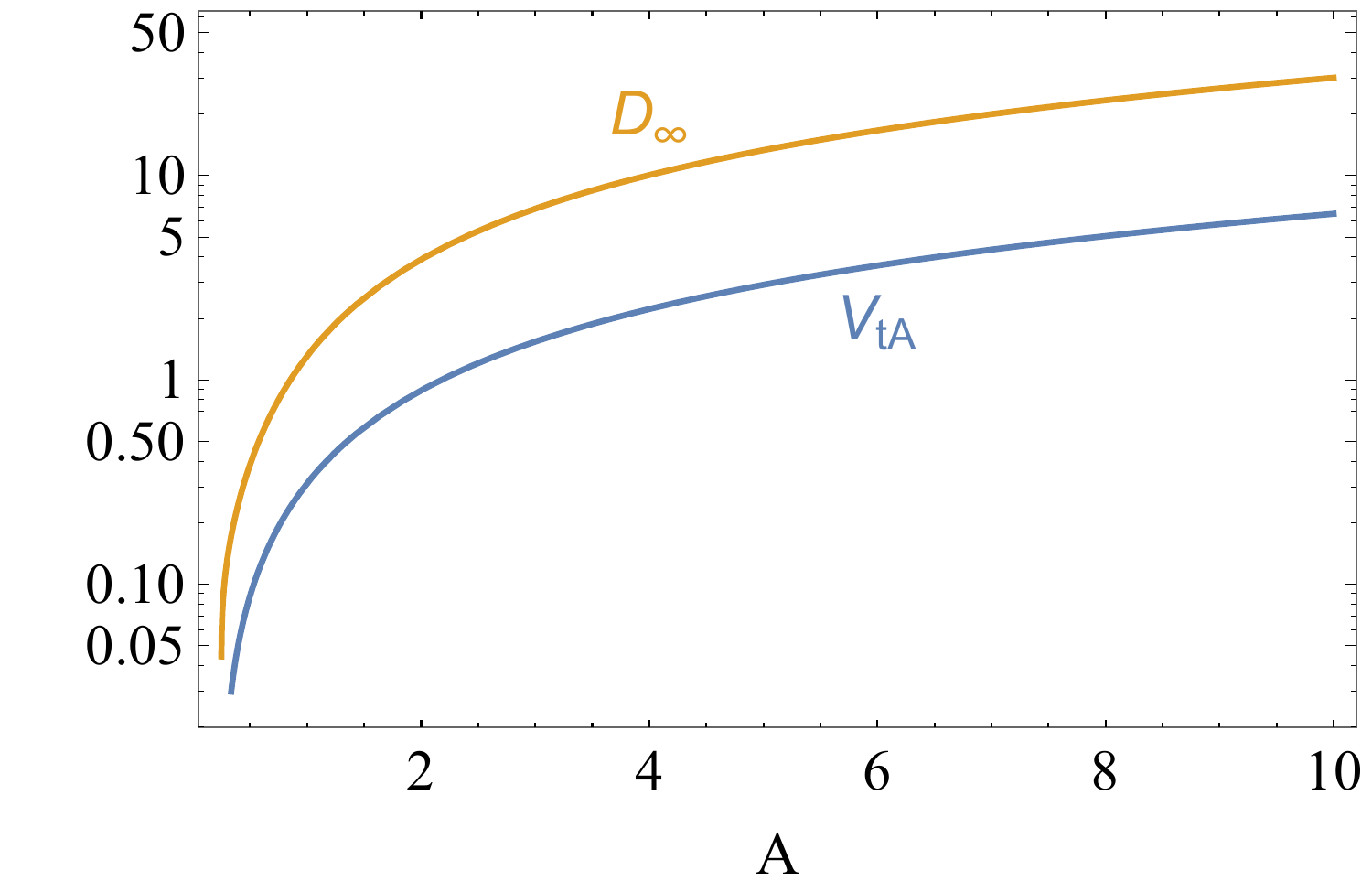}
\end{center}
\caption{
Upper left: Potential (\ref{VBoNm}) and tunneling potentials $V_t(A;\phi)$ of different types: critical $\overline{V_t}$ (black dashed); pseudo-bounces for $0<A<A_\mm$ (green); type $\mm$ BoN for $A=A_\mm$ (brown) and type-0 BoNs for $A>A_\mm$ (orange). 
Upper right: logarithmic view of the higher field structure of the previous solutions.
Lower left:Tunneling action $S$ with labels/colors indicating different types. Lower right:
For the BoN range of $A$, prefactors $V_{tA}, D_{\infty}$ that control the asymptotic $\phi\to\infty$ behaviors $V_t\sim V_{tA}e^{\sqrt{6\kappa}\phi}$ and  $D\sim D_\infty e^{\sqrt{8\kappa/3}\phi}$.
\label{fig:BoNm}
}
\end{figure}

\subsection{ \texorpdfstring{Type \bma{-} BoNs}{Type - BoNs}\label{TypeminusBoNs}}

In order to examine the behaviour of type $-$ BoNs and their interplay with other instantons, we take now a potential that goes to negative values exponentially, $V\sim - e^{a\sqrt{6\kappa}\phi}$, with $a<1$. We supplement it with a second subleading exponential term (as we know simple exact BoN  solutions of type $-$ with just two exponentials) and add further polynomial terms to get a proper minimum and barrier. For our concrete example we take (setting $\kappa=1$)
\be
V(\phi)=V_\pp -e^{\sqrt{8/3}\phi}+3e^{\sqrt{3/2}\phi}-2-\frac{5}{\sqrt{6}}\phi-\frac{5}{12}\phi^2-\frac{17}{36\sqrt{6}}\phi^3\ ,
\label{VBoNm}
\ee
which corresponds to $a=2/3$ and has a low-field expansion of the form
\be
V(\phi)=V_\pp +\frac12 m^2 \phi^2 -\frac14\lambda\phi^4+...
\ee
with $m=1$, $\lambda=13/216$ and we choose $V_\pp=-0.001$. Figure~\ref{fig:BoNm}, upper left plot, shows this potential and several different tunneling potentials as in previous examples. In this case, see upper right logarithmic plot, we find that pseudo-bounces continue indefinitely\footnote{In section~\ref{sec:Vt}, we argued that decay is possible whenever $\overline{V_t}$ intersects $V$. This case is not a counterexample but rather a limiting case with the CdL solution pushed to infinity \cite{PS}. Normal vacuum decay can certainly take place but mediated by pseudo-bounces.} and below them we find a type $-$ BoN,
with $V_t\sim -(9/5)e^{\sqrt{8/3}\phi}$ followed by a family of type 0 BoNs,
with $V_t\sim V_{tA}(A)e^{\sqrt{6}\phi}$. 

In this case there are no strong deflections of the $V_t$ solutions below the CdL one, but rather a smooth transformation of pseudo-bounce solutions into the type $-$ BoN and then on type 0 BoNs. As a result, the action is continuous and monotonic (lower left plot of figure~\ref{fig:BoNm}) and $V_{tA}$ and $D_\infty$ are monotonic as well. We also see that 
$V_{tA}$ goes to zero at the lowest end of the BoN regime, when the $V_{tA}(A)e^{\sqrt{6}\phi}$ term of type 0 BoNs switches off, leaving the type $-$ term $V_t\sim -(9/5)e^{\sqrt{8/3}\phi}$ of the BoN $-$ solution as the dominant one. From this example we see that one can think of type $-$ BoNs as a particular case of type 0 (with exponential potential having $a<1$) for which $V_{tA}$ (the prefactor of $e^{\sqrt{6\kappa}\phi}$) vanishes. This also explains why one does not find families of type $-$ BoNs, which rather appear as single solutions at the boundaries of type 0 BoN families. This is consistent with the fact that in higher dimensional theories that admit type $-$ BoNs the values of $V_A$ and $V_{tA}$ (the prefactors of $e^{a\sqrt{6\kappa}\phi}$ in $V$ and $V_t$) are both fixed in terms of $R_{KK}$, see section~\ref{sec:topdown}.
Type $-^*$ BoNs have a similar behaviour and also appear as limiting cases of type 0 BoN families, see subsections~\ref{sect:CdLPlateau} and \ref{sect:Counter}.

\section{Analytic Examples\label{sec:anex}}

Besides relying on numerical analyses, as was done in the previous sections, it is often useful to have exactly solvable examples and the tunneling potential formalism is particularly well suited to this purpose. Refs. \cite{Eg,EFH} show how one can construct pairs of analytic $V$ and $V_t$ which satisfy the EoM (\ref{EoM}) for conventional CdL decays, by postulating a simple  $V_t$ and solving (\ref{EoM}) for $V$. Using the same technique, it is also possible to find analytically tractable examples of $V_t$'s for BoN decays. In this subsection we present a few examples, illustrating the four different types of asymptotic behaviours discussed in section~\ref{sec:BotUp}. Further examples and details can be found in the appendices \ref{App:Vexp}-\ref{App:morex}. We set $\kappa=1$ in the rest of this section. All examples below can be rescaled by a constant, with $V\to A V$ and $V_t\to A V_t$ as this rescaling leaves the EoM for $V_t$ (\ref{EoMVt}) invariant. Under this rescaling  the tunneling action (\ref{SVt}) is rescaled as $S\to S/A$.

\subsection{A Type 0 Example \label{sect:simple0}}

A very simple type 0 example is given by
\be
V=\frac89 e^{\sqrt{2/3}\phi}\ ,\quad
V_t=-e^{\sqrt{6}\phi}+ e^{\sqrt{2/3}\phi}\ ,
\label{VtV0}
\ee
which has $a=1/3$. It can be checked that $V$ and $V_t$ above satisfy eq.~(\ref{EoM}). We use this first example to discuss some general features that are common to the examples we present.

The expressions above are assumed to hold for $\phi\geq \phi_{0\pp}= -\sqrt{3/2}\log 3$, the field value at which $V=V_t$. We assume that the potential has a dS minimum for $\phi<\phi_{0\pp}$ but the shape of $V$ in that region is not important.
One can simply assume $V$ is parabolic, $V=V_c + m^2 (\phi-\phi_\pp)^2/2$ with a minimum at some $\phi_\pp<\phi_{0\pp}$ and the two constants $V_c$ and  $m^2$ fixed to get a continuous $V$ and $V'$ at $\phi_{0\pp}$. This kind of construction is similar in the rest of examples we discuss.

Figure~\ref{fig:type0}, left plot, shows such $V$ and $V_t$ as described above. The  dS minimum of the potential occurs at $\phi_\pp=-2$. The tunneling potential $V_t$ coincides with $V$ between $\phi_\pp$ and $\phi_{0\pp}$, and takes the form (\ref{VtV0}) for $\phi\geq \phi_{0\pp}$.

The potential is positive without apparent signs of any instability: $V$ cannot decay at all via the usual HM or CdL channels. However, the BoN decay of the dS vacuum at $\phi_\pp$ is possible with finite action. The Euclidean action for this decay  can be obtained analytically  and consists of the usual two contributions: a Hawking-Moss-like part from
$\phi_\pp$ to $\phi_{0\pp}$ and a CdL-like part from $\phi_{0\pp}$ to $\infty$, see \eqref{SHMCdL}. One gets
\be
S_{\rm BoN}=24\pi^2\left[\frac{1}{V(\phi_{\pp})}-\frac{27}{8}\right]+\frac{81\pi^2}{2}\ ,
\ee
where we leave the value of $V(\phi_\pp)$ unspecified.

It is interesting to note that, due to the simple way we generate this effective $4d$ potential, the instanton only explores the region described by Eq. (\ref{VtV0}). However, this is exactly the form of the potential one would find by compactifying a $5d$ universe with a pure cosmological constant. In turn, this means that the instanton solution can be actually uplifted to a locally  anisotropic description of pure de Sitter space, namely, a solution of the form 
\be
ds_{5d}^2 = dr^2 + H^{-2}\cos^2(H r) d \Omega_3^2 + H^{-2}\sin^2(H r) d\theta_5^2~.
\ee
Seen from $4d$ the solution inside the horizon looks like the type 0 BoN. The relation between the BoN solutions in de Sitter space and the
anisotropic slicing of de Sitter have been already discussed in \cite{BRS}.

\begin{figure}[t!]
\begin{center}
\includegraphics[width=0.49\textwidth]{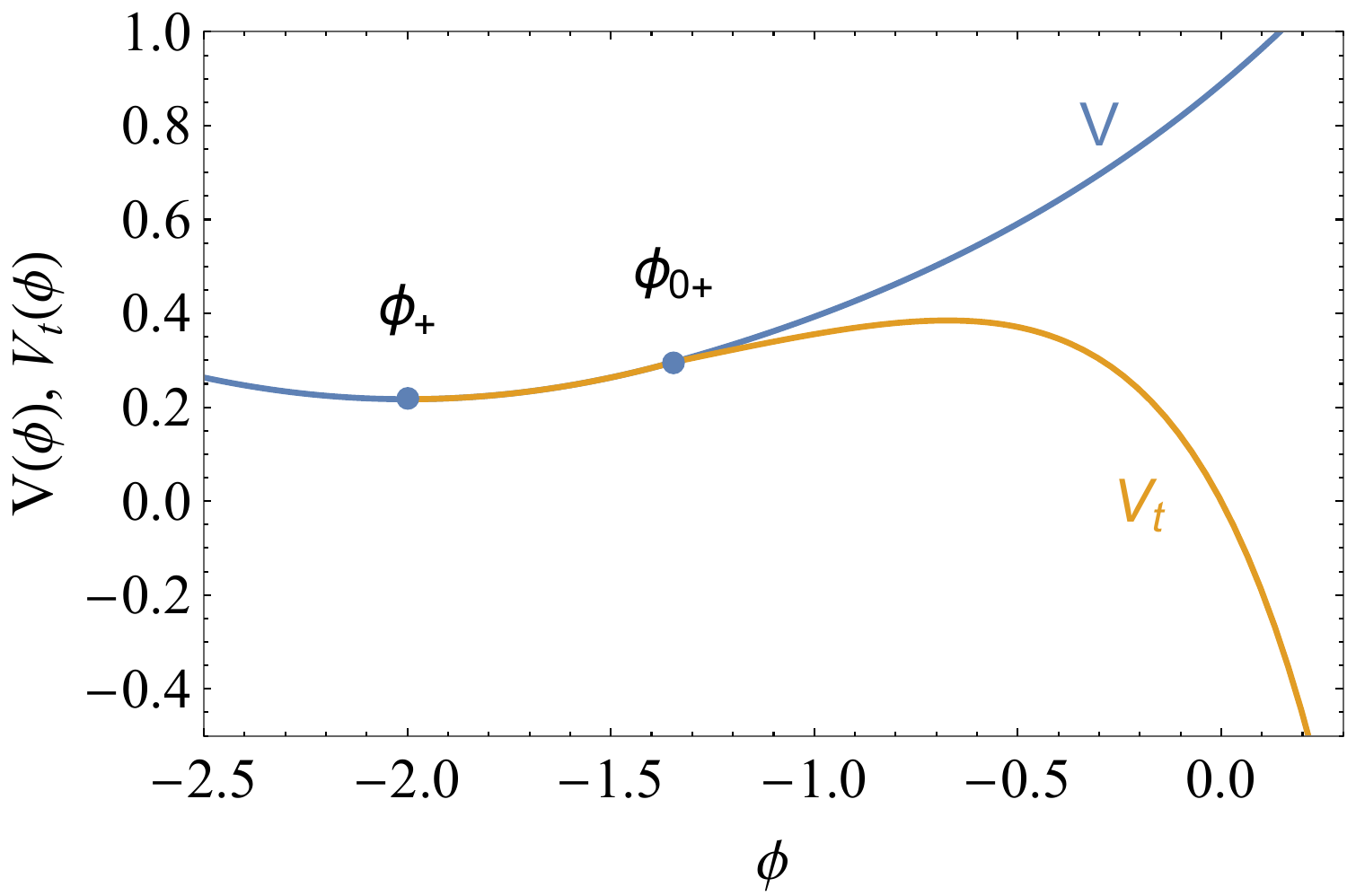}
\includegraphics[width=0.49\textwidth]{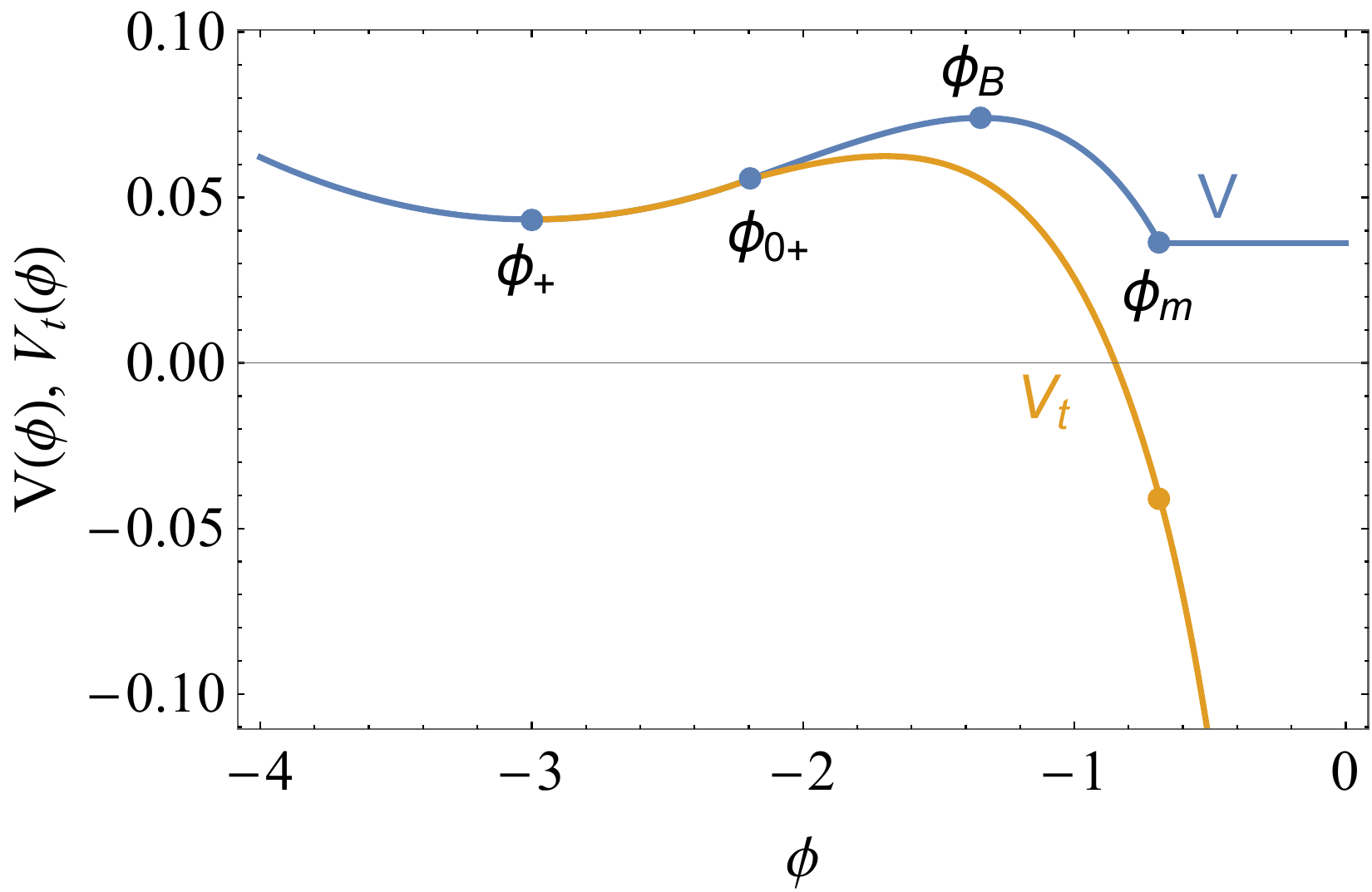}
\end{center}
\caption{Potential, $V$, and tunneling potential, $V_t$, for the type $0$ examples of subsections~\ref{sect:simple0} (left) and \ref{sect:type1C} (right). Dots mark the dS minimum at $\phi_{\pp}$, the starting point of the CdL part of the BoN at $\phi_{0\pp}$ as well as, for the right plot, the barrier maximum at $\phi_B$, and the matching point $\phi_m$ beyond which $V$ is constant.
\label{fig:type0}
}
\end{figure}

\subsection{A Type \texorpdfstring{$\bma{0}$}{0} Example for \texorpdfstring{$\bma{V(\phi>\phi_m)=}$}{V(phi>phim=)} Constant\label{sect:type1C}}

It is possible to find $V_t$ corresponding to a constant $V$ although the solution is not simple. We give the details of this derivation in Appendix~\ref{App:Vconstant}.
It is then easy to construct an example with $V(\phi>\phi_m)=V_\infty$, with $V_\infty$ a constant, for some $\phi_m$ by matching the $V_t$ solution obtained in Appendix~\ref{App:Vconstant} for a constant potential to some other solution for $\phi<\phi_m$, for instance the $V$ and $V_t$ given in subsection~\ref{sect:CdLPlateau}. Matching at $\phi_m$ should impose continuity of 
$V$, $V_t$ and $V_t'$. The complete $V_t$ obtained in this way should feature a maximum at some field value and therefore we should choose $\phi_m>\phi_T=-\sqrt{3/2}\log 4$, the value at which $V_t'=0$ in this example. We also require $V_\infty>0$ and so $\phi_m<\phi_x\equiv\sqrt{3/2}\log(2/3)$, value at which $V(\phi_x)=0$. 
Figure~\ref{fig:type0}, right plot, shows $V$ and $V_t$ after performing such matching,
choosing  $\phi_m=\sqrt{3/2}\log(4/7)\simeq -0.69$, for which $V_\infty=V(\phi_m)\simeq 0.036$. For more details about the matching procedure see Appendix~\ref{App:Vconstant}.

\subsection{An Example of Type \texorpdfstring{$\bma{-}$}{-}\label{sect:minus}}
For this example we take
\be
V=V_A e^{2\phi} +10A e^{\phi}\ ,\quad
V_t(\phi) = 3V_A e^{2\phi} +12A e^\phi\ ,
\label{exminus}
\ee
with $V_A<0, A>0$, which indeed is a type $-$ example, with $a=\sqrt{2/3}$. 
Figure~\ref{fig:typeminus}, left plot, shows such $V$ and $V_t$ with $A=1/2, V_A=-1$, completed for $\phi<\phi_{0\pp}=\log(-A/V_A)$ as in previous examples. The  dS minimum of the potential has been fixed at $\phi_\pp=-3$ and its maximum occurs at $\phi_B=\log(-5A/V_A)$. The tunneling potential $V_t$ coincides with $V$ between $\phi_\pp$ and $\phi_{0\pp}$, and takes the form (\ref{exminus}) for $\phi\geq \phi_{0\pp}$.

\begin{figure}[t!]
\begin{center}
\includegraphics[width=0.49\textwidth]{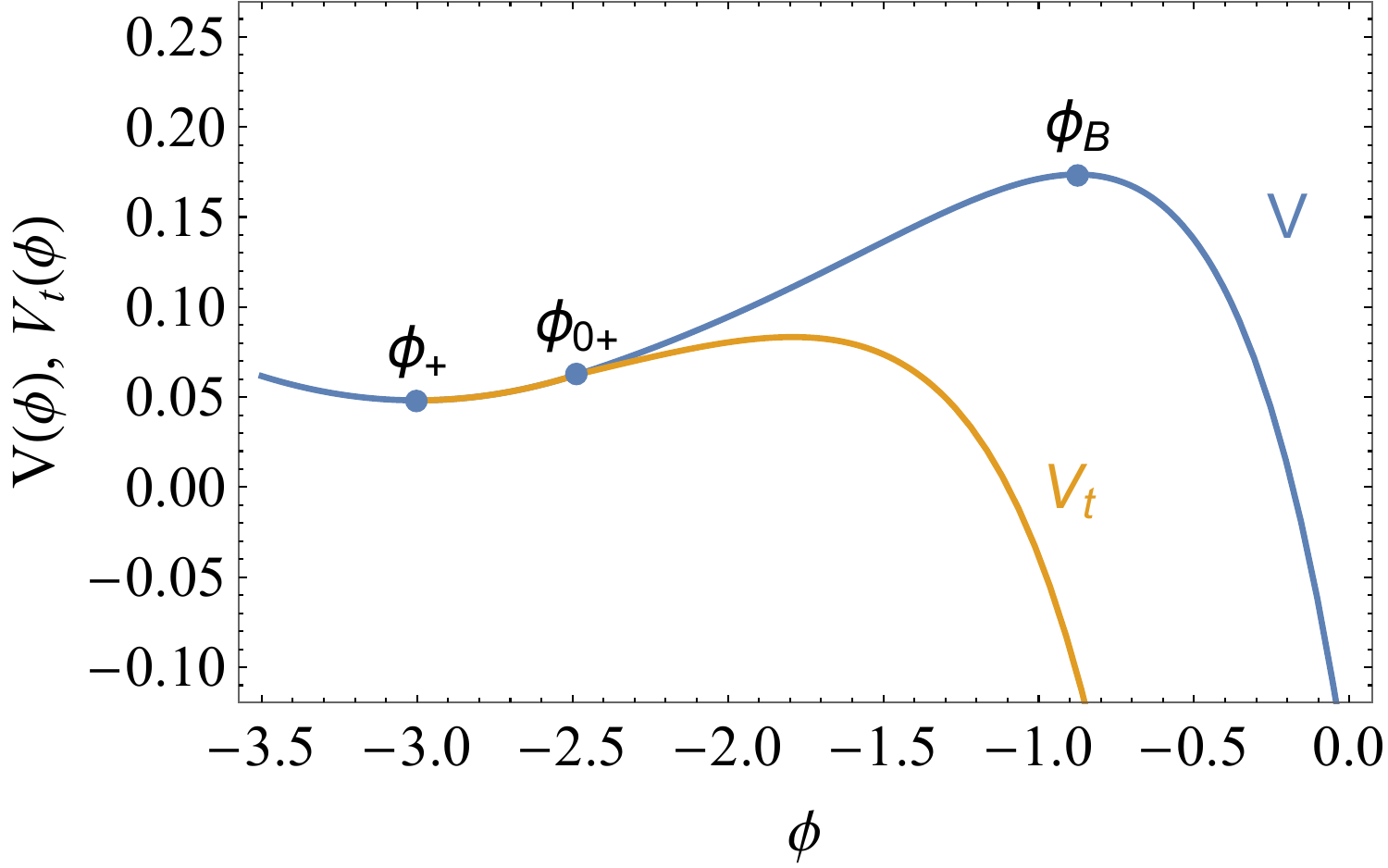}
\includegraphics[width=0.46\textwidth]{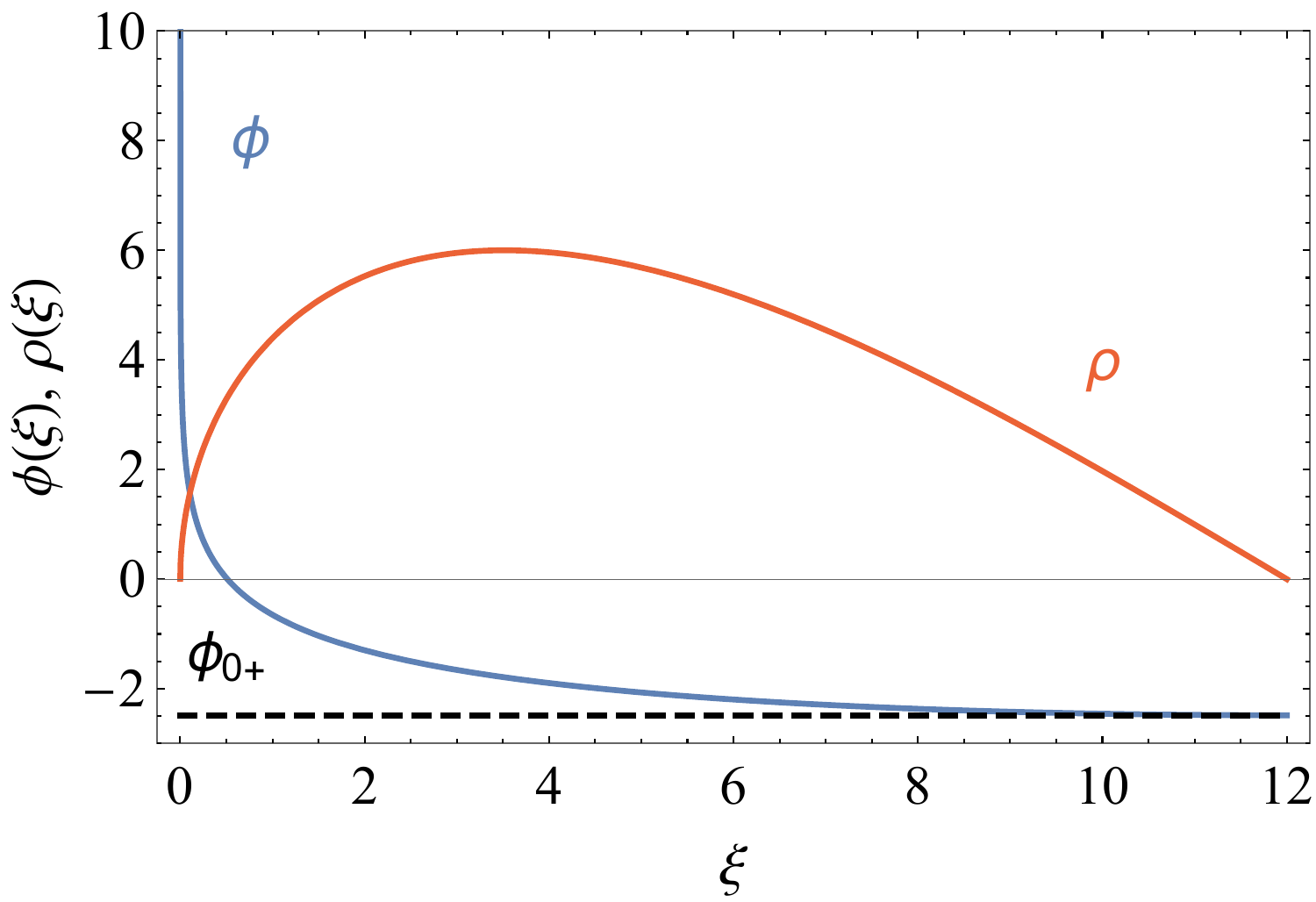}
\end{center}
\caption{
For the type $-$ example of subsection~\ref{sect:minus}: (Left) Potential, $V$, and tunneling potential, $V_t$. Dots mark the dS minimum at $\phi_{\pp}$, the barrier maximum at $\phi_B$ and the starting point of the CdL part of the BoN at $\phi_{0\pp}$. (Right) Profiles of the field, $\phi(\xi)$, and the metric function, $\rho(\xi)$, corresponding to the (compact) CdL part of the BoN instanton. 
\label{fig:typeminus}
}
\end{figure}

\begin{figure}[t!]
\begin{center}
\includegraphics[width=0.5\textwidth]{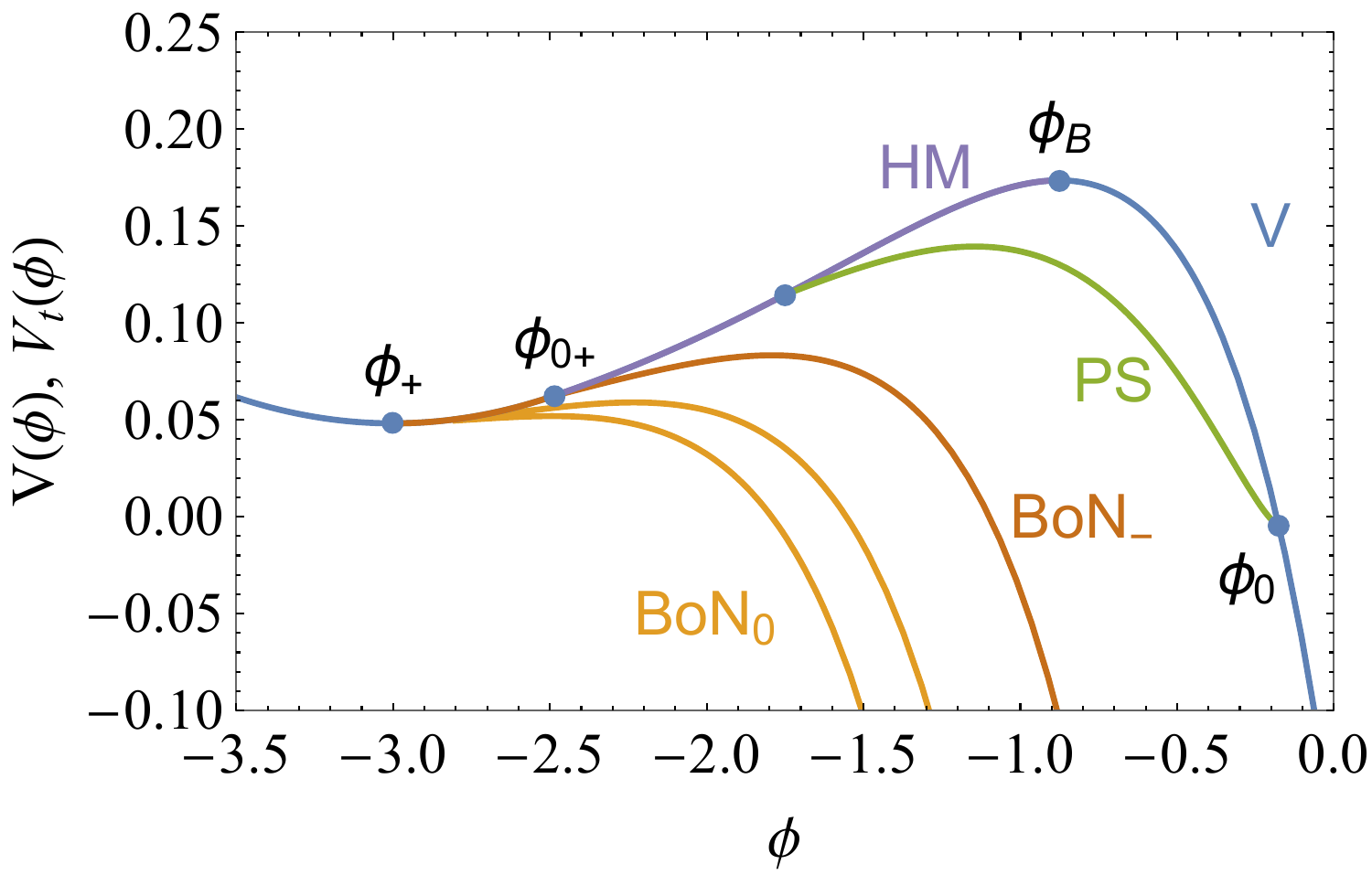}
\includegraphics[width=0.49\textwidth]{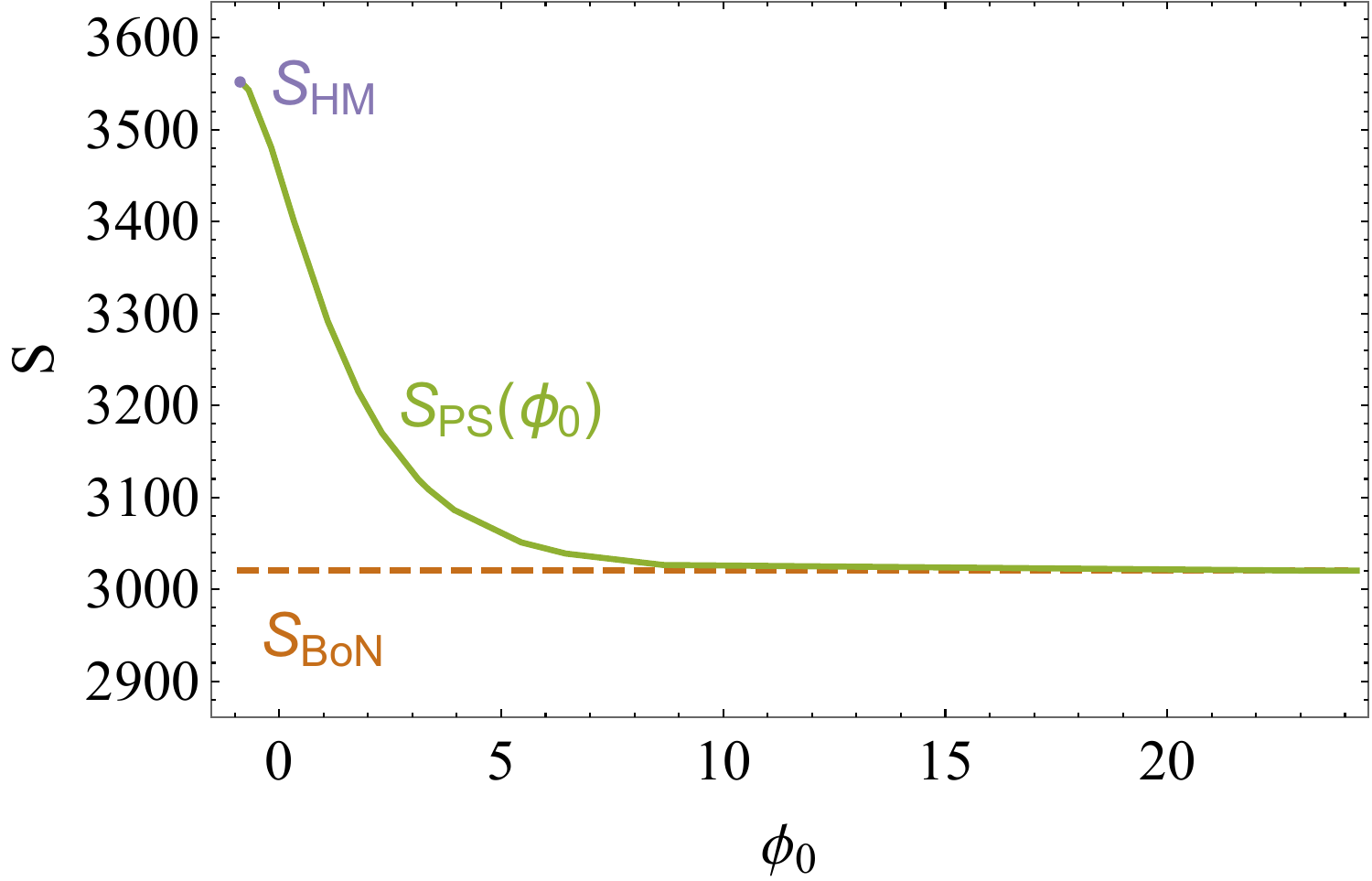}
\end{center}
\caption{
For the type $-$ example of subsection~\ref{sect:minus},  as for figure~\ref{fig:typeminus}: (Left) Potential, $V$, and tunneling potentials for decay via bubble of nothing (BoN), Hawking-Moss (HM) and pseudo-bounce (PS) towards some $\phi_0$. (Right)  Pseudo-bounce action $S_{PS}(\phi_0)$ as function of the endpoint $\phi_0$, compared with the Hawking-Moss action $S_{HM}$ and the BoN action $S_{BoN}$.
\label{fig:BoNPSHM}
}
\end{figure}

From the relations (\ref{Vtlink}) one can get the Euclidean field profile and metric function as
\be
\phi(\xi)=-\log[A\xi(2\xi_e-\xi)]\ , \quad
\rho(\xi)=\frac{(\xi_e-\xi)}{\xi_e}\sqrt{\xi(2\xi_e-\xi)}\ ,
\ee
where $\xi_e=\sqrt{-V_A}/A$. As expected for a dS decay, the instanton is compact, with $\xi\in (0,\xi_e)$. The two profiles are shown in fig.~\ref{fig:typeminus}, right plot. The asymptotic behavior of these profiles for $\xi\to 0$ are as expected from the discussion in the previous subsection, see (\ref{phirhotypepm}) and (\ref{betapm}).

Now there is a Hawking-Moss instanton that can mediate vacuum decay with action
\be
S_{HM}=\frac{24\pi^2}{V(\phi_\pp)}+\frac{24\pi^2 V_A}{25 A^2}\ ,
\ee
while the BoN has action
\be
S_{BoN}=\frac{24\pi^2}{V(\phi_\pp)}+\frac{4\pi^2 V_A}{3 A^2}\ .
\ee
As $V_A<0$, we find $S_{BoN}<S_{HM}$, so that vacuum decay proceeds preferentially via the BoN.

Figure~\ref{fig:BoNPSHM} shows one example of pseudo-bounce $V_t$ (left) and the pseudo-bounce action, $S_{PS}(\phi_0)$, (right) calculated numerically  as a function of the end-point $\phi_0$. When $\phi_0\to\phi_B$, $S_{PS}$ reproduces $S_{HM}$ as it should. When $\phi_0\to\infty$, we recover $S_{BoN}$. The analytic type $-$ BoN is the upper limit of a family of type 0 BoN's, two of which we also show in the left plot of fig.~\ref{fig:BoNPSHM}.

\subsection{An Example of Type \texorpdfstring{$\bma{+}$}{+}
\label{sect:plus}}
In this case, we take 
\be
V(\phi)= V_A e^{3\phi}+2A e^{11\phi/6}\ ,\quad
V_t(\phi) = -2V_A e^{3\phi}+3A e^{11\phi/6}\ ,
\label{Vtplus}
\ee
with $V_A,A>0$, which is an example of type $+$, with $a=\sqrt{3/2}$. For this example we get $\phi_{0\pp}=(6/7)\log[A/(3V_A)]$ and $\phi_{tT}=(6/7)\log[11A/(12V_A)]$ for the field value at which $V_t$ has its maximum.  The BoN action is
\be
S_{BoN}=24\pi^2\left[\frac{1}{V(\phi_\pp)}-\frac{1}{V(\phi_{0\pp})}\right]+\frac{648\pi^2}{77A}\left(\frac{3V_A}{A}\right)^{11/7}\ .
\ee
In figure \ref{fig:typeplus}, left plot, we show the tunneling action for this case, calculated numerically as a function of $\phi_i$, the value at which $V_t$ deviates from $V$. We take $V_A=A=1$ and complete the potential below $\phi_{0\pp}$ with a parabolic potential with minimum at $\phi_\pp=-3/2$. We find a family of BoNs of type $\pp$ of which the analytic solution (\ref{Vtplus}) is a member. The action of this solution is indicated by the arrow.  

In the family of BoN solutions found, $V_{tA}=V_A/(1-a^2)$ is fixed, and the free parameter describing the family is $V_{tX}$, the coefficient of the subleading term
$V_{tX}\exp[(a+1/a)\sqrt{6}\phi/2]$. The function $V_{tX}(\phi_i)$ is given in the right plot of \ref{fig:typeplus} and the analytic example found corresponds to the value $V_{tX}=0$.

\begin{figure}[t!]
\begin{center}
\includegraphics[width=0.49\textwidth]{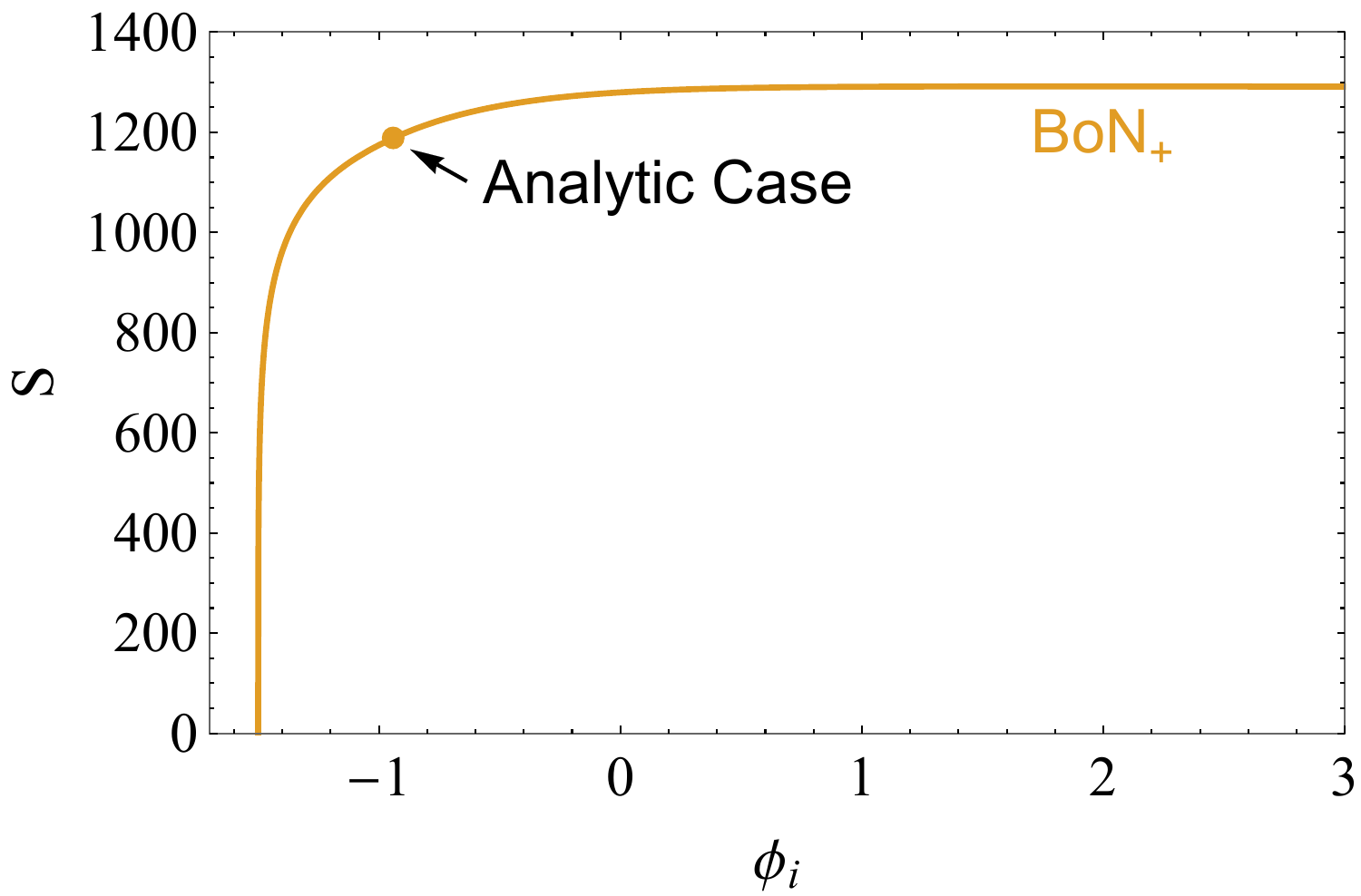}
\includegraphics[width=0.48\textwidth]{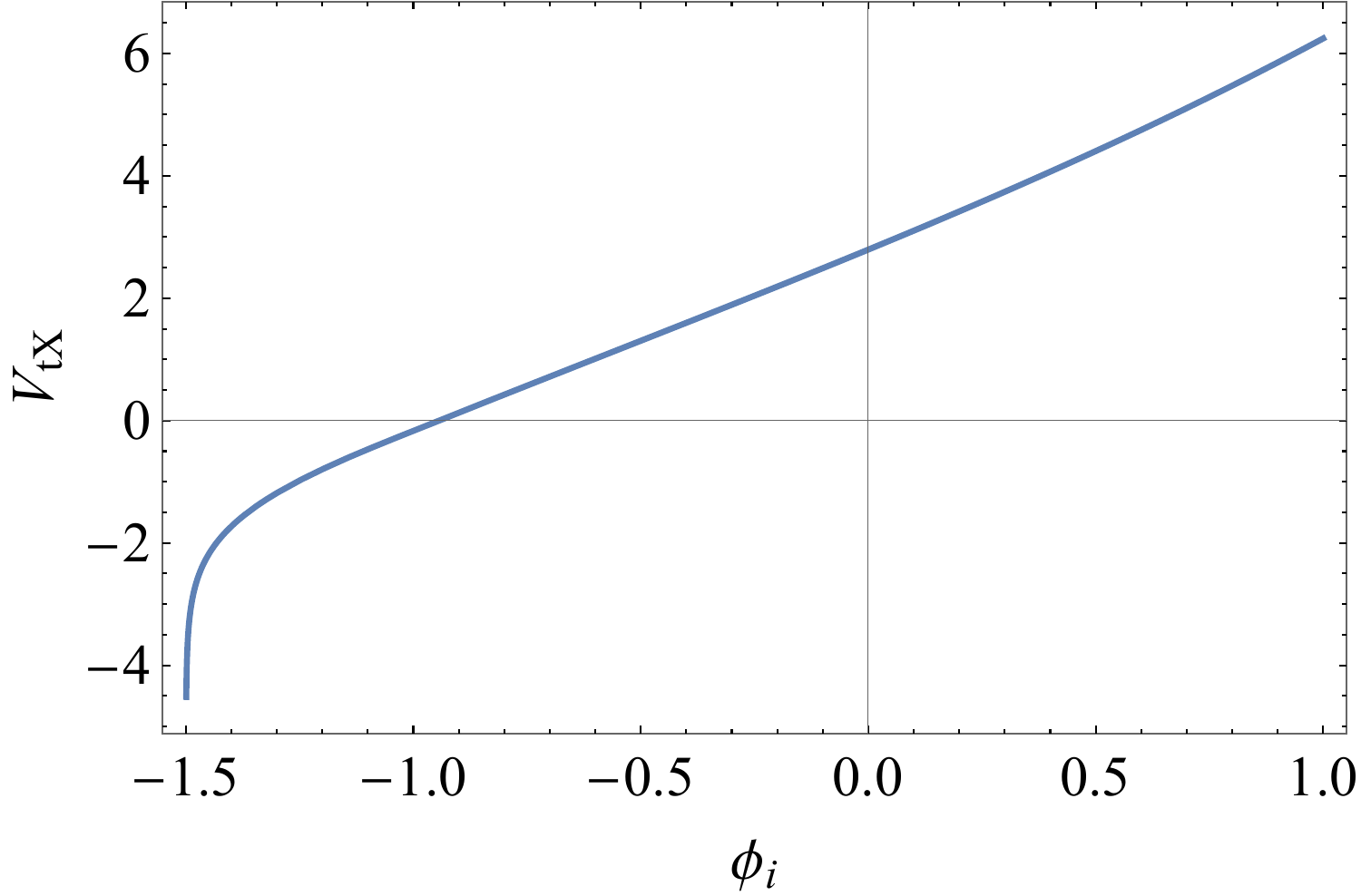}
\end{center}
\caption{For the example of subsection~\ref{sect:plus}, action structure (left) and coefficient of the subleading term in $V_t$ (right) as a function of $\phi_i$.
\label{fig:typeplus}
}
\end{figure}

\subsection{A Type 
\texorpdfstring{$\bma{-^*}$}{-*} Example\label{sect:CdLPlateau}}
In this example we have 
\be
V(\phi)=\frac49 e^{\sqrt{2/3}\phi}-\frac23 e^{2\sqrt{2/3}\phi}\ ,
\quad
V_t(\phi)=\frac12 e^{\sqrt{2/3}\phi}- e^{2\sqrt{2/3}\phi}\ ,
\label{ex1}
\ee
with $a=2/3$ and $V_{tA}/V_A=3/2$, which corresponds to a type $-^*$ case\footnote{This  example was  discussed in \cite{EFH} (see section 7.5 there) as a curious case of tunneling potential with an infinite CdL field interval, in spite of which the tunneling action was finite. The physical interpretation of such case as a possible BoN decay was unknown at the time of writing \cite{EFH}.}. This case has $\phi_{0\pp}=-\sqrt{3/2}\log 6$ and $\phi_B=\sqrt{3/2}\log(1/3)$.

\begin{figure}[t!]
\begin{center}
\includegraphics[width=0.5\textwidth]{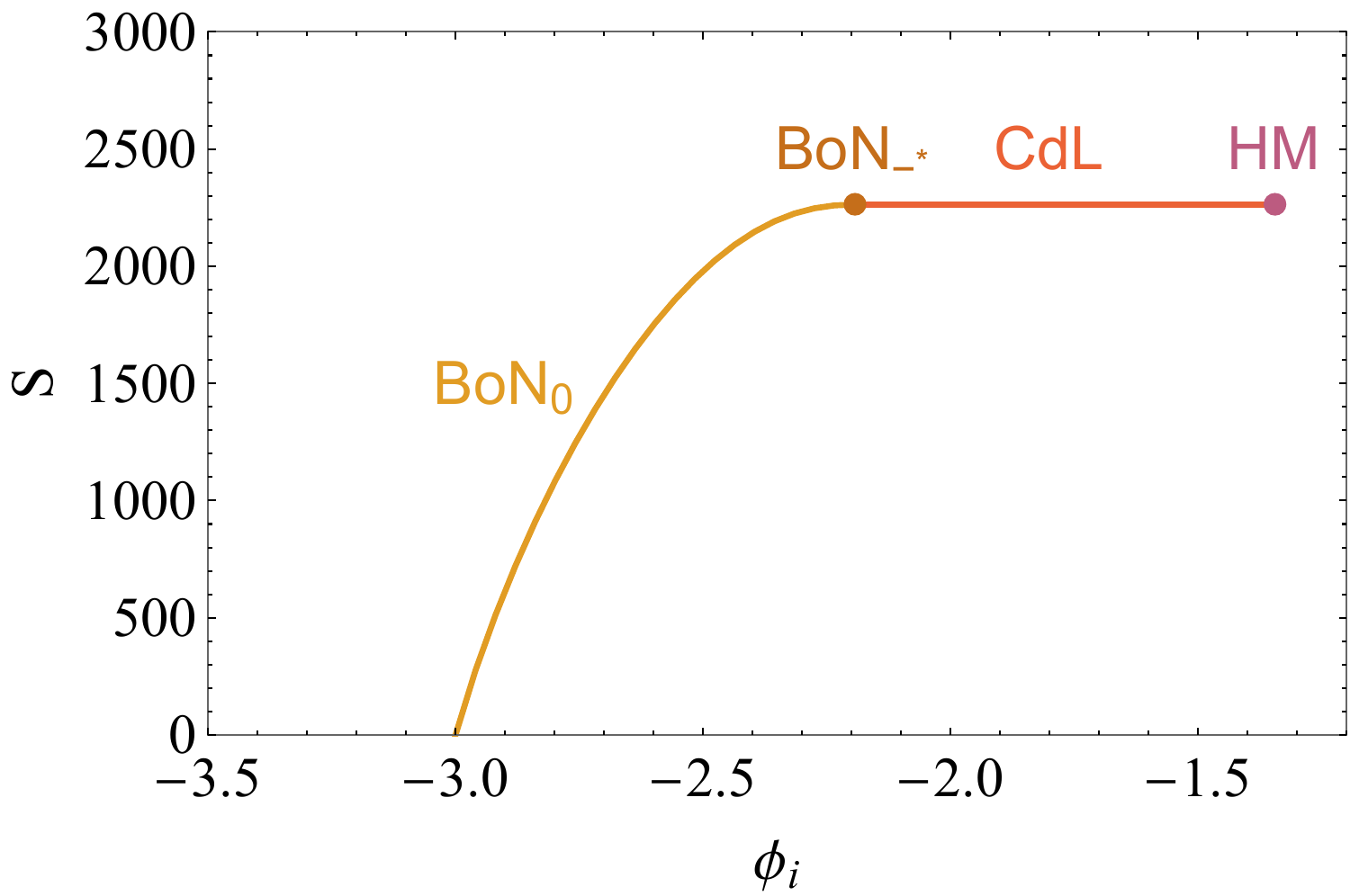}
\end{center}
\caption{Action structure for the example of subsection~\ref{sect:CdLPlateau}.
\label{fig:CdLPlateau}
}
\end{figure}

The Euclidean action for this BoN decay of the dS vacuum at $\phi_\pp$ is
\be
S_{\mathrm{BoN}}
=\frac{24\pi^2}{\kappa^2}\left[\frac{1}{V(\phi_{\pp})}-\frac{27}{2}\right]\ ,
\ee
where we leave $V(\phi_\pp)$ unspecified.
It is instructive to compare this decay action with the action for Hawking-Moss decay,  which in this example is exactly equal to $S_{\mathrm{BoN}}$.\footnote{
As explained in Section~\ref{sec:topdown}, some UV object on top of type $-^*$ BoNs might be necessary to avoid a singularity and this would contribute also to the total action. Moreover, HM and type $-^*$ BoN rates would be the same only up to differences coming from the rate prefactor, and it is not clear which one would dominate. } As consequence, this implies that the $V_t$ solutions interpolating between them are not pseudo-bounces but proper CdL solutions, see figure~\ref{fig:CdLPlateau}.  
It is interesting that this CdL plateau appears in a case in which the necessary condition to have a CdL solution, $-V''(\phi_B)>4\kappa V(\phi_B)/3$ \cite{HW}, is saturated and $-V''(\phi_B)=4\kappa V(\phi_B)/3$ holds.\footnote{The structure of the action shown in figure~\ref{fig:CdLPlateau} is quite remarkable. The only other example that we know of in which a CdL plateau for the action appears is the potential $V(\phi)=-\lambda\phi^4/4$. In that well-known case, there is an infinite family of bounces with arbitrary size and equal action. While in such example scale invariance is at the root of the CdL plateau, there is no such mechanism at work in the example of this section.}
In the same plot, to the left of the analytical BoN solution of type $-^*$ we expect to find a family of type 0  BoN solutions, with shape and action being model-dependent as we have the freedom to choose the shape of the potential for $\phi<\phi_{0+}$. For the plot,
we have completed the potential below $\phi_{0\pp}$ with a parabolic potential with the dS minimum at $\phi_\pp=-3$.

\subsection{Another Type \texorpdfstring{$\bma{-^*}$}{-*} Example\label{sect:Counter}}
For the second type $-^*$ example we take 
\be
V=\frac{11A}{12}e^{\phi/\sqrt{2}}-\frac23 e^{5\phi/(2\sqrt{2})}\ ,
\quad 
V_t = A e^{\phi/\sqrt{2}}-  e^{5\phi/(2\sqrt{2})}\ ,
\label{Vcounter}
\ee
which has $a=5/(4\sqrt{3})$. 
As in the previous example these solutions hold for $\phi>\phi_{0\pp}=(\sqrt{8}/3)\log(A/4)$, with $V(\phi_{0\pp})=V_t(\phi_{0\pp})$, but we can extend them to $\phi<\phi_{0\pp}$ as in previous cases. A plot of $V$ and $V_t$ would look quite similar to the figure~\ref{fig:typeminus} (left).

The HM and BoN actions can be computed analytically as in the previous example, with
\be
S_{\rm BoN}=\frac{24\pi^2}{\kappa^2V(\phi_{\pp})}+\delta S_{\rm BoN}\ ,\quad
S_{\rm HM}=\frac{24\pi^2}{\kappa^2V(\phi_{\pp})}+\delta S_{\rm HM}\ ,
\ee
with
\be
\delta S_{\rm BoN} =\frac{64\pi^2}{5}\left(\frac{2}{A^5}\right)^{1/3}\ ,\quad \delta S_{\rm HM}= 5\left[1-3\left(\frac{5}{11}\right)^{5/3}\right]\delta S_{\rm BoN} \simeq 0.97 \delta S_{\rm BoN}\ ,
\label{counterexample}
\ee
so that $ S_{\rm HM}<S_{\rm BoN} $ and HM would dominate over BoN. In figure~\ref{fig:Special} we show the type $-^*$ BoN solution (left plot) as well as members of a family of type 0 BoNs that interpolate between HM and the type $-^*$ BoN. The right plot shows the action of all these solutions as a function of $\phi_i$ (the starting field value of the CdL part of the instanton). To the left of $\phi_i=\phi_{0\pp}$ (for the type $-^*$ BoN action) we expect a second family of type 0 BoNs that would depend on how $V$ is completed below $\phi_{0\pp}$.


\begin{figure}[t!]
\begin{center}
\includegraphics[width=0.49\textwidth]{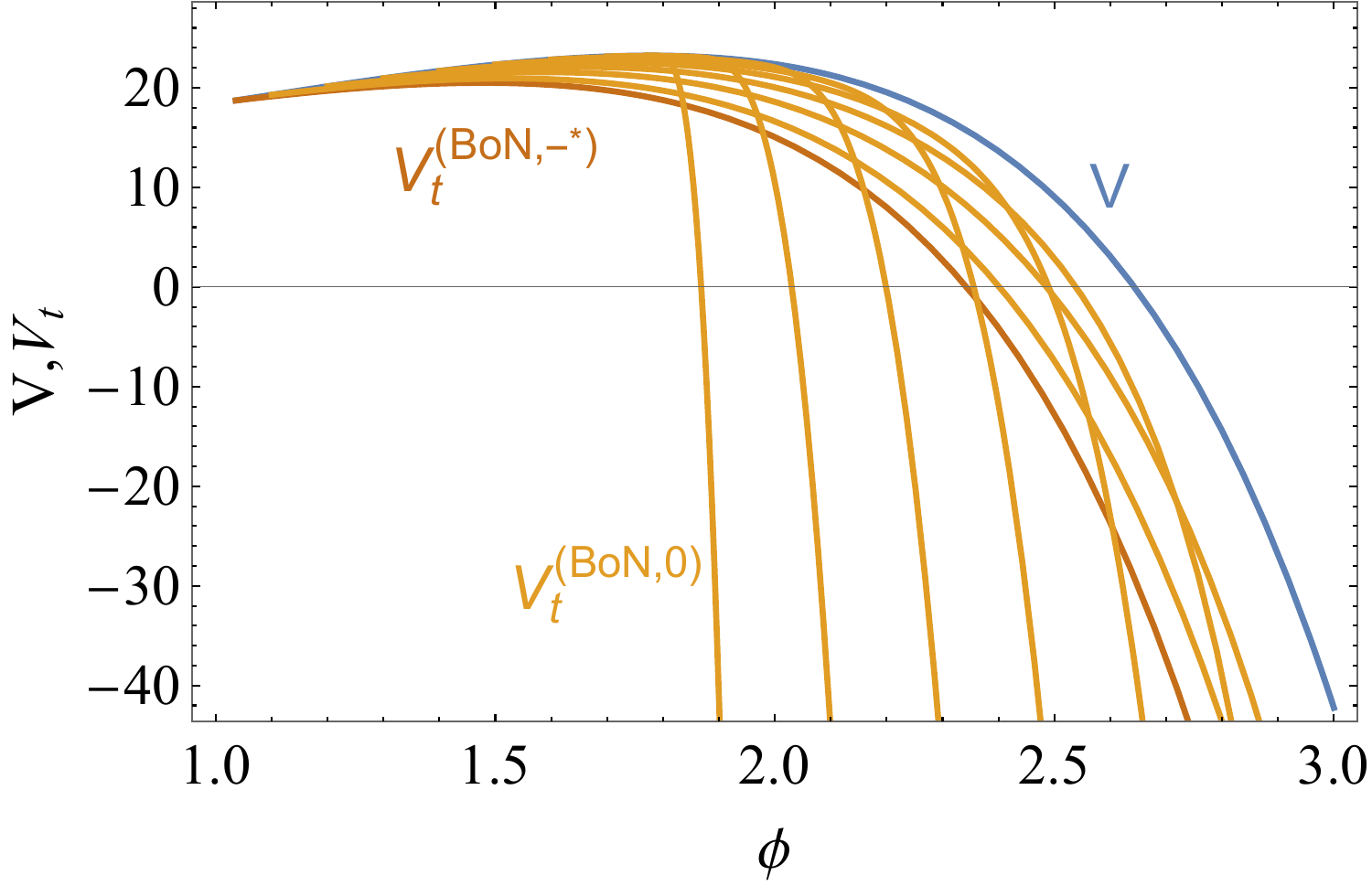}
\includegraphics[width=0.49\textwidth]{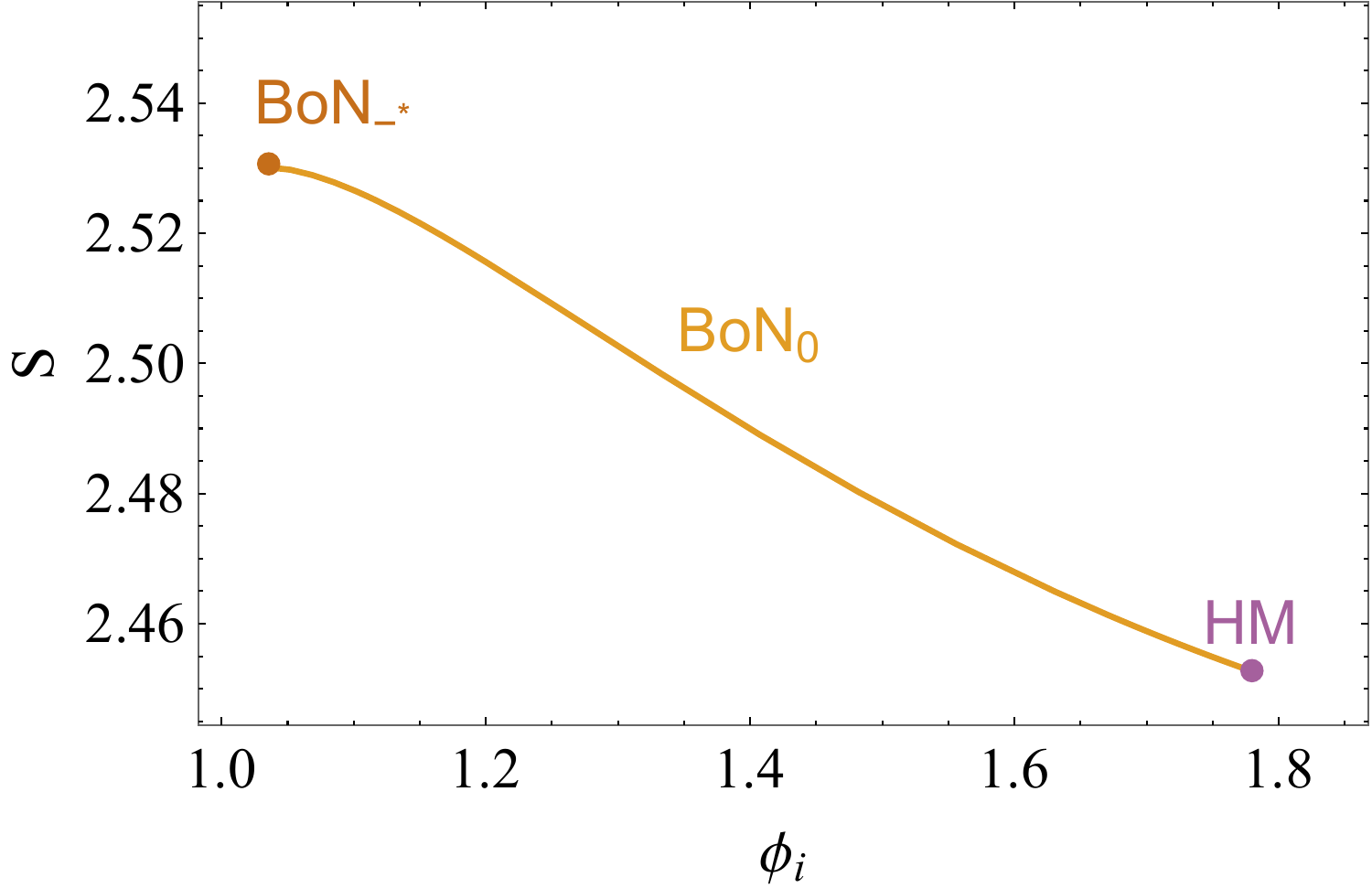}
\end{center}
\caption{For the example of subsection~\ref{sect:Counter}, $V$ and $V_t$'s (left) and action structure as a function of $\phi_i$ (right).
\label{fig:Special}
}
\end{figure}

This example is another counterexample to the general expectation, already conjectured by \cite{DGL}, that BoNs dominate decay. However, the decay rate depends on a non-exponential prefactor that can compensate for the small difference in tunneling actions found above.  Nevertheless, one should keep in mind that it is not clear whether type $-^*$ solutions can be realized as consistent $4d$ truncations of a proper BoN in the higher-dimensional theory, see Section~\ref{sec:topdown}. Moreover, as already mentioned in the previous example, a UV object on top of theses BoNs might be needed to avoid a singularity and this would also modify the total action.

\section{BoNs with Nonzero Potential. Top-down Analysis\label{sec:topdown}}

In this section we take as starting point BoN geometries in $D=4+d$ dimensions and integrate out the $d$ compact extra dimensions to get an effective $4d$ description in terms of a modulus field $\phi$ with a potential $V(\phi)$ \cite{DFG,DGL}. In this way, one describes the original BoN in terms of a singular CdL bounce in $4d$, or alternatively, a divergent tunneling potential, $V_t$. By performing this top-down analysis we can explore what is the higher-dimensional origin of the parameters entering in the different types of $V_t$ solutions discussed in previous sections, and in particular of $V_{tA}$, which determines  the boundary condition at $\phi\to \infty$ for the tunneling potential. 
As we show below, although the $4d$ description of the BoN instanton is always singular,  we can  obtain constraints on the parameter $V_{tA}$ from smoothness conditions on the $4+d$ BoN.

Let us consider first the case of a BoN in a spacetime with the compactified space being the $d$-dimensional sphere, $S^d$. 
The most general ansatz for an $\mathrm{O}(4)$ symmetric BoN instanton, which also preserves the symmetry of the $S^d$ compact space, can be written in the gauge \eqref{eq:WittenWgauge} (with the replacement $\alpha \to r$) as follows
\be
ds^2=dr^2+{\cal R}^2B(r)^2d\Omega_3^2 + R_{KK}^2C(r)^2 d\Omega_d^2\ ,
\label{dsUV}
\ee
where $d\Omega_d^2$ is the line element of the $d$-dimensional unit sphere.
The BoN is located at $r=0$ and $r\to\infty$ corresponds to the vacuum geometry. The boundary conditions at the BoN location which ensure the regularity of the metric are
\be
B(0)=1\ ,\quad B'(0)=0\ ,\quad C(0)=0\ ,\quad C'(0)=1/R_{KK}\ ,
\label{eq:BoNcoreBC}
\ee
Here ${\cal R}$ is the bubble nucleation radius and $R_{KK}$ is the Kaluza-Klein radius.

The choice of boundary conditions for the metric functions far from the bubble depend on the character of the false vacuum \cite{Blanco-Pillado:2010vdp}. When the vacuum energy of the false vacuum vanishes, the metric of the non-compact space should tend to Minkowski space-time far from the bubble ($r\to \infty$) and, therefore
\be
B'(\infty)\to 1/{\cal R}\ , \quad C(\infty)\to 1\ .
\label{eq:BoNinfBC}
\ee
When the energy of the false vacuum is negative $V(\phi_+)<0$ the geometry of the non-compact directions should be asymptotically anti-de Sitter, $AdS_4 \times S^d$, and therefore we must impose
\begin{equation}
\left.\frac{B'}{B}\right|_\infty \to \frac{1}{L_{AdS}}, \quad C(\infty)\to 1\ ,
\end{equation}
where $L_{AdS}\equiv \sqrt{-3/(\kappa V(\phi_+))}$ is the AdS scale of the vacuum. Finally, when the false vacuum has a positive energy $V(\phi_+)>0$, the geometry of the non-compact space should be asymptotically de Sitter $dS_4 \times S^d$. In this case the instanton is compact, with the radial coordinate taking values in $0<r<r_h$, and the boundary conditions at the cosmological horizon at $r=r_h$ are given by
\begin{equation}
B(r \to r_h) \approx -(r-r_h)/\mathcal{R} + \ldots. 
\end{equation}
The value of the of the KK radius at the cosmological horizon, $R_{KK} C(r_h)\neq R_{KK}$, is in the basin of attraction of the radius of the compactification vacuum solution, $dS_4 \times S^d$, and it is determined by the equations of motion and boundary conditions.

As for Witten's BoN, one can integrate over the compact dimension to get the reduced $4d$ metric, introducing  a modulus field that tracks the size of the extra dimensions and can be made canonical with a convenient Weyl rescaling. 
The $4+d$ BoN metric can be rewritten in terms of the canonical modulus field,  $\phi$ and the $4d$ CdL metric as
\be
ds^2=e^{\gamma d\phi}ds^2_4 +e^{-2\gamma\phi}R_{KK}^2d\Omega_d^2\ ,
\label{eq:4dGauge}
\ee
with 
\be
ds_4^2=d\xi^2+\rho(\xi)^2d\Omega_3^2\ ,\quad \quad \gamma = \sqrt{\frac{2\kappa}{d(d+2)}}\  .
\ee
Comparing with (\ref{dsUV}), we get
\be
\rho=C^{d/2}B{\cal R}\ , \quad C=e^{-\gamma\phi}\ ,\quad
\frac{dr}{d\xi}=C^{-d/2}\ ,\quad \frac{dC}{d\phi}=-\gamma C\ .
\ee

As the $4+d$ BoN solution is smooth at $r\to 0$ with a flat metric ($ds^2\simeq dr^2+r^2 d\Omega_d^2$) at a fixed point on the $S^d$, 
this implies the small $r$ behaviour
\be
\frac{d\xi}{dr}\simeq e^{\gamma d\phi/2}\ , \quad r^2\simeq e^{-2\gamma\phi}R_{KK}^2\ .
\ee
From this we obtain the $\xi\to 0$ scaling
\be
\phi \simeq -\sqrt{\frac{2d}{\kappa(d+2)}}\log\frac{(d+2)\xi}{2R_{KK}}\ ,\quad\rho\simeq {\cal R}\left[\frac{(d+2)\xi}{2R_{KK}}\right]^{d/(d+2)}\ ,
\label{phirhotop}
\ee
which agrees with the results presented in \cite{DGL}.  For later convenience we also give the asymptotic dependence of the metric profile function $\rho(\phi)$  near the BoN core
\be
\rho(\phi \to \infty)\simeq  \mathcal{R} e^{-\sqrt{\frac{d\kappa}{2(d+2)}}\phi}\ ,
\label{rhoCore}
\ee
as well as
\be
\dot\rho(\phi\to\infty)\simeq \frac{d}{2}\frac{\cal R}{R_{KK}}e^{\sqrt{\frac{2\kappa}{d(d+2)}}\phi}\ ,\quad
\dot\phi(\phi\to\infty)\simeq -\frac{1}{R_{KK}}\sqrt{\frac{d(d+2)}{2\kappa}}e^{\sqrt{\frac{(d+2)\kappa}{2d}}\phi}\ .
\ee
We can now compare with the scalings found in section~\ref{sec:BotUp} using the bottom-up approach and the tunneling potential, see (\ref{phirhotype0}), (\ref{phirhotypepm}) and (\ref{betapm}). Comparing $\rho$, we get
\be
c_\rho = {\cal R}\left(\frac{d+2}{2R_{KK}}\right)^{d/(d+2)}
\ ,\quad \beta = \frac{d}{d+2}\ ,
\label{UVresultsrho}
\ee
while, for the different constants of the $V_t$ formalism, we find that
\be
a=\sqrt{\frac{d+2}{3d}}\ ,\quad V_A-V_{tA} =\frac{d(d+2)}{4\kappa R_{KK}^2}\ , \quad 
D_\infty = \frac{3}{R_{KK}{\cal R}}\sqrt{\frac{d(d+2)}{2\kappa}}
\ .
\label{UVresults}
\ee
Therefore, the type 0 case is realized for $d=1$, while $d>1$ corresponds to type
$-$ [as $1/\sqrt{3}<a=\sqrt{(d+2)/(3d)}<1$]. Types + and $-^*$ cannot be obtained from such simple extra compact spaces and would require a more complicated geometry (if they can be realized at all). The same applies to type $-$ examples with 
$a^2\neq(d+2)/(3d)$.

When the regularity conditions \eqref{UVresults} are substituted in the $4d$ equations of motion \eqref{E2}, it can be shown that any compatible scalar potential should have the following limiting behaviour for $\phi\to \infty$ \cite{DGL}
\be
V(\phi \to \infty) \simeq -\frac{d(d-1)}{2\kappa R_{KK}^2}e^{\sqrt{\frac{2  (d+2) \kappa}{d}}\phi}\ + \ldots \; .
\label{VUVminus}
\ee
We also get the asymptotic behaviour for $V_t=V-\dot\phi^2/2$ as
\be
V_t(\phi \to \infty) \simeq -\frac{3d^2}{4\kappa R_{KK}^2}e^{\sqrt{\frac{2(d+2) \kappa}{d}}\phi}\ + \ldots \; .
\label{VtUV}
\ee
These formulas tells us how $V_A$ and $V_{tA}$ are determined by the high-dimensional theory.

Interestingly, when the compact dimensions are integrated out, the effective $4d$ Euclidean action receives a contribution from the curvature of the internal space to the potential which is precisely of the form (\ref{VUVminus}).  In other words, in order for the BoN geometry to be smooth, the scalar potential should be dominated by the curvature contribution  to $V$ in the limit $\phi \to \infty$. The $d=1$ case does not pick up such contribution, which is compatible with the fact that $V$ is subleading for type 0 cases.
On the other hand, for $d>1$ we get a contribution that is precisely of the form $V_A e^{a\sqrt{6\kappa}\phi}$ expected for the type $-$ cases. 

As reviewed in \cite{DGL}, there are  well known sources of moduli potentials. 
\begin{itemize}
\item The potential (\ref{VUVminus}) is one instance of the general result
\be
\delta V(\phi) = -\frac{{\cal R}_d}{2\kappa}e^{\sqrt{\frac{2(d+2)\kappa}{d}}\ \phi}\ ,
\ee
where ${\cal R}_d$ is the curvature scalar of the compactified space. Deforming the geometry, e.g. by having compact dimensions of different radii, modifies the prefactor while the exponent is fixed by $d$.

\item A non-zero cosmological constant in the $4+d$ theory, $\Lambda_{4+d}$ would also produce a contribution 
\be
\delta V(\phi) = \frac{\Lambda_{4+d}}{\kappa }e^{\sqrt{\frac{2 d\kappa}{d+2}}\ \phi}\ ,
\label{Vcc}
\ee 
where $M$ is the higher dimensional Planck mass. If this is the dominant term in $V$, then the $a$ parameter of our effective $V_t$ description (see table~\ref{table:types}) would be
$1/3\leq  a=\sqrt{d/[3(d+2)]}<1/\sqrt{3}$,  which can only correspond to a type 0 case (as $a<1$).

\item Finally, a $d$-form flux wrapped around the $d$-dimensional compact space,   $\int_{S^{d}} F_{d} =Q$, leads to
\be
\delta V(\phi)= \frac{Q^2}{2 g^2 \mathcal{V}_{(d)}} e^{3\sqrt{\frac{2d\kappa}{d+2}}\ \phi}\ ,
\label{Vflux}
\ee
where $g$ is the gauge coupling and $\mathcal{V}_{(d)}$ is the volume of the $d-$sphere. This gives $1\leq  a=\sqrt{3d/(d+2)}<\sqrt{3}$: the scaling of type + cases (provided $d>1$). 
\end{itemize}

However,  for the contributions \eqref{Vcc} and \eqref{Vflux} to dominate when $\phi\to \infty$, we would need the compactification to be different from the simple cases discussed above, since for the compactifications on $S^d$ the regularity conditions \eqref{phirhotop} require the potential to behave as \eqref{VUVminus} in this limit. Nevertheless, the presence of additional scalar fields besides the modulus field could also modify the scalar potential probed asymptotically by the BoN field configuration. An example of this effect is presented in section~\ref{sec:BoNFlux}, which discusses a BoN solution in a flux compactification model.

The bottom-up analysis, therefore, motivates the question: what are the possible geometries  of the compactified space (or field content in the effective $4d$ theory) that can realize BoN solutions of types $+$ or $-^*$ [or type $-$ with $a^2\neq(d+2)/(3d)$]? A different avenue to realise these more exotic type of solutions would be to embrace the possibility of singular BoNs (see \cite{CobConj,Horowitz:2007pr,Bomans:2021ara} and, more recently, \cite{Hebecker}). Indeed,  the presence of a singularity might be signalling the need to dress the BoN with a brane, or another UV object, whose properties (tension and charge) could be inferred from the behaviour of the solution in the limit $\phi\to\infty$. Such a study would require a case by case analysis which is out of the scope of the present discussion. 

\section{BoNs and End-of-the-World branes\label{sec:quench}}

As we discussed in sections \ref{sec:BCs} and \ref{sec:BoNvsOther}, when a dynamical constraint is saturated, the BoN nucleation rate may be suppressed by the CdL mechanism for  Minkowski or AdS decay. In particular, if standard CdL decay is quenched ($\overline V_t$ does not intersect $V$), the critical tunneling potential $\overline V_t$ corresponds to an unbounded BoN solution, which sets a dynamical constraint on the tunneling potential, namely $V_t \le \overline V_t$ (see section \ref{sec:Vt}). For type 0 BoN solutions, using the asymptotic form of $V$ and $\overline{V_t}$, this constraint can be written [see \eqref{eq:dyConst1}] as
\be
V_{tA}\le V_{tA*} = \overline V_{tA}\ .
\label{dyConst}
\ee
If the potential is deformed so that the constraint is saturated ($V_t \to \overline V_t$), the BoN becomes infinite and static, an ETW brane, with the metric given by
\be
ds^2 = d\xi^2 + \rho_c^2(\xi) (dx_1^2+dx_2^2+dx_3^3), 
\label{eq:ETW}
\ee
where $\rho_c(\xi=0)=0$ and $\xi\ge 0$. This line element represents a space-time which ``ends'' at $\xi=0$, and which approaches $AdS_4$ or $M_4$ [depending on $V(\phi_+)$] far from $\xi=0$.  
In this critical limit the tunneling rate becomes exponentially suppressed, as a consequence of the Coleman-de-Luccia mechanism, that is,  due to the divergence of the tunneling action. 

The description of ETW branes in the tunneling  action formalism was previously used in \cite{DynCo}. Here we study two analytical examples to illustrate the interplay between BoN and ETW branes, and the onset of the CdL mechanism when the dynamical constraint is saturated.  
In this section we assume that the asymptotic parameter $V_{tA}$ has been fixed by the higher dimensional theory by requiring the regularity of the BoN spacetime for internal geometries of the form $S^d$, as derived in the relation \eqref{VtUV}. Therefore, in contrast with the analysis in section \ref{sec:BoNvsOther}, here the boundary condition for $V_t$ is kept fixed to
\be
V_{tA}=-\frac{3d^2}{4\kappa R_{KK}^2}\ ,
\label{VtUV2}
\ee
and instead we study how BoN solutions depend on the shape of the scalar potential as we deform it around criticality [i.e. when it takes the form \eqref{Vc}]. Such solutions can be found in fake supergravity models and inherit some nice properties of those supergravity solutions.


\subsection{Type 0 BoN
}

The first example we consider is a scalar potential compatible with the (near-critical) decay of a vacuum $AdS_4 \times S^1$, which according to our discussion in section \ref{sec:topdown} corresponds to a type 0 tunneling potential. More specifically, we study the following family of potentials 
\be
V=-M^4\,  (1+\epsilon) \cosh \left(\sqrt{2\kappa/3} \, \phi \right),
\ee
where $M^4 \equiv 6/(\kappa R_{KK}^2)$ and $\epsilon$ is a parameter that controls the size of the deformation away from criticality, as we show below.  This potential has a perturbatively stable AdS vacuum at $\phi_+=0$ which, in spite of being a maximum, has  no tachyonic instabilities, since the mass squared  $m^2=-2 L_{AdS}^{-2}$ respects the Breitenlohner-Freedman bound, $m^2>m_{BF}^2 = -(9/4)\, L_{AdS}^{-2}$.\footnote{Although AdS backgrounds with $m^2_{BF}<m^2<m^2_{BF}+L_{AdS}^{-2}$ admit more general boundary conditions\cite{Breitenlohner:1982jf,Klebanov:1999tb,Henneaux:2002wm,Hertog:2004ns,Hertog:2004dr}, our example is consistent with the standard boundary conditions for AdS, where the field $\phi$ approaches the vacuum  near the AdS boundary at $\rho\to \infty$ as $\phi \sim  \rho^{-2}$, with $\rho$ being an asymptotic area coordinate.} 

The computation of the tunneling potential is straightforward and illustrates how one can use the critical tunneling potential $\overline{V_t}$ and the bound \eqref{dyConst} to determine if a dynamical obstruction to the decay is present or not. The EoM for $\overline{V_t}$, eq.~\eqref{Vtc}, can be solved exactly (to all orders in $\epsilon$), giving 
\be
\overline{V_t} =-M^4\, (1 + \epsilon) \cosh ^3(\sqrt{2\kappa/3} \, \phi)\ ,
\ee
with the asymptotic behaviour 
\be
\overline{V_t}(\phi\to \infty)=\overline{V}_{tA}  e^{\sqrt{6 \kappa}\phi}+\ldots = -\frac{3}{4\kappa R_{KK}^2} \left( 1 + \epsilon \right) e^{\sqrt{6 \kappa}\phi}+\ldots\, .
\ee
Plugging this $\overline{V}_{tA}$ in the bound \eqref{dyConst}, taking $V_{tA}$ for $d=1$ from
\eqref{VtUV2}, the BoN decay is dynamically forbidden if $\epsilon>0$ (for which $V_t>\overline{V_t}$).

In the subcritical case ($\epsilon<0$) we can solve the tunneling potential EoM imposing the boundary condition \eqref{VtUV2}, as required for the higher dimensional spacetime to be regular, finding  (to all orders in $\epsilon$)
\be
V_t =-M^4 (1+\epsilon) \cosh ^3(\sqrt{2\kappa/3} \, \phi) + \epsilon M^4 \sinh^3(\sqrt{2\kappa/3} \, \phi)\ .
\ee
As a consistency check, near the critical limit $|\epsilon| \ll 1$
\be
D^2 =-6\kappa M^8 \epsilon \cosh(\sqrt{2\kappa/3} \, \phi)\sinh^3(\sqrt{2\kappa/3} \, \phi) + \mathcal{O}(\epsilon^2),
\ee
which is  positive only when $\epsilon<0$.

Using the dictionary between $V_t$ and Euclidean formalisms discussed in section~\ref{sec:Vt}, it is immediate to get the metric profile function [from  \eqref{rho}] as 
  \be
\rho(\phi) = \frac{\sqrt{3}}{\sqrt{-\kappa M^4 \epsilon  \sinh(\sqrt{2\kappa/3}\, \phi)} }\ .
\ee
This $\rho$ diverges in the limit $\phi\to \phi_+= 0$, where the space-time approaches the $AdS_4$ geometry of the vacuum. The BoN nucleation radius can be obtained from the limiting behaviour of $\rho(\phi)$ at the BoN core, $\phi\to\infty$. Indeed, setting $d=1$ in \eqref{rhoCore}, and comparing with the previous expression we find $\mathcal{R} \simeq  R_{KK}/\sqrt{-\epsilon}$, which diverges as  $\epsilon\to 0^-$. In this limit the BoN becomes static and of infinite  radius, that is, an End-of-the-World brane, whose four dimensional line element is given by eq. \eqref{eq:ETW}, with the metric profile function given by
\be
\rho_c(\phi) \equiv \lim_{\epsilon\to 0} \rho/\mathcal{R} =\frac{1}{\sqrt{2 \sinh(\sqrt{2\kappa/3}\, \phi)}}\ . 
\label{eq:ETWprofile}
\ee
Using the expression for the action, $S[V_t]$ from \eqref{SVt}, we get $S[V_t] = 2(\pi m_P R_{KK})^2/\sqrt{-\epsilon}$, which indeed diverges at $\epsilon\to 0$ and thus, the BoN nucleation probability is exponentially suppressed in that limit,  as anticipated: in the critical limit the compactification is protected against the decay to nothing  by the Coleman-de Luccia mechanism.

\subsection{Type \texorpdfstring{$\bma{-}$}{-}  BoN}

We consider next a scalar potential compatible with the decay of an $AdS_4 \times S^6$  vacuum, described by a type $-$ tunneling potential (see section \ref{sec:topdown}). The potential is  
\be
V=-\frac{M^4}{18} \left[13+5 \cosh (\sqrt{8\kappa/3}\, \phi )\right]+\epsilon M^4  \sinh ^{3/2}(\sqrt{2\kappa/3}\, \phi ) \tanh ^{2}(\sqrt{2\kappa/3}\, \phi ),
\label{eq:AdS4S6}
\ee
where $M^4 \equiv 108/(\kappa R_{KK}^2)$. We consider the parameter $|\epsilon| \ll 1$ to be small, with the critical case  given by $\epsilon=0$, see below. We solve the equations of motion perturbatively in $\epsilon$, and study the instanton solutions both numerically and analytically.
As in the previous example, the potential has a perturbatively stable AdS maximum  at $\phi_+=0$ (since the tachyonic mass is above the BF bound $m^2 = -(20/9) \, L_{AdS}^{-2}>m_{BF}^2=-(9/4)\, L_{AdS}^{-2}$).\footnote{Since $m_{BF}^2<m^2<m_{BF}^2+L_{AdS}^{-2}$ this background admits more general boundary conditions than the usual AdS reflective ones. Our example corresponds to a theory where the field approaches the vacuum at the AdS boundary $\rho\to \infty$ as $\phi \sim \alpha \rho^{-\Delta_-} + 0 \cdot \rho^{-\Delta_+}$, where $\rho$ is an asymptotic area coordinate and $\Delta_{\pm}= 3(1\pm \sqrt{1-m^2/m_{BF}^2})/2$ (see \cite{Klebanov:1999tb,Hertog:2004ns}).} 
\begin{figure}[t!]
\begin{center}
\includegraphics[width=0.475\textwidth]{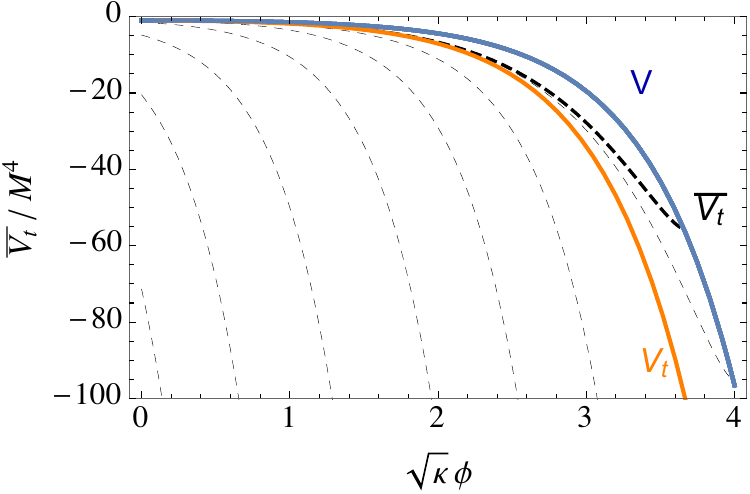} \hspace{.4cm}
\includegraphics[width=0.475\textwidth]{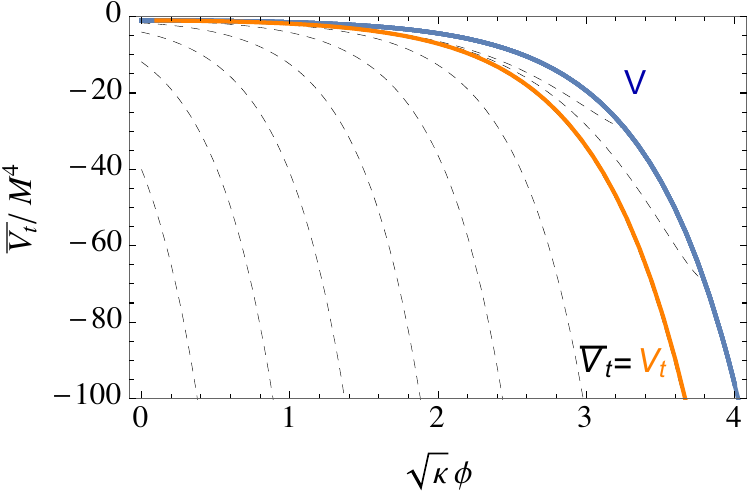}
\includegraphics[width=0.475\textwidth]{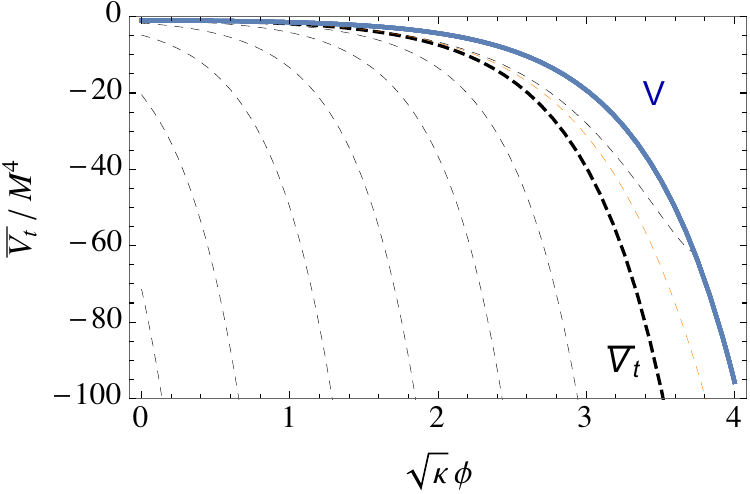}
\end{center}
\caption{Scalar potential \eqref{eq:AdS4S6} with a perturbatively stable AdS vacuum  at $\phi_+=0$.  \emph{Upper left plot:} subcritical case ($\epsilon= -0.01$).  \emph{Upper right plot:} critical case ($\epsilon=0$). \emph{Lower plot:} supercritical case ($\epsilon=0.01$). Dashed lines are integral curves of $D=0$ and the thick black dashed line is the critical tunneling potential \eqref{Vtc} with $\overline{V_t}(\phi_+)=V(\phi_+)$. In the subcritical and critical cases the orange line is the tunneling potential, with  asymptotic behaviour \eqref{VtUV2}. In the supercritical case, without viable  tunneling potential, we have indicated with an orange dashed line the solution to $D=0$ satisfying \eqref{VtUV2} as a reference.
\label{plot:quenching}}
\end{figure}

In order to analyze gravitational quenching, we have computed numerically the critical tunneling potential $\overline{V_t}$ for different values of the deformation parameter $\epsilon$ (see figure \ref{plot:quenching}). As a reference, we have also indicated with dashed lines other solutions to the $D=0$ equation \eqref{Vtc} different from the critical tunneling potential, i.e. with boundary conditions $\overline{V_t}(\phi_+) \neq V(\phi_+)$. Fig. \ref{plot:quenching} illustrates the three cases described above: 
\begin{itemize}
\item {\bf Subcritical decay} (top left plot). When $\epsilon<0$, the critical tunneling potential (thick black dashed line) reaches the potential (solid blue line) at some finite value of the field, and therefore it leaves room for $V_t$ (solid orange line) to meet the boundary condition \eqref{VtUV}. Therefore there is no dynamical constraint for the decay to nothing. 

\item {\bf Critical case} (top right plot). When $\epsilon=0$, the critical tunneling potential diverges for large values of $\phi$, and satisfies the regularity condition \eqref{VtUV2}, and therefore  $\overline{V_t}= V_t$. The tunneling action is infinite and the decay is forbidden. 

\item {\bf Supercritical case} (lower plot). When $\epsilon>0$, the BoN decay is dynamically forbidden. The critical tunneling potential (thick black dashed line) violates the bound \eqref{dyConst}, since it lies below the limiting behaviour imposed by regularity at the BoN core \eqref{VtUV2}. For reference, the $D=0$ line asymptoting as \eqref{VtUV2} is given by the orange dashed line. 
\end{itemize}
In the critical and subcritical cases, the tunneling potential can be easily computed up to  leading order in $\epsilon$, as 
\be
V_t =-M^4 \cosh ^2(\sqrt{2 \kappa/3} \, \phi)+ \mathcal{O}(\epsilon^2)\ .
\ee
As a consistency check, we get 
\be
D^2 =- 6 \,\kappa  M^8 \epsilon \sinh^{7/2}(\sqrt{2\kappa/3}\, \phi)>0\ ,
\ee
which is positive for $\epsilon<0$.

Using $V$, $V_t$  and $D$ above in \eqref{rho}, we find that  the metric profile function $\rho(\phi)$ in the subcritical case is given by
\be
\rho(\phi) =  \frac{2}{\sqrt{-3\kappa \epsilon} \, M^2}\sinh^{-3/4}(\sqrt{2\kappa/3}\, \phi)  + \mathcal{O}(|\epsilon|^{1/2}),
\ee
and thus, the bubble nucleation radius is $\mathcal{R} \simeq  R_{KK}\frac{2^{3/4}}{9\sqrt{-\epsilon}}$.  As the deformation is tuned down $\epsilon \to 0^-$ the bubble nucleation radius increases as expected, and in the strict limit $\epsilon=0$ the bubble size diverges $\mathcal{R}\to \infty$. In other words, when $\epsilon=0$   the bubble of nothing becomes infinite and static, i.e. and End of the World brane, with the  space-time geometry described by the line element  \eqref{eq:ETWprofile}, and the corresponding metric profile function given by 
\be
\rho_c(\phi) =\lim_{\epsilon\to 0} \rho/\mathcal{R} =\left[2 \sinh(\sqrt{2\kappa/3}\, \phi) \right]^{-3/4}.
\ee
As expected, the tunneling action also diverges in this limit, $S[V_t] \propto (m_P R_{KK})^2/\sqrt{-\epsilon}\to \infty$ and, therefore, the vacuum is protected against decay  by the  CdL mechanism when $\epsilon\geq 0$.

\section{BoN in Flux Compactifications\label{sec:BoNFlux}}

As reviewed in the introduction, in String Theory compactifications the scalar potential is partly generated by the presence of fluxes, which are turned on along the internal space. This introduces two difficulties for BoNs: on the one hand, the instanton must include a source for the flux so that it is absorbed at the point where the extra dimension pinches off and, on the other hand, the p-form fields associated to the fluxes involve additional degrees of freedom which complicate the resolution of the equations of motion.  

In order to show that the tunneling potential approach is also appropriate to discuss these situations, in this section we consider a simple model of a flux compactification first discussed in \cite{BS}, and study its instability by nucleation of bubbles of nothing.  The model presented in \cite{BS} admits an  $AdS_4 \times S^1$ vacuum, where the Kaluza-Klein circle is stabilised by the winding of a complex scalar field. In this case the corresponding source for the flux is a global solitonic string, a codimension 2 object in $d=5$ dimensions which wraps the surface of the BoN. The coupling of this object to the flux allows the Kaluza-Klein circle to shrink to zero size while keeping the instanton smooth. This solution involves  additional degrees of freedom besides the KK radius: those of the complex scalar field. As we describe next, this BoN can also be described using an appropriate generalisation of the tunneling potential approach to the multifield case \cite{EK,EGrav}.

The $5d$ action (with Minkowski signature) studied in \cite{BS} is
\be
S=\int d^5x\sqrt{-G}\left[\frac{1}{2\kappa_5}R_5-\frac12\partial_M\bar\Phi\partial^M\Phi-\frac{\lambda_5}{4}(\bar\Phi\Phi-\eta_5^2)^2-\Lambda_5\right]\ ,
\ee
with $\lambda_5>0$ and $\Lambda_5<0$. The $5d$ metric $G_{MN}$ describing the BoN geometry can be read from the line element
\be
ds^2=dr^2+\mathcal{R}^2 B^2(r)(-dt^2+\cosh^2t\, d\Omega_2^2)+ R_{KK}^2 C^2(r)dy^2\ ,
\ee
and the scalar field configuration is given by
\be
\Phi(x^M)=f_5(r)e^{iny}\ ,
\ee
where $n$ is  an integer.
Here $y \in [0,2 \pi)$ parametrises the KK circle, and the metric functions $C(r)$ and $B(r)$ satisfy the boundary conditions \eqref{eq:BoNcoreBC} and \eqref{eq:BoNinfBC}, so that $\mathcal{R}$ is the BoN nucleation radius and $R_{KK}$ the KK radius.

Integrating over the 5th dimension\footnote{This leads to $1/\kappa=2\pi R_{KK}/\kappa_5$. We set set $\kappa=1$ below.}, $y$, followed by a Weyl rescaling of the $4d$ metric:
\be
g_{\mu\nu}\to g_{\mu\nu}e^{\sqrt{2/3}\phi}\ ,
\ee
where the field $\phi$ is defined by the relation
\be
C(r)=\mathrm{e}^{-\sqrt{2/3}\phi}\ ,
\label{rC}
\ee
gives the $4d$ reduced action
\be
S=\int d^4x\sqrt{-g}\left\{
\frac{R_4}{2}-\frac{(\partial\phi)^2}{2}-\frac{(\partial f)^2}{2}-V(\phi,f)
\right\} ,
\ee
where 
\be
V(\phi,f)=e^{\sqrt{2/3}\phi}\left[\frac{n^2}{2R_{KK}^2}f^2e^{2\sqrt{2/3}\phi}+\frac{\lambda}{4}(f^2-\eta^2)^2+\Lambda\right]\ .
\label{Vphif}
\ee
The relation between $4d$ and $5d$ quantities is
\be
f\equiv \sqrt{2\pi R_{KK}}\, f_5\ ,\quad
\eta \equiv \sqrt{2\pi R_{KK}}\, \eta_5\ ,\quad
\Lambda \equiv 2\pi R_{KK}\, \Lambda_5\ ,\quad
\lambda \equiv \lambda_5/(2\pi R_{KK})\ .
\label{5d4d}
\ee
The  $4d$ (Euclidean) metric can be written in CdL form $ds^2=d\xi^2+\rho(\xi)^2d\Omega_3^2$ with
\be
\frac{dr}{d\xi}=e^{\sqrt{2/3}\phi/2}\ ,\quad\quad
\rho(\xi)=\mathcal{R} B(r)e^{-\sqrt{2/3}\phi/2}\ .
\label{xiB}
\ee
In terms of these variables, and using relations like $
(\partial \phi)^2=g^{\mu\nu}\partial_\mu\phi\partial_\nu\phi=e^{\sqrt{2/3}\phi}(\partial_r\phi)^2=(\partial_\xi\phi)^2$,
the Euclidean action for the $O(4)$ symmetric BoN takes the form 
\be
S_E=2\pi^2\int d\xi \rho^3\left\{-\frac12 R_4+\frac12 \dot\phi^2+\frac12 \dot f^2+V(\phi,f)\right\}\, 
\label{SEBoN}
\ee
where $\dot x\equiv dx/d\xi$ and
\be
R_4=\frac{6}{\rho}(1-\rho \ddot\rho-\dot\rho^2)\ .
\ee
The action (\ref{SEBoN}) takes the form of the action for a multifield bounce. The bounce equations of motion read
\bea
\ddot \phi + 3\frac{\dot\rho}{\rho}\dot\phi  &=& \partial V/\partial\phi\ ,\\
\ddot f + 3\frac{\dot\rho}{\rho}\dot f  &=& \partial V/\partial f\  ,
\label{eomf}\\
\dot\rho^2-1 &=& \frac13 \kappa\rho^2\left(\frac12 \dot\phi^2+\frac12 \dot f^2-V\right). 
\eea
It can be checked that the $5d$ BoN equations of motion derived in \cite{BS}
coincide with the bounce equations above, once $5d$ quantities are translated to $4d$ quantities following (\ref{5d4d}), $r$-derivatives are expressed in terms of $\xi$-derivatives and the relations (\ref{rC}) and (\ref{xiB}) are used. 

In terms of the four dimensional fields $\rho$ and $\phi$, the regularity conditions \eqref{eq:BoNcoreBC} at the BoN surface, $\xi\to 0$, are given by the expressions \eqref{phirhotop} which, in the case of a circle compactification reduce to   
\be
\phi(\xi) \simeq -\sqrt{\frac{2}{3}}\log\left(\frac{3\xi}{2R_{KK}}\right)\to \infty\ ,\qquad \rho(\xi)\simeq {\cal R}\left[\frac{3\xi}{2R_{KK}}\right]^{1/3} \to 0\ .
\label{eq:BCfluxBoN1}
\ee
As a consequence of these boundary conditions and \eqref{eomf}, it is also possible to show that   $f(\xi) \sim   \xi^{2n/3} \to 0$ in the same limit. The boundary conditions \eqref{eq:BoNinfBC} at infinity, $\xi\to\infty$,  take the form 
\be
\phi(\infty)=\phi_m=0, \quad \rho(\infty)\to\infty, \quad  \rho'(\infty)=1 \ ,
\label{eq:BCfluxBoN2}
\ee
with  $f(\infty)=f_m$, a constant. The scalar potential has an AdS critical point at $(\phi_m,f_m)$  provided we set
\be
f_m^2= \eta^2 - \frac{n^2}{\lambda R_{KK}^2}= \frac{2 \eta^2}{5} (1+ \frac{3}{2} \Delta) \le \eta^2, \quad \phi_m=0\ ,
\ee
and we impose the KK radius $R_{KK}$ to satisfy
\be 
R_{KK}^2 = -\frac{3 n^2\eta^2}{4\Lambda} (1+ \Delta)\le -\frac{3 n^2\eta^2}{2\Lambda}\ ,
\ee
where\footnote{We follow the notation of \cite{BS} except for the KK radius (which we denote by $R_{KK}$) and the size of the KK circle $R_{KK} C(r)$ (given by the function $r C(r)$ in \cite{BS}).} $\Delta\equiv \sqrt{1+20\Lambda/(9\lambda \eta^4)}\in [0,1]$. The value of the potential at the critical point is
\be
V_{m}\equiv 
V(\phi_m,f_m)=-\frac{9}{25} \left(\frac{2}{3} + \Delta \right) (1-\Delta) \lambda \eta^4 \le 0\ ,
\ee
which agrees with the value of $H^2$ given in \cite{BS}. The following quantities are the same in $4d$ or $5d$: $\Lambda/(\lambda\eta^4)=\Lambda_5/(\lambda_5\eta_5^4)$, $\kappa \Lambda=\kappa_5\Lambda_5$ and $\eta^2/\Lambda=\eta_5^2/\Lambda_5$.

In the tunneling potential formalism, the multifield BoN can be described by a $V_t$ that is related to the Euclidean quantities by
\be
V_t = V -\frac12 \dot\phi^2-\frac12\dot f^2\ ,
\label{eq:FluxVt}
\ee
and can be expressed as a function of a single field $\varphi$ which is
defined by
\be
d\varphi^2 \equiv d\phi^2+ df^2\ ,
\label{varphi}
\ee
with $\varphi(\xi=0)=0$. Note that $\phi'^2+f'^2=1$ follows from (\ref{varphi}).
We therefore have 
\be
V_t(\varphi) = V -\frac12 \dot\varphi^2\ .
\ee
The EoM for $V_t(\varphi)$ is of the same form as for the single-field case, eq.~(\ref{EoM}), but with $x'\equiv dx/d\varphi$. As now we have two fields, in order to determine the trajectory in field space followed by the instanton solution, there is an additional equation, which reads \cite{EK,EGrav}
\be
2(V-V_t)\phi''=\nabla_T V\equiv\partial V/\partial\phi -\phi'V'\ ,
\label{EoMT}
\ee
where $V'=\partial V/\partial \varphi=(\partial V/\partial\phi) \phi'+ (\partial V/\partial f) f'$.

\begin{figure}[t!]
\begin{center}
\includegraphics[width=0.7\textwidth]{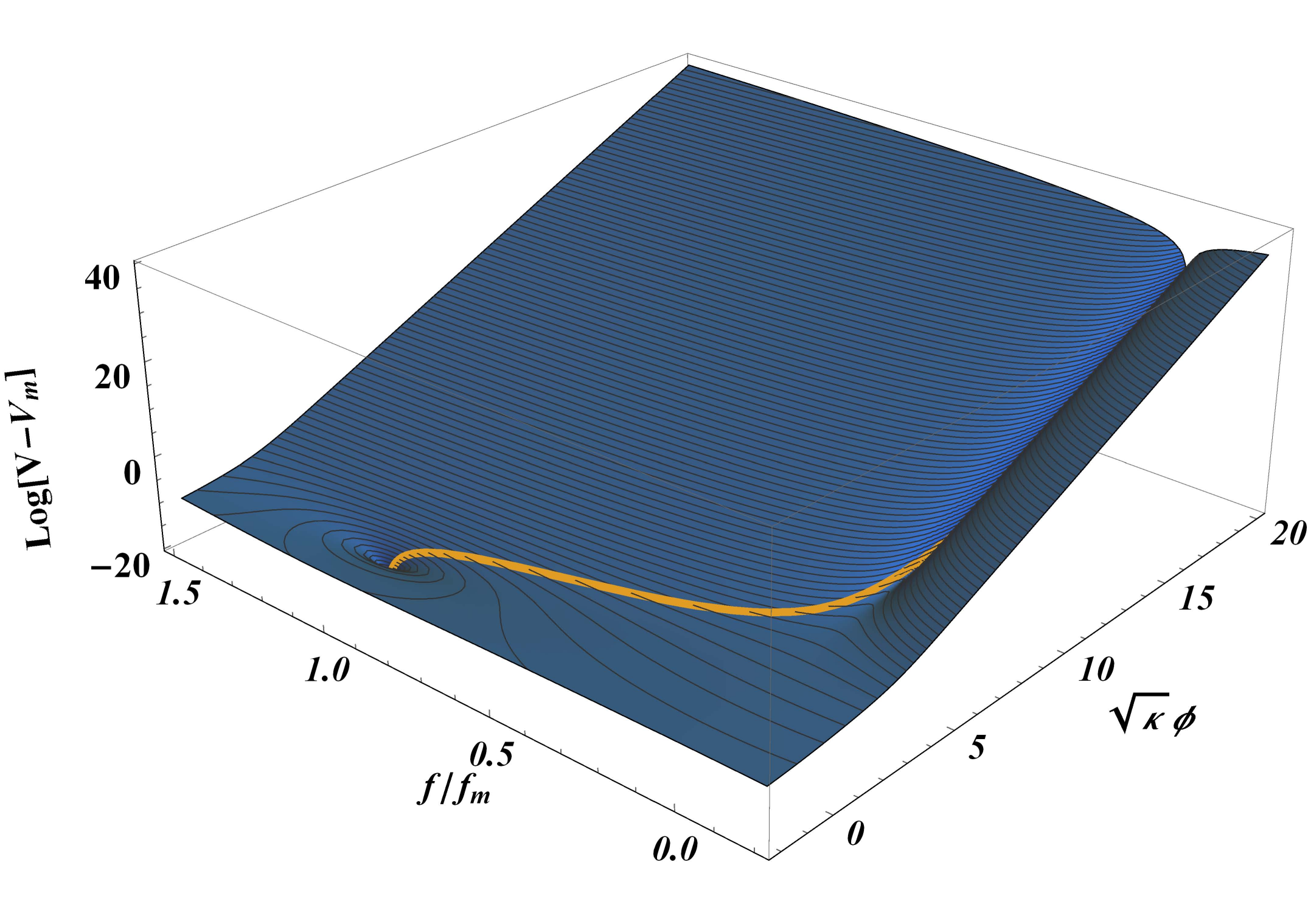}\\
\includegraphics[width=0.45\textwidth]{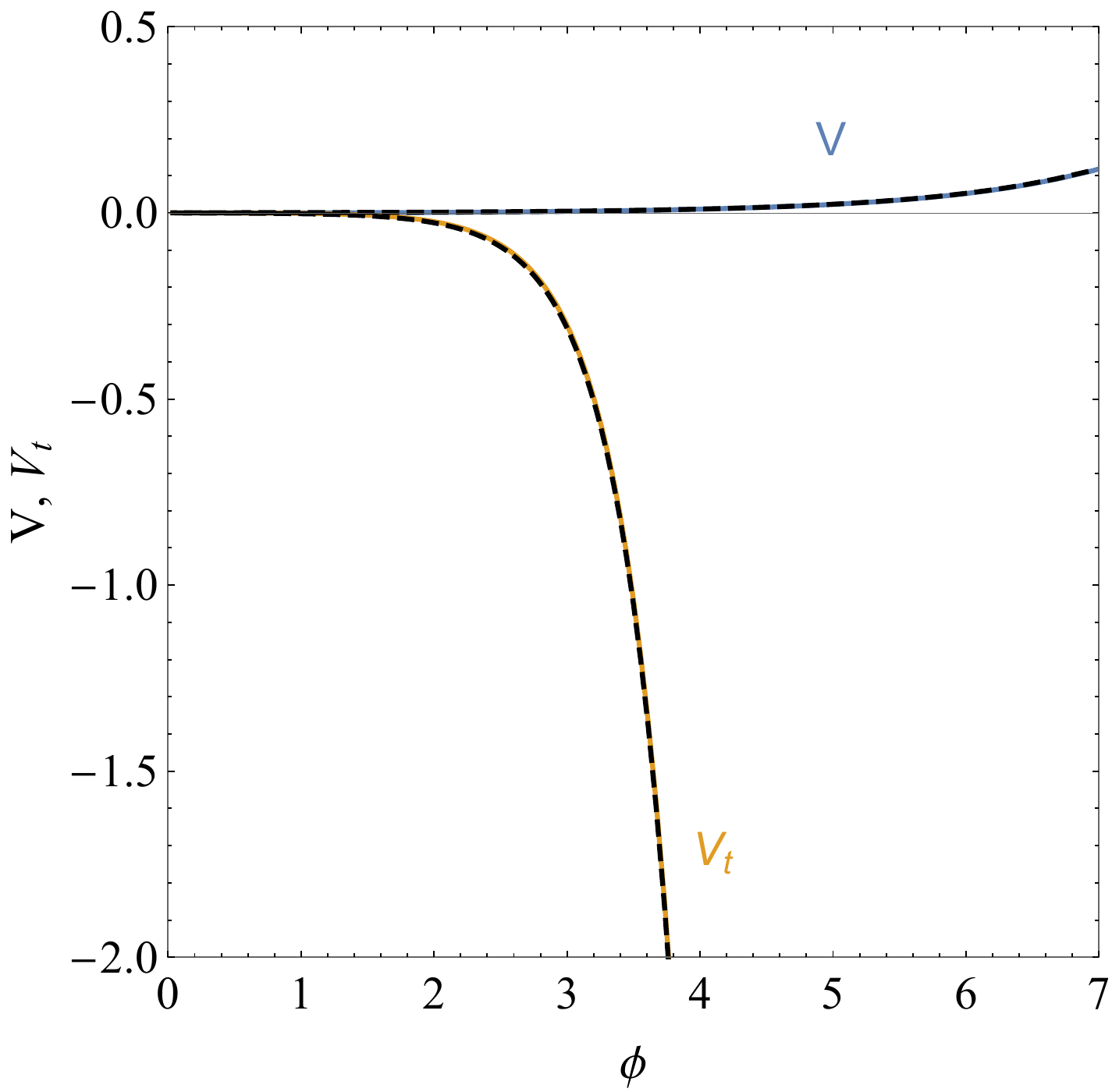}
\end{center}
\caption{Upper plot: Potential $V(\phi,f)$ from (\ref{Vphif}), plotted logarithmically, showing the minimum and the potential defile along $f=0$. The orange line follows the trajectory for BoN vacuum decay. Lower plot: $V$ and $V_t$ along the BoN, using $\phi$ as parameter. The dashed lines show the asymptotic approximations from (\ref{Vasymptotic}) and (\ref{Vtasymptotic}).
\label{fig:VFlux}
}
\end{figure}

According to the boundary conditions \eqref{eq:BCfluxBoN1},  in the regime of small $\xi$ (or large $\phi$) we have $\phi\to\infty$ and $f\to 0$ 
so that we can take $f$ as a small correction and identify $\phi$ and $\varphi$. It is immediate to check that this limiting behaviour is consistent with \eqref{EoMT}, since this equation is trivially satisfied when we set $\varphi\simeq \phi$. The scalar potential has the following asymptotic behaviour for $\varphi \simeq \phi \to \infty$
\be
V (\varphi\to\infty)= \left(\Lambda+\frac{\lambda\eta^4}{4}\right)e^{\sqrt{2/3}\varphi} + \frac{n^2}{2R_{KK}^2}f^2(\varphi) \, e^{3\sqrt{2/3}\varphi} +{\cal O}\left(f^2e^{\sqrt{2/3}\varphi}\right)\ .
\label{Vexp}
\ee
Although $f$ is expected to be exponentially suppressed, it appears in the potential enhanced by $e^{3\sqrt{2/3}\varphi}$ so that we should check if
its contribution is small or not compared with the first term in the potential above.  The limiting behaviour of $f$ when $\xi\to0$  implies $f\simeq f_0 e^{-n\sqrt{2/3}\varphi}$ for some $f_0>0$ [see \eqref{eq:BCfluxBoN1} and the discussion below] and, therefore, for $n=1$ both terms in (\ref{Vexp}) are of the same order, while for $n>1$ the flux contribution becomes subleading. Substituting  the $n=1$ form of the scalar potential 
\be
V(\varphi) \simeq \left(\Lambda+\frac{\lambda\eta^4}{4}+\frac{f_0^2}{2R_{KK}^2}\right) \mathrm{e}^{\sqrt{2/3} \varphi}\ ,
\label{Vasymptotic}
\ee
in equation \eqref{EoM} we find that the tunneling potential should be of the form\footnote{We discard a second solution of type $-$ with $a = 1/3$, outside the 
consistency range $1/\sqrt{3}<a<1$, see the discussion in section \ref{sec:BotUp}.}
\be
V_t(\varphi) \simeq -\frac{3}{4R_{KK}^2}e^{\sqrt{6}\varphi}\ ,
\label{Vtasymptotic}
\ee
which, according to our classification in section \ref{sec:BotUp} corresponds to a type $0$ case (see table \ref{table:types}). This is as expected on general grounds from the general discussion in section~\ref{sec:topdown}: from eq.~(\ref{UVresults}), for a compactification from $5d$ to $4d$ we get $\rho\sim \xi^{1/3}$, which is the behaviour corresponding to type $0$ cases (see table~\ref{table:types}). The equation of motion for the tunneling potential \eqref{EoM} leaves the coefficient in front of the exponential as a free parameter, which we have fixed using relation \eqref{eq:FluxVt} and the regularity conditions \eqref{eq:BCfluxBoN1}.

Figure~\ref{fig:VFlux}, upper plot, shows the potential (\ref{Vphif}) for
a choice of parameters already considered in \cite{BS}: $n=1$, $\Lambda_5=-(0.347 M_{P5})^5$, $\eta_5=(0.630 M_{P5})^{3/2}$ and $\lambda_5=1/(0.995M_{P5})$ where $M_{P5}=1/\kappa_5^{1/3}$. The trajectory of the BoN, marked in orange, goes out of the potential minimum and heads towards the narrow potential defile along $f=0$ (with $f_0\simeq 1.2$). The values of $V$ and $V_t$ along that BoN trajectory, as functions of $\phi$, are given in the lower plot of the same figure. The dashed lines correspond to the analytical asymptotic approximations in (\ref{Vasymptotic}) and (\ref{Vtasymptotic}), which are excellent. 

As anticipated at the end of section~\ref{sec:topdown}, this example shows how the presence of an additional field, such as $f$, can modify the asymptotic scaling behaviour of the BoN solution for $\phi\to\infty$: For a fixed value of $f$, one has $V\sim e^{\sqrt{6}\phi}$ [see (\ref{Vphif})], which corresponds to the expected scaling of a flux-generated potential from 5 dimensions [see (\ref{Vflux})]. However, when one takes into account how $f$ responds to the BoN $\phi$ profile, the effective potential along the BoN has a different scaling, $V\sim e^{\sqrt{2/3}\phi}$, which is subleading and corresponds to a type 0 BoN. In physical terms, this phenomenon can be related to the presence of the global  string wrapping the BoN surface,  mentioned at the beginning of this section. The flux and its induced potential, if not neutralised, are incompatible with a smooth BoN geometry [see discussion in section \ref{sec:topdown}], and thus they prevent the collapse of the KK circle to zero size. The role of the global string soliton, whose position is precisely the vanishing locus of $f$, is to absorb the flux, and thus, to cancel its contribution to the scalar potential.  

As we mentioned in the introduction, other smooth BoN solutions have been studied in the context of field theory models of flux compactification, see \cite{Blanco-Pillado:2010vdp}. The situation in these other models seem to resemble the one presented here. The presence of the p-form flux allows for the perturbatively stable compactification to exist and their smooth solitonic source is the key ingredient to be added to the bare BON to permit the quantum instability of the type we are discussing here. Furthermore the solution is such that deep in the core of the source the flux contribution is turned off leading to a solution of the form given by Eqs. (\ref{phirhotop}). In other words, the solitonic nature of these solutions allow the solutions to be always of the type 0 BoN.

End of the world type of solutions have also been obtained before in field theory models of flux compactification \cite{Blanco-Pillado:2010vdp}. It would be interesting to investigate these type of solutions from a $4d$ point of view to compare with the solutions obtained in the potential tunneling formalism similarly to what was done in previous sections. We leave this comparison for future work.

Finally, let us also note that in models of flux compactifications one can think of the BoN as a limiting case of the flux transitions \cite{BS}. Therefore in order to understand the most likely decay channel one should in principle consider these type of tunneling processes as well \cite{Brown:2010mf}. The $4d$ description of those transitions would also involve the presence of a brane coupled to the moduli field \cite{Bandos:2023yyo,Draper:2023ulp}. This should also be incorporated in the tunneling potential formalism.

\section{Summary and Outlook\label{sec:concl}}

Bubble of nothing (BoN) decays are of fundamental importance for the stability of vacua in theories 
with compactified extra dimensions, like String Theory, and they are closely connected to 
the cobordism conjecture of the Swampland program. Their study is greatly facilitated by dimensional reduction to an effective theory in 4 dimensions, with the size of the compact dimension(s) described by a modulus scalar field, stabilized by some potential $V(\phi)$. In this language, the BoN instanton reduces to a Coleman-De Luccia (CdL) instanton with a singular behaviour at its core (where the field diverges, $\phi\to\infty$) but having a finite action (that agrees with the action of the BoN instanton in the high dimensional theory).

The $4d$ reduced setting is very convenient to study the properties of BoNs in the presence of a modulus potential, $V(\phi)$, and has been used in the past for this purpose, most recently in \cite{DGL}. In this paper we have followed the $4d$ approach but using the tunneling potential formalism rather than traditional Euclidean methods. In this formalism, vacuum decay is described by a tunneling potential function $V_t(\phi)$ that can be compared directly with $V(\phi)$ without having to deal with the profiles of the modulus field  nor the Euclidean metric. The analysis is further facilitated by the fact that all the possible decay channels of a given potential [BoNs, Coleman-de Luccia or Hawking-Moss (HM) instantons as well as the so-called pseudo-bounces] can be described in the same $V_t$ language, with different types of decay  connected with the others in a continuous way. Moreover, the tunneling action for the different decay modes can be calculated using the simple universal formula (\ref{SVt}). The fact that this general formula reproduces the Euclidean BoN action is an important result proven in Appendix \ref{App:ActionAgrees}.

Using the $V_t$ formalism we have performed an analysis exploring which type of modulus potentials $V(\phi)$ allow for BoN decays and which different types of BoN exist. This study is similar in spirit to the one of \cite{DGL}. We confirm many of the findings of \cite{DGL} and extend that work in several directions. 

We identify four different types of BoN, with different asymptotic behaviour in the compactification limit ($\phi\to\infty$, corresponding to the BoN core) and different possible higher dimensional origin, as summarized in table~\ref{table:types}. Type 0 BoNs can appear when the compact manifold is a circle (these, for dS vacua, were the main subject of \cite{DGL}). Type $-$ BoNs can appear instead when the compact manifold is an $S^d$ sphere, with $d>1$. The type $+$ or $-^*$ BoNs can be relevant vacuum decay channels for more complicated compact geometries, that require the presence of some UV object or defect to allow the shrinking to zero of the internal manifold in the BoN core. 

For BoNs of types 0 or $-$, there is a simple link between the asymptotic $\phi\to\infty$ behaviour of $V$, $V_t$ and important quantities derived from them, like $D^2$ in (\ref{D2}), and the geometric properties of the BoN in the higher dimensional theory (like the KK radius, $R_{KK}$, and the BoN radius, ${\cal R}$). Assuming the compact manifold is $S^d$, we find for $\phi\to\infty$,
\be
V\simeq  -\frac{d(d-1)}{2\kappa R_{KK}^2}e^{\sqrt{\frac{2(d+2)\kappa}{d}}\, \phi}+...\ ,\quad\quad
V_t\simeq  -\frac{3d^2}{4\kappa R_{KK}^2}e^{\sqrt{\frac{2(d+2)\kappa}{d}}\, \phi}+...\ ,
\label{UVasym}
\ee
and
\be
D^2\simeq  \frac{9d(d+2)}{2\kappa^3 R_{KK}^2{\cal R}^2}e^{\sqrt{\frac{8\kappa}{d(d+2)}}\, (d+1)\phi}+...
\label{UVasym2}
\ee

These relations, which capture information from the higher-dimensional theory, can be confronted with the asymptotic behaviour of the $V_t$ solutions for BoN decay that we find by solving the equation of motion for $V_t$ in (\ref{EoMVt}) from low field to high field values.
Typically, for a given potential that does not grow as fast as $e^{\sqrt{6\kappa}\phi}$, we find a continuous family of possible type 0 BoN solutions $V_t(p;\phi)$, labeled by some parameter $p$, with the $\phi\to\infty$ asymptotics
\be
V_t(p;\phi)\simeq V_{tA}(p) e^{\sqrt{6\kappa}\phi}+...\ ,\quad
D(p;\phi)\simeq D_\infty(p) e^{\sqrt{8\kappa/3}\phi}+...
\label{4dasym}
\ee
with $V_{tA}(p)<0$ and $D_\infty(p)>0$. For a fixed higher-dimensional theory (thus fixed $R_{KK}$), matching (\ref{4dasym}) to (\ref{UVasym}) and (\ref{UVasym2}) selects a finite number of BoNs from the family [each with different radius ${\cal R}$, determined by $D(p;\phi)$, and different tunneling action]. 
The number of such selected BoN solutions is model dependent in the following way. 

If the modulus has a single vacuum (or if gravity forbids its decay to other minima), the selected BoN solution is unique (for a given $R_{KK}$). Moreover, when the vacuum is a Minkowski or AdS one, there is an upper critical limit $R_{KK}^*$ [corresponding to a lower limit on $-V_{tA}(p)$] for which the BoN has infinite action and radius and corresponds to an end-of-the-world brane. For theories with $R_{KK}>R_{KK}^*$, BoN decay  is not allowed (this is quenching by a CdL mechanism). Figures \ref{fig:BC0Mink} and \ref{fig:BC0AdS} show particular examples of this and section~\ref{sec:quench} is fully devoted to discussing this mechanism. 

However, when the modulus potential has additional vacua and admits standard decay channels to them (via either Coleman-de Luccia or Hawking-Moss instantons), there are at least two $V_t$ BoN solutions, the BoN solution with lowest action being the relevant one. Also in this case there is a critical value $R_{KK}^*$, which again corresponds to the minimum of $-V_{tA}(p)$. However, this time criticality corresponds to the merging of the two BoN solutions into one which is a saddle-point with finite tunneling action. In any case, for $R_{KK}>R_{KK}^*$ again BoN decay is forbidden (a dynamical quenching). Figures \ref{fig:PSCdLBoN}, \ref{fig:PSCdLBoNdS} and \ref{fig:BoNHMdS} show some examples of this  when the false vacuum is Minkowski and dS. Interestingly, we found  that BoN decay of dS vacua is dynamically forbidden only when the EFT and KK energy scales are comparable (without scale separation), that is, in regimes of parameter space where the validity of the EFT is questionable.

When BoN solutions coexist with CdL or HM instantons (the second case just described), the same parameter $p$ that labels the family of BoNs in some range of values, naturally describes (outside the BoN range) the CdL/HM instanton solutions as well as the so-called pseudo-bounces that connect CdL and HM (an action valley in configuration space). The associated tunneling actions can be easily calculated in the $V_t$ formalism and we find an action $S(p)$ which is continuous across the boundaries between different classes of $V_t$ solutions. End-of-the-world branes
correspond to $S(p^*)\to\infty$, while the critical cases with dynamical quenching and 
finite action correspond to a maximum of $S(p)$. Again, figures \ref{fig:PSCdLBoN}, \ref{fig:PSCdLBoNdS} and \ref{fig:BoNHMdS} illustrate this behaviour. 

The previous results for type 0 BoNs are in agreement with the findings of \cite{DGL} when there is overlap between both works. We believe that the powerful $V_t$ approach sheds light from a new angle on the whole topic. Moreover, we have extended the analysis to any kind of vacua (Minkowski, dS or AdS), clarified the nature of the hybrid branches of solutions found in \cite{DGL}, and studied in detail the two types of quenching of BoN decays.  Concerning other types of BoN, not studied in \cite{DGL}, we find that type $-$ BoN solutions for a fixed potential (with the right asymptotic behaviour) appear singly, at the boundary of type-0 BoN families. 
This is consistent with the fact that now a given $R_{KK}$ fixes the asymptotic behaviour of both $V$ and $V_t$, see (\ref{UVasym}). The BoN radius is as usual determined by the asymptotics of $D$. Figures \ref{fig:BoNm} and \ref{fig:BoNPSHM} show examples that illustrate the behaviour just described.

Using the $4d$ approach we have also found (both numerically and analytically) examples of potentials that admit BoN solutions of types $+$ and $-^*$. While the type $+$ can appear in a continuous family (like the type 0 BoNs, see figure~\ref{fig:typeplus}),
the type $-^*$ BoNs appear as limiting cases of type 0 BoN families (like the type $-$ Bons, see figures~\ref{fig:CdLPlateau} and \ref{fig:Special}). 
Nevertheless, for type $+$ and $-^*$ BoNs the picture is less complete as the higher dimensional theory has a more complicated geometry for the compact space and shrinking it to zero requires a defect or some other UV object, which we have not considered in detail. 

BoNs in flux compactifications are also of particular interest as their existence is nontrivial: shrinking the compact space in the presence of flux requires a charged object at the BoN core to absorb the flux. We have shown that a $4d$ description of such a BoN (in a $5d$ flux compactification) is also possible. Besides the modulus field, the $4d$ theory contains  an additional scalar field descended by dimensional reduction from the charged object.  
In this case, a two-field $V_t$ can describe the BoN. Interestingly, the $4d$ counterpart of the flux absorption is realized by a $V_t$ that selects a direction in the two-field space with the right (type 0) asymptotic behaviour rather than the generic type $+$ asymptotics expected in flux compactifications, see figure~\ref{fig:VFlux} and the discussion in section~\ref{sec:BoNFlux}.

There is a number of further directions for future work. The study of singular BoNs (types $+$ and $-^*$) and how they are regulated by some UV object or defect is of clear interest. Applying a $V_t$ approach to their description seems possible and our first explorations look rather promising. Indeed, we find that the $V_t$ description of Witten's BoNs with defects \cite{Blanco-Pillado:2016xvf,Hebecker} is rather straightforward via a simple rescaling of the prefactor of Witten's BoN in (\ref{WBoN}) to take into account the deficit angle associated to the singular behaviour of the BoN. It would be interesting to see how this generalizes in the presence of a nonzero potential for the modulus field.

Our results indicate that in some limits the existence of the BoN decay channel is suppressed by some dynamical obstruction. It would be interesting to understand this effect in the higher dimensional theory, in particular in models of flux compactification. In some cases this quenching can be understood as a CdL suppression where the bubble becomes static and infinite. This type of objects have been found numerically in simple higher dimensional theories in \cite{Blanco-Pillado:2010vdp}. On the other hand, in other cases the suppression does not have a limiting static wall and it would be interesting to explore such solutions, and study if they can be consistently obtained within the regime of validity of the $4d$ EFT. Finally, we would also like to investigate the existence of the End of the World type of BoNs in connection with supersymmetric vacua, similarly to what was done in \cite{Blanco-Pillado:2016xvf}.

\section*{Acknowledgments\label{sec:ack}} 
The work of J.R.E. and J.H. has been supported by the following grants: IFT Centro de Excelencia Severo Ochoa CEX2020-001007-S and by PID2022-142545NB-C22 and PID2021-123017NB-I00 by MCIN/AEI/10.13039/501100011033 funded by ``ERDF; A way of making Europe''. The work by J.H. is supported by the FPU grant FPU20/01495 from the Spanish Ministry of Education and Universities. J.J.B.-P. has been supported in part by the PID2021-123703NB-C21 grant funded by MCIN/ AEI /10.13039/501100011033/ and by ERDF; ”A way of making
Europe”; the Basque Government grant (IT-1628-22) and the Basque Foundation for Science (IKERBASQUE).

\appendix

\section{Energy of the BoN\label{App:E}}
Consider the BoN decay of a false Minkowski vacuum. A nucleated BoN has zero energy (so that energy is conserved in the decay process). In the case of Witten's BoN, for the decay of the $5d$ KK vacuum, this follows immediately from the asymptotic behaviour of the metric \eqref{BoNmetric} at $r\to\infty$ which does not have $1/r$ terms (see e.g. \cite{Brown:2014rka}). This key property should be inherited by the $4d$ description of the BoN. 

Let us consider the $4d$ description of a general BoN mediating the decay of a compactification with vanishing vacuum energy $V(\phi_+)=0$ (i.e. the non-compact component of space-time is  Minkowski space). The usual energy integral 
\be
E=4\pi\int_0^\infty d\rho \rho^2\left[\frac12\dot\phi^2+V(\phi)\right]\ ,
\label{eq:energyIntegral}
\ee
which vanishes for regular CdL bubbles (see \cite{MPW}) diverges for BoNs, even for Witten's one\footnote{In that case $V(\phi)\equiv 0$ and then the integrand is positive definite, making it impossible to have $E=0.$}. To understand the origin of this discrepancy, first note that the integrand in (\ref{eq:energyIntegral}) is a total derivative so that 
\be
E=\left.\frac{4\pi}{3}\rho^3\left[V(\phi)-\frac12\dot\phi^2\right]\right|_{\xi=0}^{\xi=\infty}\ ,
\label{eq:Eboundary}
\ee
which diverges at $\xi=0$ (the $\xi=\infty$ term vanishes).
One can connect this expression with the asymptotic behaviour of the metric by using the constraint equation (\ref{E2}) to get
\be
E=\left.\frac{4\pi}{\kappa}\left[\rho(1-\dot\rho^2)\right]\right|_{\xi=0}^{\xi=\infty}\ .
\ee
The $\xi=\infty$ contribution corresponds to the usual ADM mass, $M_{ADM}$,  which should vanish, while the $\xi=0$ contribution vanishes provided the metric is smooth at the origin. However, since in the $4d$ description the  BoN space-time is singular at $\xi=0$, the integral \eqref{eq:energyIntegral} cannot be identified with the ADM mass. 
A careful derivation of $M_{ADM}$ in $4d$ starting from $5d$ (similar to the calculation of the Euclidean action in Subsection~\ref{sec:4dReduction}) gives
\be
M_{ADM} = \frac{4\pi}{\kappa}\left[\rho(1-\dot\rho^2)\right]_{\xi=0} + 4\pi\int_0^\infty d\rho \rho^2\left[\frac12\dot\phi^2+V(\phi)\right]. 
\ee
Although both terms in the r.h.s. of the previous expression are separately divergent, the divergences cancel exactly. Using \eqref{eq:Eboundary}, we get 
\be
M_{ADM} =\frac{4\pi}{3}\rho^3\left[V(\phi)-\frac12\dot\phi^2\right]_{\xi=\infty}\ ,
\ee
Plugging the asymptotic behaviour of $\rho$ and $\dot\phi$ for the $4d$ reduced Witten's BoN it can be checked that indeed $M_{ADM}=0$. 

In more general situations, but still imposing that $V(\phi_+)=0$, we find that the expression for the BoN energy, written in the $V_t$ formalism, reads
\be
M_{ADM} =\left. 36 \pi  \frac{\left[2 (V-V_t)\right]^{3/2}}{D^3} V_t \right|_{\phi\to \phi_+}
\label{MADM}
\ee
For Witten's BoN, it can be readily checked that using $V_t$ from (\ref{WBoN}) leads to $M_{ADM}=0$. For other cases of Minkowski vacuum decay, we can get the limit (\ref{MADM}) by using the low-energy expansion ($\phi\to\phi_\pp=0$) of $V_t$ obtained in (\ref{eq:prodLog}), which gives $V\sim \phi^2$, $V_t,D\sim\phi^2/\log\phi$ and leads to $M_{ADM}=0$. This result can be extended to the decay of AdS vacua (for dS vacua one cannot define energy).

\section{\texorpdfstring{\bma{S[V_t]=\Delta S_E}}{S(Vt)=Delta SE}\label{App:ActionAgrees}}
The agreement between the action $S[V_t]$ of the $V_t$ formalism and the Euclidean action difference $\Delta S_E$ of Coleman-De Luccia for regular CdL transitions was proven in \cite{Eg}  (for the proof in general dimension, see \cite{EF}). In this section we extend this proof to BoN decays. 
 
Let us first review briefly the proof for regular CdL solutions.
In \cite{Eg,EF} the Euclidean action is rewritten in the form
\be
S_E=-2\pi^2\int_0^{\xi_{\rm max}}\rho V d\xi + S_{\rm GHY}\ ,
\ee 
where $\xi_{\rm max}=\infty$ for AdS or Minkowski false vacua and finite for dS, while $S_{\rm GHY}$ is the Gibbons-Hawking-York boundary action \cite{GHY} (the value of this term is irrelevant to show the agreement between the actions in both formalisms).
Let us consider first the case of Minkowski or AdS vacuum decays.
The proof rewrites $\Delta S_E=S_E[\phi_B]-S_E[\phi_\pp]$ for CdL decay as an integral in field space, using the dictionary between formalisms, and then establishes the identities 
\bea
s(\phi)-s_{E,B}(\phi) &=&\frac{dH_B}{d\phi}\ ,
\label{slink1}\\
s_{E,\pp}(\phi) &=&\frac{dH_\pp}{d\phi}\ ,
\label{slink2}
\eea
where the action densities are defined by 
\bea
S[V_t]&=&\int_{\phi_\pp}^{\phi_0} s(\phi)d\phi=\frac{6\pi^2}{\kappa^2}\int_{\phi_\pp}^{\phi_0} \frac{(D+V_t')^2}{D V_t^2}d\phi\ ,\\ 
\Delta S_E&=&\int_{\phi_\pp}^{\phi_0} \left[s_{E,B}(\phi) - s_{E,\pp}(\phi)\right] d\phi=
-108\pi^2\int_{\phi_\pp}^{\phi_0}\frac{V-V_t}{D^3}\left(V+\frac{V_\pp V_t'}{D_\pp}\right)d\phi\ ,
\eea
where $D_\pp^2\equiv V_t'{}^2+6\kappa(V-V_t)(V_t-V_\pp)$ and the term proportional to $V_\pp$ comes from subtracting the false vacuum action (see \cite{Eg} for details).
The functions $H_B$ and $H_\pp$ are given by
\be
H_B=\frac{216\pi^2(V-V_t)^2}{D^3}\left[\frac{V_t(V_t'-2D)}{(D-V_t')^2}\right]\ ,\quad
H_\pp=\frac{216\pi^2(V-V_t)^2}{D^3}\left[
\frac{V_\pp(D_\pp+2D)}{(D+D_\pp)^2}\right]\ .
\ee
The identities (\ref{slink1}) and (\ref{slink2})  hold on-shell, that is, for any $V_t$  solution of the EoM (\ref{EoMVt}) and thus holds both for CdL and BoN solutions, so that we can still rely on them for the BoN proof below. Integration of the sum of the identities (\ref{slink1}) and (\ref{slink2})  in the interval $(\phi_+,\phi_0)$ gives 
\be
S[V_t]=\Delta S_E+H_B(\phi_0)+H_\pp(\phi_0)-H_B(\phi_+)-H_\pp(\phi_+)\ ,
\ee
and the equality of the actions results from $H_B(\phi_0)=H_\pp(\phi_0)=H(\phi_+)=H_\pp(\phi_+)=0$, which follows from the behaviour of $V$ and $V_t$ at $\phi_+$ and $\phi_0$ for CdL solutions describing $V_\pp\leq 0$ decays, see \cite{Eg}.

In the case of dS vacuum decay we have
\be
\Delta S_E = \int_{\phi_{0\pp}}^{\phi_0}s_{E,B}(\phi)d\phi + \frac{24\pi^2}{\kappa^2 V_\pp}\ , 
\ee
where the first term is the bounce action and the second (minus) the false vacuum action,
which is now finite. Integrating (\ref{slink1}) in the interval $(\phi_{0\pp},\phi_0)$ one gets
\be
\int_{\phi_{0\pp}}^{\phi_0}s(\phi)d\phi=\Delta S_E-\frac{24\pi^2}{\kappa^2 V_\pp}+H_B(\phi_0)-H_B(\phi_{0\pp})\ .
\label{dSagree}
\ee
It is straightforward to show that $H_B(\phi_0)=0$ and $H_B(\phi_{0\pp})=-24\pi^2/[\kappa^2 V(\phi_{0\pp})]$. This $H_B(\phi_{0\pp})$ combines with $-24\pi^2/[\kappa^2 V(\phi_{\pp})]$
to give the HM part of the $S{V_t}$ action missing in the l.h.s. of (\ref{dSagree}) resulting in $S[V_t]=\Delta S_E$.

To extend the previous proof to BoNs, we have to care about two differences with respect to the standard case. First note that now $\Delta S_E$ has an extra piece associated to the BoN core as given by (\ref{FinalSE}), 
\be
\delta_{BoN}\Delta S_E=-\left.\pi^2\sqrt{\frac{2}{3\kappa}}\rho^3\dot\phi\,\right|_{\xi=0}\ ,\label{deltaBoNSE}
\ee
which will appear for any type of BoN vacuum decay, which has an additional term from the extra-dimensional origin of the BoN.  Second, for BoN solutions the behaviour of $V$ and $V_t$ as $\phi\to\phi_\pp$ or $\phi\to\phi_{0\pp}$ is similar to that for regular CdL bounces so that $H_B$ and $H_\pp$ take the same previous values as well, and would lead to the equality between $\Delta S_E$ and $S[V_t]$ up to the extra BoN term (\ref{deltaBoNSE}). However, the asymptotic behaviour of $V$ and $V_t$ as $\phi\to\phi_0$ is different than for regular CdL solutions (as $\phi_0\to\infty$ for BoNs) and, it is remarkable that the asymptotics of $H_{B,\pp}(\phi_0\to\infty)$ is such that one gets precisely the term needed to reproduce (\ref{deltaBoNSE}), as we show below, thus completing the proof of $\Delta S_E=S[V_t]$.

It can be shown that $H_\pp(\phi\to\infty)=0$ for any type of BoN,  simply using the different asymptotics summarized in Table~\ref{table:types}. For $H_B(\phi\to\infty)$, using the same table, we find the following.

For type~0 solutions, $D$ is subleading compared to $V_t'$ and we find 
\be
H_B(\phi_0\to\infty)= \frac{216\pi^2(V-V_t)^2V_t}{D^3V_t'}=\frac{216\pi^2(V-V_t)^2}{D^3\sqrt{6\kappa}}\ ,
\ee
where we have used $V_t\sim e^{\sqrt{6\kappa}\phi}$ to derive the last expression. Using the $V_t$-to-Euclidean dictionary to rewrite this in terms of Euclidean quantities, we immediately find 
\be
H_B(\phi_0\to\infty)=-\sqrt{\frac{2}{3\kappa}}\pi^2\rho^3\dot\phi,
\label{HtoE}
\ee
evaluated at $\xi=0$. This is precisely the piece $\delta_{BoN}\Delta S_E$ of (\ref{deltaBoNSE}) that should be added to the usual Euclidean action integral to get the action for BoNs, and thus $S[V_t]$ agrees with it, as anticipated. In this case it can also be checked that (\ref{HtoE}) gives a finite nonzero result.

For type $-$ and type $-^*$ solutions, it can be readily checked from the asymptotic behaviours of $V$, $V_t$ and $D$ given in  section~\ref{sec:BotUp} that $H_B(\phi_0\to\infty)=0$. One also has $\rho^3\dot\phi\to 0$ for $\xi=0$ and thus $\delta_{BoN}\Delta S_E=0$ and again we find that $S[V_t]$ reproduces the Euclidean action.

For type $+$ solutions $D$ is also subleading compared to $V_t'$ at large field values, so that (\ref{HtoE}) also holds and this reproduces once again $\delta_{BoN}\Delta S_E$ so that $S[V_t]$ agrees with the Euclidean result. In this case, however, (\ref{HtoE}) diverges. Indeed we have 
\be
H_B(\phi_0\to\infty)\sim \frac{e^{2a\sqrt{6\kappa}\phi}}{e^{3\sqrt{\kappa/6}(3a+1/a)\phi}}\to\infty \ ,
\ee
as $a>1$. In the Euclidean formulation this corresponds to
\be
\rho^3\dot\phi\sim -\xi^{1/a^2-1}\ ,
\ee
which also diverges for $\xi\to 0$ and $a>1$. In spite of this divergence we find cases of this type with a finite total action for $a>1/\sqrt{3}$. In the Euclidean approach the divergence is cancelled by a contribution in $\Delta S_E$.

In summary, for all types of BoN we find agreement between the Euclidean action [supplemented by the needed extra term (\ref{deltaBoNSE}) from the higher dimensional theory] and the usual action $S[V_t]$ (without any added terms).

\section{Large field expansion of \texorpdfstring{$\bma{V_t(\phi)}$}{Vt} for \texorpdfstring{$\bma{V(\phi)=V_A e^{a\sqrt{6\kappa}\phi}}$}{V=VA Exp(a sq(6k)phi)}\label{App:Vexp}} 
In this appendix we derive the large field expansion of the tunneling potential if we assume that
$V(\phi)=V_A e^{a\sqrt{6\kappa}\phi}$ and consider the four possible different types of asymptotic behaviour discussed in the text. Although subleading terms in $V$ can have an influence on $V_t$, it is still instructive to just study this simple $V$. Also, we consider this $V$ not as being the whole potential, but just as valid in some field range extending to infinity.

Before solving for $V_t$, it is convenient to discuss first the solutions of the differential equation $D(\phi)^2=0$ [which we call $\overline{V_t}(\phi)$]. In the context of regular CdL bounces for Minkowski or AdS false vacua, $\overline{V_t}(\phi)$ with boundary conditions $\overline{V_t}(0)=V_\pp$ and $\overline V'_t(0)=0$ plays a role in determining whether the false vacuum decay is quenched by gravity or not \cite{EGrav}. Moreover, $\overline{V_t}$ solutions joining two vacua describe domain wall solutions. In the context of BoN solutions, $\overline{V_t}(\phi)$ solutions correspond instead to dynamical cobordisms/ end of the world branes \cite{DynCo}.

Figure \ref{fig:types} shows the ranges of $a$ and ${\rm sign}(V_A)$ for which the different types of asymptotic behaviour discussed below apply.

\begin{figure}[t!]
\begin{center}
\includegraphics[width=0.475\textwidth]{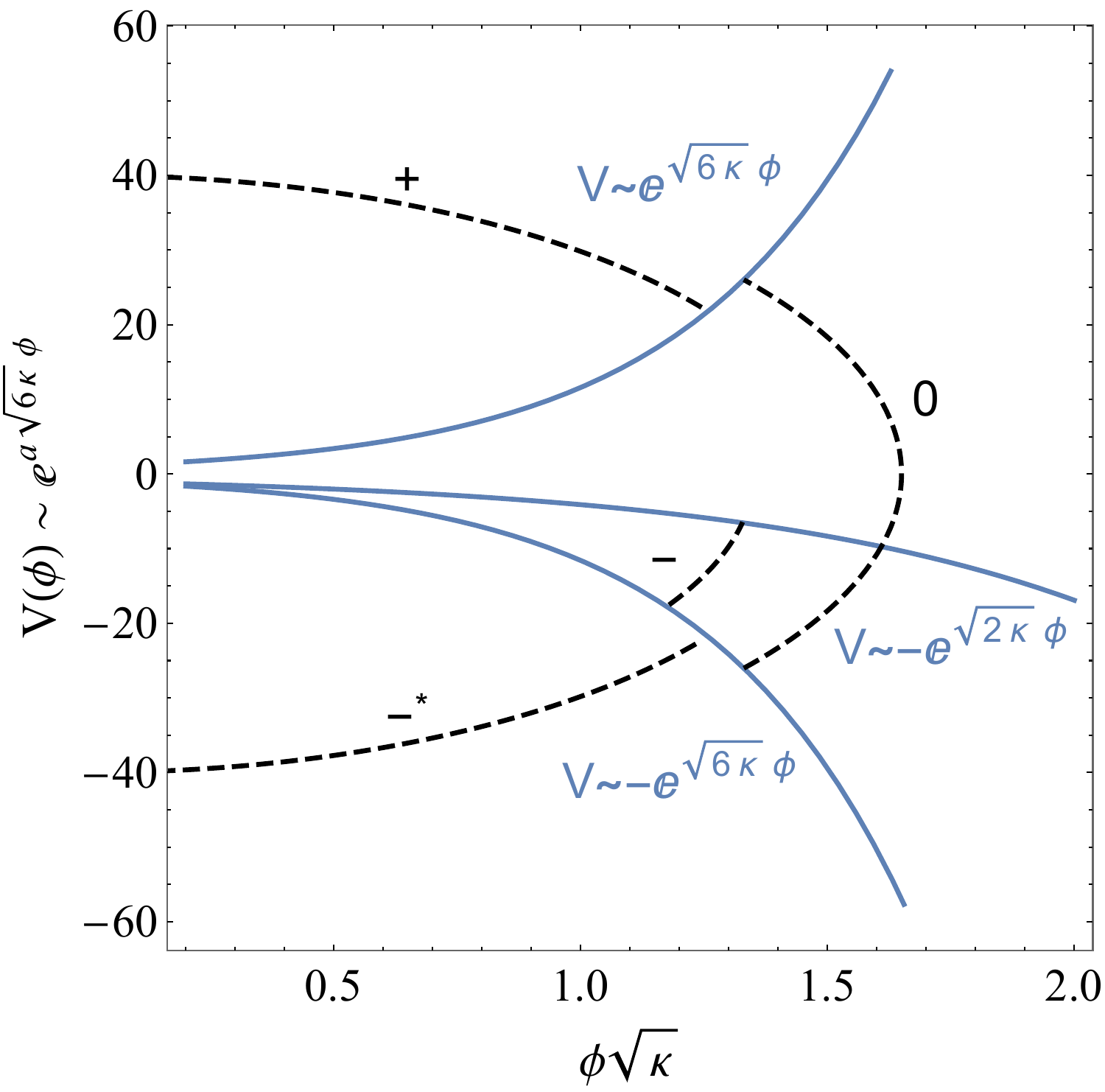}
\end{center}
\caption{For $V=V_A e^{a\sqrt{6\kappa}\phi}$ with varying $a$ and both signs of $V_A$, the plot shows the ranges for the four different types of asymptotic behavior of $V$ and $V_t$ discussed in the text. 
\label{fig:types}
}
\end{figure}

\subsection{\texorpdfstring{$\bma{\overline{V_t}}$}{bar Vt} Solutions}

The asymptotic behaviour of $\overline{V_t}(\phi\to\infty)$ in the four different types considered in the text can be obtained directly from the differential equation
$D(\phi)=0$ with $\overline{V_t}'<0$, which gives
\be
\overline{V_t}'=-\sqrt{-6\kappa(V-\overline{V_t})\overline{V_t}}\ .
\label{D0}
\ee
The solutions of this equation do not cross each other away from the points with 
$\overline{V_t}=V$, and they foliate the region of the $(V,\overline{V_t})$ plane with $\overline{V_t}<0$ and  $V-\overline{V_t}>0$.
The asymptotic behaviour of $V_t$ must be at least as strong as that of $\overline{V_t}$, as $D^2>0$ forces $V_t$ to have a slope more negative than that of the $\overline{V_t}$'s.

Below, we also present some exact solutions, to obtain which it is convenient to
write
\be
\overline{V_t}(\phi) = - e^{a\sqrt{6\kappa}\phi}\frac{[F(\phi)-V_A]^2}{4F(\phi)}\ .
\ee
Eq.~(\ref{D0}) fixes $F(\phi)$ to be given implicitly by
\be
\frac{[(a-s)F(\phi)-(a+s)V_A]^{2a}}{[F(\phi)]^{a+s}}=M^{4(a+s)}e^{(1-a^2)\sqrt{6\kappa}\phi}\ ,
\label{Fsol}
\ee
where $s\equiv {\rm sign}(V_A)$ and $M$ is an arbitrary integration constant (with dimensions of mass). This relation can be solved for $F(\phi)$ for some particular values of $a$, or used also to get the asymptotic behaviour of $\overline{V_t}$ at $\phi\to\infty$.
The results are as follows.

{\bf Type $\bma{0}$} ($a<1$ and $V_A$ of either sign): The expansion of $\overline{V_t}$ is
\be
\overline{V_t} = \overline V_{tA} e^{\sqrt{6\kappa}\phi} +\frac{V_A}{2(1-a)} e^{a\sqrt{6\kappa}\phi} + ...
\ee
Here, $\overline V_{tA}$ is the free parameter (which should be negative). 
An explicit example of exact solution of this type can be obtained {\it e.g.} for $a=1/2$, $V_A=s=\pm 1$ and $\kappa=1$, for which
\be
\overline{V_t}=-e^{\sqrt{3/2}\phi}\frac{(1-3sx+x^2)^2}{4x(x-s)^2}\ ,
\ee
with
\be
x\equiv \left[s+2P+2\sqrt{P(P+s)}\right]^{1/3}\ ,\quad
P\equiv M^{12}e^{(3/2)\sqrt{6}\phi}\ .
\ee
The large $\phi$ expansion of this exact solution agrees with the general expectation for this type.

{\bf Type $\bma{-}$} ($V_A<0$ and $1/\sqrt{3}<a<1$): There are three qualitatively different classes of $\overline{V_t}$. One is the exact solution
\be
\overline V_{t,c} = \frac{V_A}{1-a^2}e^{a\sqrt{6\kappa}\phi} \ .
\label{Vtesp}
\ee
The second class is formed  by $\overline{V_t}$'s lying above $\overline V_{t,c}$, which do not extend to $\phi=\infty$ and intersect $V$ at some finite value of $\phi$. The third class is formed by  $\overline{V_t}$'s lying below $\overline V_{t,c}$, which do extend to $\phi=\infty$ and have the expansion 
\be
\overline{V_t} = V_{tA}e^{\sqrt{6\kappa}\phi}+
\frac{V_{A}}{2(1-a)} e^{a\sqrt{6\kappa}\phi}+...
\label{Vtbexp}
\ee
where $V_{tA}<0$ is the free parameter. The asymptotic behaviour of $\overline{V_t}$ in (\ref{Vtbexp}) is too steep for a type $-$ solution and it indeed is exactly the same found above for type 0 solutions.
For an exact example of this class take $a=3/5$, $V_A=-1$ and $\kappa=1$. We get
\be
\overline{V_t}=-e^{\sqrt{3/5}\phi}\frac{(8+z+q)^2}{8(6+z+q)}\ ,
\ee
with
\bea
q &\equiv & \sqrt{12-z^2-16/z}\ ,\quad 
z\equiv \sqrt{5-25P/x+5x/6}\ ,\nonumber\\
P &\equiv & M^{16}e^{(8/5)\sqrt{6}\phi}\ ,\quad 
x\equiv 6\left[-P(1+\sqrt{1+P(5/6)^3)})\right]^{1/3}\ ,
\eea 
and its large $\phi$ expansion gives (for $M\neq 0$)
\be
\overline{V_t} = \overline V_{tA} e^{\sqrt{6}\phi}-\frac{5}{4} e^{(3/5)\sqrt{6}\phi} + ...
\ee
which corresponds to the expansion in (\ref{Vtbexp}) for $a=3/5$. For $M\to 0$, one gets instead
\be
\overline{V_t} = -\frac{25}{16} e^{3\sqrt{6}\phi/5} \ ,
\ee
which is the exact case of (\ref{Vtesp}) for $a=3/5$.

{\bf Type $\bma{+}$} ($a>1$ and $V_A>0$): The expansion of $\overline{V_t}$ is
\be
\overline{V_t} = \frac{V_{A}}{1-a^2} e^{a\sqrt{6\kappa}\phi} + B e^{[(a+1/a)/2]\sqrt{6\kappa}\phi}-\frac{(1-a^2)^2B^2}{4a^2 V_A}e^{\sqrt{6\kappa}\phi/a}+... 
\label{barVtA}
\ee
Here, $B$ is a free parameter [related to $M$ in (\ref{Fsol})].
An explicit example of exact solution of this type can be obtained {\it e.g.} for $a=3$, $V_A=1$ and $\kappa=1$, for which
\be
\overline{V_t}=-e^{3\sqrt{6}\phi}\frac{[12x+6P x^2+P^2]^2}{48x(24x+6P x^2+P^2)}\ ,
\ee
with
\be
P\equiv M^{8/3}e^{-(4/3)\sqrt{6}\phi}\ ,\quad
x\equiv \left[1+\sqrt{1-(P/6)^3}\right]^{1/3}\ .
\ee
It can be checked that the expansion of this exact solution conforms to the general expression in (\ref{barVtA}).

{\bf Special Case} ($V_A>0$ and $a=1$): The expansion of type $+$ above is not valid. However, this case can be solved exactly to get
\be 
\overline{V_t} = -V_A e^{\sqrt{6\kappa}\phi}\frac{[W(e^{2\sqrt{6\kappa}(\phi-\phi_C)})-1]^2}{4W(e^{2\sqrt{6\kappa}(\phi-\phi_C)})}\ ,
\ee 
where $\phi_C$ is arbitrary and $W(z)$ is Lambert's (or product log) function satisfying $We^W=z$. The large $\phi$ expansion of this solution gives (setting $\kappa=1$)
\be
\overline{V_t} = -V_A e^{\sqrt{6}(\phi-\phi_C)}\left\{\sqrt{3/2}(\phi-\phi_C)-\frac14\left[2+\log(2\sqrt{6}(\phi-\phi_C))\right]+...\right\}+...
\ee

{\bf Type $\bma{-^*}$}  ($a>1$ and $V_A<0$): 
For this case there are no $\overline{V_t}$ solutions  extending to $\phi\to\infty$ ({\it i.e.} no cobordism solutions). Indeed, from (\ref{D0}) and the fact that $V,\overline{V_t}<0$ at large $\phi$, one has the inequality $-\overline{V_t}'< \sqrt{6\kappa} |\overline{V_t}|$ and this shows that $\overline{V_t}$ cannot drop faster than $\overline{V_t}\sim -e^{\sqrt{6\kappa}\phi}$. In other words, 
all $\overline{V_t}$'s hit $V$ at some finite field value. In the context of regular CdL bounces, this implies \cite{EGrav} that such potentials are necessarily 
unstable against CdL decay. Nevertheless, the possible existence of BoNs is still open, as is discussed below.

\subsection{\texorpdfstring{$\bma{V_t}$}{Vt} solutions}

We can next derive the large $\phi$ expansion of the tunneling potentials directly from their equation of motion. As this EoM is a second order differential equation, we expect two arbitrary constants in the expansions of $V_t$. In all types except type $-^*$, which is special, the $V_t$ expansions consist of the corresponding expansion for $\overline{V_t}$  discussed above (which already depends on one free parameter) plus additional terms that depend on a second free parameter. This second free parameter gives a positive $D^2$ (and setting it to zero one would recover $\overline{V_t}$, which has $D^2=0$). We obtain the following results:

{\bf Type $\bma{0}$} ($a<1$ and $V_A$ of either sign): The expansion of $V_t$ is
\be
V_t = V_{tA} e^{\sqrt{6\kappa}\phi} +\frac{V_A}{2(1-a)} e^{a\sqrt{6\kappa}\phi} +...+ B e^{\sqrt{6\kappa}\phi/3}+... 
\label{Vt0exp}
\ee
The two free parameters are $V_{tA}$ (which has to be negative) and $B$ (which should be positive). Again we find that $V_t$ is given by $\overline{V_t}$ plus additional terms proportional to the free parameter B, which makes $D(\phi)$ non zero.\footnote{The intermediate ellipsis in (\ref{Vt0exp}), and similar $V_t$ expansions below, stands for terms  more relevant than the $B$ term (the number of such terms depends on $a$). Calculating $D^2$ directly from expression (\ref{D2}) requires to keep all these terms up to the $B$ one as they are crucial for the cancellations that end up giving $D^2\propto B$.} Indeed, one has
\be
D(\phi)= \sqrt{-8V_{tA}B\kappa}\, e^{\sqrt{8\kappa/3}\phi}+...
\ee

{\bf Special Case} ($V_A>0$ and $a=1$): For this case, at the boundary between types $+$ and 0, the expansion of $V_t$ takes the following form 
\bea
V_t &=& -V_A e^{\sqrt{6\kappa}\phi}\left\{\sqrt{\frac{3\kappa}{2}}\phi-\frac14\left[2+\log(2\sqrt{6\kappa}(\phi-\phi_c))\right]+{\cal O}\left(\frac{\log\phi}{\phi}\right)\right\}\nonumber\\
&+&\frac{A}{\phi^{1/3}} e^{\sqrt{6\kappa}\phi/3}\left[1+\frac{3+\log(2\sqrt{6\kappa}(\phi-\phi_c))}{6\sqrt{6\kappa}\phi}+{\cal O}\left(\frac{\log^2\phi}{\phi^2}\right)\right]\ .
\eea
The two free parameters are $\phi_c$ and $A$, with $V_t$ given by $\overline{V_t}$ (first line) plus terms proportional to $A$ that make $D(\phi)$ nonzero. One has
\be
D(\phi)= 2\sqrt{\sqrt{6\kappa}\kappa AV_A}\,\phi^{1/3} e^{\sqrt{8\kappa/3}\phi}+...
\ee

{\bf Type $\bma{-}$} ($V_A<0$ and $1/\sqrt{3}<a<1$): The expansion of $V_t$ is
\be
V_t =\frac{V_{A}}{1-a^2} e^{a\sqrt{6\kappa}\phi}+... +A e^{\sqrt{6\kappa}\phi/(3a)} +...
\ee
The only free parameter is $A$ (which has to be negative). In this case, $V_t$ is given by the special $\overline V_{tc}$ of (\ref{Vtesp}) plus terms proportional to $A$ that make $D(\phi)$ nonzero
\be
D(\phi)= \sqrt{\frac{2(1+3a^2)AV_A\kappa}{a^2-1}}\, e^{(3a+1/a)\sqrt{\kappa/6}\phi}+...
\ee

{\bf Type $\bma{+}$} ($a>1$ and $V_A>0$): The expansion of $V_t$ is
\be
V_t = \frac{V_{A}}{1-a^2} e^{a\sqrt{6\kappa}\phi} +
A e^{[(a+1/a)/2]\sqrt{6\kappa}\phi}+...
+ B e^{\sqrt{6\kappa}\phi/(3a)} +... 
\ee
The two free parameters are $A$ and $B$ (which should be positive). One recognizes the expansion of type $+$ $\overline{V_t}$ solutions plus terms proportional to the parameter $B$. One gets
\be
D(\phi)= \sqrt{\frac{2(1+3a^2)BV_A\kappa}{a^2-1}}\, e^{(3a+1/a)\sqrt{\kappa/6}\phi}+...
\ee
which indeed would give zero if $B=0$ (limit in which $V_t\to\overline{V_t}$).
We see that, for $a\to 1^+$, both $V_t$ and $D$ above  blow up.

{\bf Type $\bma{-^*}$}  ($a>1$ and $V_A<0$): The expansion of $V_t$ is
\be
V_t = \frac{3V_A}{2} e^{a\sqrt{6\kappa}\phi} +A e^{[(1+a)/2]\sqrt{6\kappa}\phi} + B e^{\sqrt{6\kappa}\phi/3}+... 
\ee
The two free parameters are $A$ and $B$. In this case there was no $\overline{V_t}$ extending to infinite field values and, therefore, $V_t$ does not take the standard form found in all the other types. One has
\be
D(\phi)=-V_A \sqrt{3\kappa/2}\, e^{a\sqrt{6\kappa}\phi}+...
\ee
and indeed we see that $D(\phi)$ is nonzero even for zero values of the free parameters $A$ and $B$.

\section{Potentials for \texorpdfstring{$\bma{V_t(\phi)=-e^{a\sqrt{6\kappa}\phi}}$}{Vt=-Exp(a sq(6k)phi)}\label{App:Vtexp}}
Instead of assuming that the potential follows a simple exponential, as done in the previous appendix, it is also instructive to assume that $V_t$ is the simple exponential
\be
V_t(\phi)=-e^{a\sqrt{6}\phi}\ ,
\ee
(we take again $\kappa=1$ and assume $a>0$) and derive the corresponding $V(\phi)$'s that would satisfy the $V_t$ equation of motion. This can be done exactly following the technique explained in \cite{EGrav} and one gets
\be
V(\phi)=e^{\sqrt{6}\phi}\left[-1+\frac{a^2}{1+\frac{3a^2-1}{1+(3a^2-1)C e^{(3a-1/a)\sqrt{2/3}\phi}}}\right]\ ,
\label{VforVtexp}
\ee
where $C$ is an integration constant. (There is only one integration constant as the EoM is a first order differential equation when solved for $V$). We then get
\be 
D(\phi)^2 = \frac{6a^2(3a^2-1)e^{(6a+1/a)\sqrt{2/3}\phi}}{3a^2 e^{\sqrt{2/3}\phi/a}+(3a^2-1)Ce^{\sqrt{6}a\phi}}\ .
\ee 
The asymptotics of $V_t(\phi)$ and $D(\phi)$  at $\phi\to\infty$  depends on the value of $a$ as follows.

{\bf Case $\bma{a<1/\sqrt{3}}$:} In this case we get
\be 
V(\phi)\simeq -\frac23 e^{\sqrt{6}\phi} +...
\ee 
which is a type $-^*$ example but leads to
\be 
D(\phi)^2=-2(1-3a^2)e^{2\sqrt{6}\phi}+...
\ee 
which is negative, so that this case is not acceptable.

{\bf Case $\bma{a>1/\sqrt{3}}$:} In this case we get
\be 
V(\phi)\simeq (a^2-1)e^{\sqrt{6}\phi}-\frac{a^2}{C}e^{[1-a+1/(3a)]\sqrt{6}\phi}...
\ee 
which leads to
\be 
D(\phi)^2= \frac{6a^2}{C}e^{2(3a+1/a)\phi/\sqrt{6}}+ ...
\ee 
So, for $a>1$ this is of type $+$ while for $1/\sqrt{3}<a<1$ it is of type $-$.

{\bf Special Case $\bma{a=1}$:} In this case we get
\be 
V(\phi)\simeq -\frac{1}{C}e^{\sqrt{6}\phi/3}...
\ee 
which leads to
\be 
D(\phi)^2= \frac{6}{C}e^{2\sqrt{8/3}\phi}+ ...
\ee 
This corresponds to type $0$.

{\bf Special Case $\bma{a=1/\sqrt{3}}$:} In this case, the result (\ref{VforVtexp}) does not apply. We get instead
\be 
V(\phi)= (C-\sqrt{2}\phi) e^{-\sqrt{2}\phi}\ .
\ee 
which leads to
\be 
D(\phi)^2= -4e^{2\sqrt{2}\phi}+ 6\sqrt{2}\phi-6C\ .
\ee 
As this is negative at $\phi\to\infty$, this case is not acceptable.

\section{Tunneling Potential for Constant \texorpdfstring{$\bma{V(\phi)}$}{V(phi)}\label{App:Vconstant}} 

In this appendix we calculate the BoN tunneling potential in a region of the potential that is constant, $V(\phi)=V_\infty$, in some region extending to $\phi=\infty$, to be matched eventually to a different potential (e.g. containing a minimum) valid in a different field range. That is, the $V_t$ solution below should be considered just as a part of a complete BoN solution. We consider here the case with $V_\infty>0$, but $V_\infty<0$ can be treated in the same way.

Solving the $V_t$ EoM (\ref{EoM}) for constant $V$ directly is too hard and we follow a different route. An alternative formulation of the EoM for $V_t$ is possible rewriting it as a first order differential equation for $D$ [defined in (\ref{D2})] as in (\ref{EoMVtD}):
\be
\frac{D'}{D}=\frac{3V'-4V_t'}{6(V-V_t)}\ .
\ee
For a constant $V=V_\infty$ this can be integrated to give
\be
D(\phi)= C \sqrt{6\kappa} V_\infty^{1/3} [V_\infty-V_t(\phi)]^{2/3}\ ,
\label{DVc}
\ee
with $C$ a dimensionless integration constant  (normalized with $\sqrt{6\kappa} V_\infty^{1/3}$ for convenience later on).\footnote{We know from the general asymptotic behaviour discussed in subsection~\ref{genasy} that $V_t(\phi\to\infty)\sim -e^{\sqrt{6\kappa}\phi}$. Plugging this in the definition of $D$ shows there is a cancellation of the leading exponential term. Equation (\ref{DVc}) shows that $D(\phi\to\infty)\sim e^{\sqrt{8\kappa/3}\phi}$, as expected for a type 0 BoN.}
Next we use the definition of $D^2$ in (\ref{D2}) to get a differential equation for $V_t'$:
\be
(V_t')^2=6\kappa\left[C^2 V_\infty^{2/3}(V_\infty-V_t)^{1/3}- V_t\right](V_\infty-V_t)\ .
\label{Vtpeq}
\ee
Before showing how to deal with this equation, we can deduce some general properties of the solutions $V_t(\phi)$. First, if $V_t(\phi)$ is a solution of (\ref{Vtpeq}), other related solutions are $V_t(\phi+c)$ (given the shift invariance of the potential) and $V_t(-\phi)$ (given that $V_t'$ appears quadratically).  Also,  the right hand side of (\ref{Vtpeq}), restricted to  positive values, is a monotonic function of $V_t$ for $V_t\leq V_\infty$ that grows when $V_t\to-\infty$. These properties allow us to take $V_t(\phi)$ symmetric around $\phi=0$, with $V_t'(\phi<0)>0$ and $V_t'(\phi>0)<0$, with $\phi=0$ corresponding to a maximum with $V_t'(0)=0$. Figure~\ref{fig:Vinfty} shows how such $V_t(\phi)$ looks. Setting $V_t'=0$ in (\ref{Vtpeq}) we find that $C$ is related to $V_t(0)\equiv V_T$ by
\be
C^2=\frac{V_T/V_\infty}{(1-V_T/V_\infty)^{1/3}} \ .
\ee
This also shows that the maximum of $V_t$ occurs at $V_T\geq 0$.

We can then try to solve (\ref{Vtpeq}) for $\phi>0$ and this can be done analytically for the inverse function $\phi(V_t)$, which is as good as solving for $V_t(\phi)$. Introducing the variable
\be
z\equiv (1-V_t/V_\infty)^{1/3}\ ,
\ee
we recast  (\ref{Vtpeq})  as
\be
\kappa \left(\frac{d\phi(z)}{dz}\right)^2=\frac{3z}{2(z^3+C^2z-1)}\ ,
\label{dphidz}
\ee
which can be integrated in terms of elliptic functions. The polynomial in the denominator has a real root, and two complex ones: 
\be
z_T\equiv (1-V_T/V_\infty)^{1/3}\ ,\quad\quad
z_\pm\equiv \left(-1/2\pm i \sqrt{1/z_T^3-1/4}\right)z_T\ ,
\label{dSroots}
\ee
with $0< z_T\leq 1$. In terms of $z_T$ we have $C^2=1/z_T-z_T^2$.

The solution of (\ref{dphidz}) can be written as
\bea
\phi(V_t)&=&\left.\pm \sqrt{\frac{6}{(2r_++1)\kappa}}
\right\{
F(\alpha(r)|z_k)\, +\, (r_+-1)\,\Pi(r_+;\alpha(r)|z_k)\, +\, iK(1-z_k)\nonumber\\
&&\left.
+i(r_++2)\left[\Pi(1-z_k/r_+|1-z_k)\,-\, K(1-z_k)\right]\frac{}{}\right\}\ ,
\label{phiVt}
\eea
where we have adjusted the constant shift so that $\phi(V_T)=0$ and
$r\equiv z/z_T$, $r_\pm\equiv z_\pm/z_T$, with
\be
 z_k\equiv \frac{r_+(r_++2)}{2r_++1}\ ,\quad
 \alpha(r)\equiv \arcsin\sqrt{\frac{1-r/r_+}{1-r}}\ ,
\ee
and $F(a|m)$ and $\Pi(n;a|m)$ are the incomplete elliptic integrals of the first and third kind, respectively, while $K(a)$ and $\Pi(a|m)$ are the corresponding complete elliptic integrals.

\begin{figure}[t!]
\begin{center}
\includegraphics[width=0.475\textwidth]{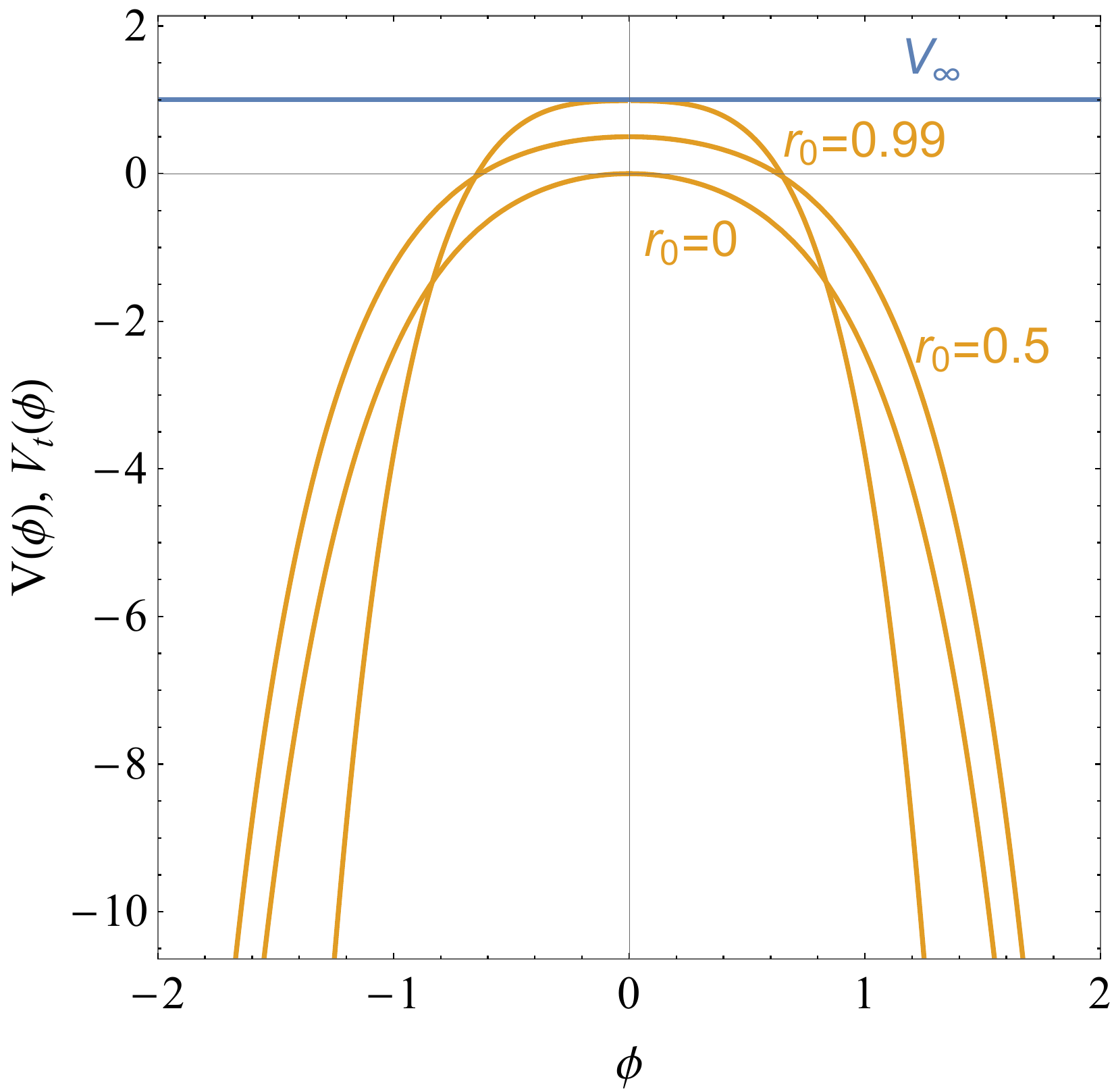}
\end{center}
\caption{Tunneling potentials $V_t(\phi)$ [the inverse of eq.~(\ref{phiVt})], for a constant potential, 
$V(\phi)=V_\infty$, for different values of $r_0\equiv V_t(0)/V_\infty$
as indicated by the labels. 
\label{fig:Vinfty}
}
\end{figure}

Figure \ref{fig:Vinfty} shows $V_t$ for $V_\infty=1$ and several values of  $r_0\equiv V_T/V_\infty$. The limiting case $r_0=0$ admits a simple expression and is given by
\be
V_t(\phi) = -V_\infty \sinh^2\sqrt{\frac{3\kappa \phi^2}{2}}\ .
\ee
It can be checked that this case leads to $D=0$ so that it can at most be part of an end-of-the-world brane rather than a bubble of nothing.

For $r_0=1-\delta^3$, with $0<\delta\ll 1$ (so that $z_T=\delta$), one gets the approximation 
\be
V_t(\phi) \simeq V_\infty\left[1-\delta^3-\frac{1}{\delta^{3/2}}\sinh^{3/2}
\left(\sqrt{\frac{2\kappa \phi^2}{3}}\phi\right)\right]\ .
\ee
This corresponds to the small $R_{KK}$ limit of a Witten-like BoN.

Let us assume then that $V=V_\infty$ for $\phi\in(\phi_m,\infty)$, while $V$ has some non trivial structure (a dS minimum, a barrier, etc.) for $\phi<\phi_m$ [call it $V_L(\phi)$]. Then the complete $V_t$ consist of two parts, one for $\phi<\phi_m$ (call it $V_{tL}$) and the solution found in this appendix for $\phi>\phi_m$ (call it $V_{tR}$), and both should be matched at $\phi_m$ to get a continuous $V_t$ and $V'_t$. If $V_L$ and $V_{tL}$ are known for $\phi<\phi_m$, the solutions for $\phi>\phi_m$ are matched as follows.
First, $V_R(\phi>\phi_m)=V_\infty=V_L(\phi_m)$. Calling $V_{tm}\equiv V_{tL}(\phi_m)$ and $V'_{tm}\equiv V'_{tL}(\phi_m)$, equation (\ref{Vtpeq}) fixes the constant $C^2$
as
\be
C^2=\frac{(V'_{tm})^2+6\kappa(V_\infty-V_{tm})V_{tm}}{6\kappa V_\infty^{2/3}(V_\infty-V_{tm})^{2/3}}\ ,
\ee
to ensure $V'_{tR}(\phi_m)=V'_{tm}$. Once $C^2$ is known, $z_T$ can be determined, fixing the solution $\phi(V_{tR})$ in (\ref{phiVt}) up to an overall shift.
To also match $V_{tR}(\phi_m)=V_{tm}$ we simply solve for $\delta\phi_m$ in 
$\phi_m=\phi(V_{tm})+\delta\phi_m$. This procedure is carried out for the example of subsection~\ref{sect:type1C}.

The tunneling action in the range $(\phi_m,\infty)$, where the solution for $V_t$ found in this appendix is assumed to hold, can be calculated analytically. In order to do this we change the integration variable in (\ref{SVt}) from $\phi$ to $V_t$ (or, equivalently $z$). In this way one gets
\be
\Delta S=
\frac{6\pi^2}{\kappa^2}\int_{V_{tm}}^{-\infty}\frac{(D+V_t')^2}{D V_t^2 V_t'}dV_t=
\frac{-18\pi^2}{\kappa^2 V_\infty}\int_{z_m}^\infty\frac{[D(z)+V_t'(z)]^2z^2}{D(z) (1-z^3)^2V'_t(z)} dz\ ,
\ee
where $z_m\equiv (1-V_{tm}/V_\infty)^{1/3}$, and
\be
D(z)\equiv C \sqrt{6\kappa}V_\infty z^2\ ,\quad
V_t'(z)\equiv -V_\infty z^{3/2}\sqrt{6\kappa[C^2 z-1+z^3]}\ .
\ee
Performing the integral, we get
\bea
\Delta S(z_m)&=&\frac{12\pi^2}{V_\infty \kappa^2\sqrt{z_T^{-3}-1}}
\left\{1+\frac{\sqrt{z_T^{-3}-1}}{1-z_m^3}-\frac{(z_T^{-1}-z_m^2)\sqrt{z_m(z_m^3+C^2 z_m-1)}}{(z_m-z_T)(1-z_m^3)}
\right.\nonumber\\
&+&\left.\frac{(1-r_+)\left[E(\gamma(\infty)|\psi)-E(\gamma(z_m)|\psi)\right]+
F(\gamma(z_m)|\psi))-F(\gamma(\infty)|\psi)}{\sqrt{-1-2z_T^2z_+}}\right\} ,\quad\quad
\eea
where
\be
\gamma(z)\equiv \arcsin\sqrt{\frac{r(1-r_+)}{r_+(1-r)}}\ , \quad
\psi\equiv \frac{1-z_+z_T^2}{1-z_-z_T^2}\ ,
\ee
and $E(a|m)$ is the elliptic integral of the second kind.


\section{More Families of Analytic Potentials\label{App:morex}}
In this appendix we give the general potential solution for
some simple families of BoN tunneling potentials. For each simple $V_t$, following the general technique presented in \cite{EFH} we write
\be
V(\phi) = V_t(\phi) + \frac{[V_t'(\phi)]^2}{6\kappa [1/F(\phi)-V_t(\phi)]}\ ,
\label{VF}
\ee
and solve 
\be
V_t'F'=2\kappa (1-FV_t)\ ,
\label{Feq}
\ee
for $F(\phi)$. We also have
\be 
D(\phi)^2=\frac{V_t'{}^2}{1-V_tF}\ .
\label{DF}
\ee 

\subsection{\texorpdfstring{\bma{V_t(\phi)=V_T - \cosh (a\sqrt{6}\phi)}}{Vt=VT-cosh(a sq(6)phi)} }
Inspired by the constant $V$ solution discussed in the previous appendix, we take the first family of solutions to be generated by
\be
V_t(\phi)=V_T - \cosh (a\sqrt{6}\phi)\ .
\ee
We take $\kappa=1$, $a>0$ and $V_T>1$. Solving (\ref{Feq}) we get
\be
F(\phi)=\frac{1}{V_T+1}\, {}_2F_1(1,p_+-p_-,1+p_+;\cosh^2(a\sqrt{3/2}\phi))+C\frac{\left[\sinh(a\sqrt{3/2}\phi)\right]^{2p_-}}{\left[\cosh(a\sqrt{3/2}\phi)\right]^{2p_+}}
\ ,
\ee
where $C$ is an integration constant and
\be
p_\pm \equiv \frac{V_T\pm 1}{6a^2}\ .
\ee
The asymptotic behaviour of $F(\phi)$ at $\phi\to\infty$  is
\be
F(\phi) \simeq 4^{1/(3a^2)}(C-C_s)
\exp\left(-\frac{\phi}{a}\sqrt{\frac23}\right)+\frac{2}{(3a^2-1)}\exp\left(-a\sqrt{6}\phi\right)+...\ ,
\ee
where 
\be
C_s= (-1)^{1-p_++p_-}\frac{\Gamma(1+p_--p_+)\Gamma(p_+)}{(V_T-1)\Gamma(p_-)}\ ,
\ee
which is real and finite only if $1-p_++p_-=1-1/(3a^2)$ is a positive integer, which cannot be accomplished.


To get a real $F(\phi)$ we therefore need to take $C=C_s$, and then $1/F(\phi)\simeq [(3a^2-1)/2]\exp(a\sqrt{6}\phi)$. Using this in (\ref{VF}), we get  a type $-^*$ case with $V(\phi)\simeq -(1/3)\exp(a\sqrt{6}\phi)$. On the other hand,  (\ref{DF}) gives $D(\phi)^2\simeq [(3a^2-1)/2]\exp(2a\sqrt{6}\phi)$ and we see that  $a>1/\sqrt{3}$ is needed to have $D(\phi)^2>0$.

The previous solution can be used in some field interval, as done in the text with other analytic solutions. If that interval includes $\phi=\phi_T=0$, point at which $V_t$ reaches its maximum, then to satisfy $V(0)>V_t(0)$ we need to set $C$ to a particular value, $C_T$, which can be found by inspecting the expansion of $F(\phi)$ around $\phi_T$ (see \cite{EFH} for details). We get
\be
C_T=(-1)^{p_-}\frac{\Gamma(p_+)\Gamma(1-p_-)}{(V_T-1)\Gamma(p_+-p_-)}\ .
\ee
In general $C_T\neq C_s$ but for the particular choice $V_T= 6a^2 n -1$ (so that $p_+=n$), with $n$ an integer, one has $C_T=C_s$ (and therefore one gets a working type $-^*$ example).

\subsection{\texorpdfstring{\bma{V_t(\phi)=-A e^{\sqrt{6}\phi}+B e^{\sqrt{2/3}\phi}}}{Vt=-A Exp(sq(6)phi)+B Exp(sq(2/3)phi))}\label{sect:morex}}

For this family of examples we take $\kappa=1$ and
\be
V_t(\phi)=-A e^{\sqrt{6}\phi}+B e^{\sqrt{2/3}\phi}\ ,
\label{eq:VtABfamily}
\ee
with $A>0, B>0$ so that $V_t$ has a maximum at $\phi_T=\sqrt{3/2}\log\sqrt{B/(3A)}$. As in previous subsections, we consider these examples as valid in some range $(\phi_m,\infty)$, where $V_t(\phi_m) = V(\phi_m)$. In order to construct  the complete $V$ and $V_t$, for some of the following examples, it might be necessary to match these functions to other solutions  valid for $\phi<\phi_m$.

From the tunneling potential $V_t$ in \eqref{eq:VtABfamily}, following the procedure mentioned at the beginning of the appendix, one gets
\be
V(\phi)=\frac{6A(4BC-1)e^{\sqrt{6}\phi}-8B(2BC-1) e^{\sqrt{2/3}\phi}}{9\left(2ACe^{\sqrt{8/3}\phi}+1-2BC\right)}\ ,
\label{eq:VABfamily}
\ee
where $C$ is an integration constant. The large field expansion gives
\be
V(\phi\to\infty)=\frac{4BC-1}{3C}e^{\sqrt{2/3}\phi}+{\cal O}(e^{-\sqrt{2/3}\phi})
\ee
(which is valid for $C\neq 0$)
so that $V$ is indeed subleading in the limit $\phi\to \infty$, and of type $0$. As we show below, the case $C=0$ is special and gives a type $-^*$ case.

Depending on the value of $C$ relative to $C_*\equiv 1/(4B)$ we can get
\be
V(\phi\to\infty)\to\left\{\begin{array}{cl}
+\infty & \mathrm{for}\quad C>C_*\quad \mathrm{or}\quad C<0\ ,\\
0 & \mathrm{for}\quad C=C_*\ ,\\
-\infty & \mathrm{for}\quad 0\leq C<C_*\ .
\end{array}
\right.
\ee

\begin{figure}[t!]
\begin{center}
\includegraphics[width=0.6\textwidth]{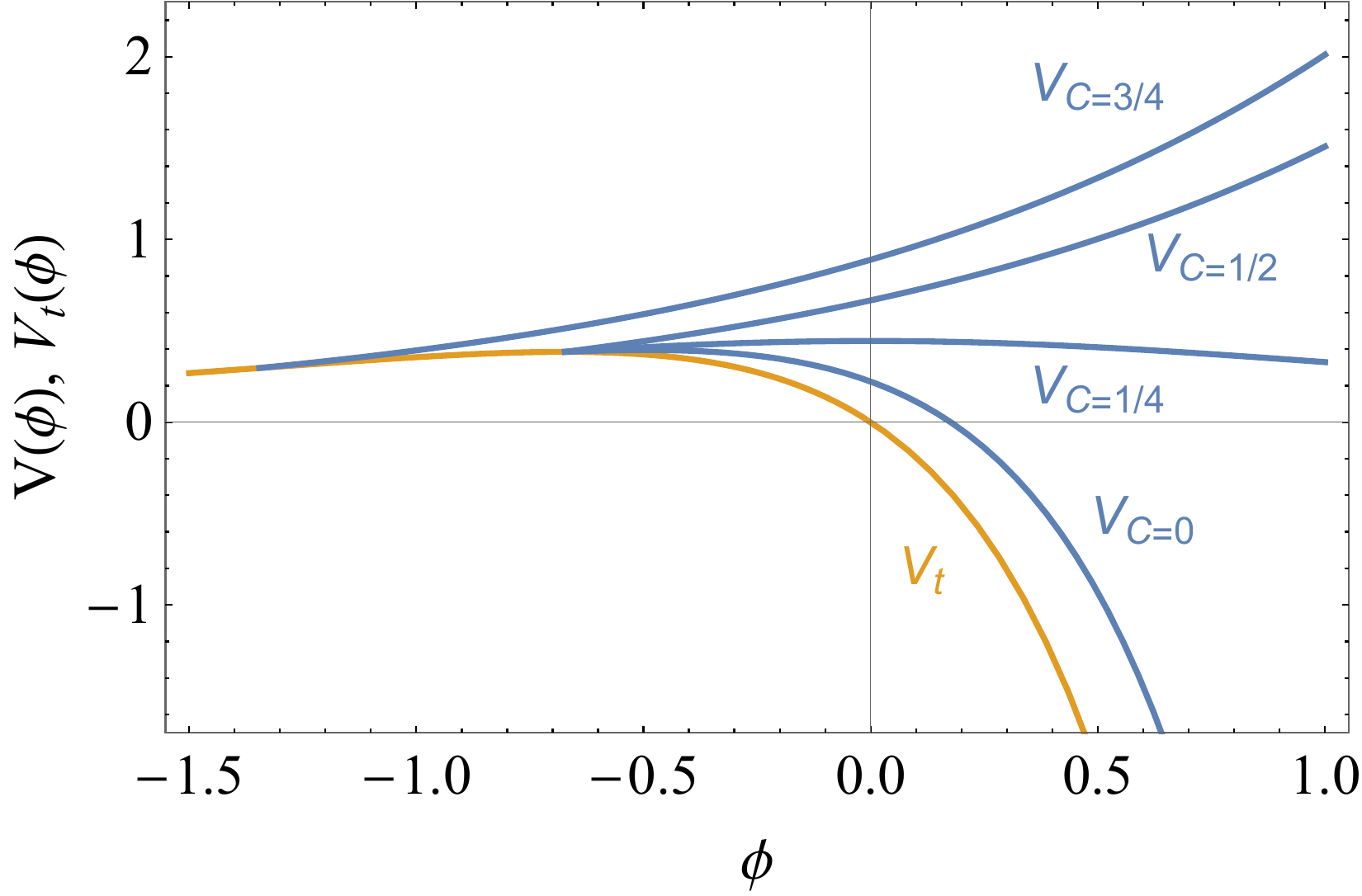}
\end{center}
\caption{Potentials and tunneling potential, $V$ and $V_t$, for the analytic examples of subsection~\ref{sect:morex}, for the indicated values of the integration constant $C$.  
\label{fig:VCs}
}
\end{figure}

We can illustrate the above cases with some numerical examples, taking $A=B=1$, and the indicated values of $C$ to get the following simple potentials
\bea
V(\phi)&=&\frac{8}{9}e^{\sqrt{2/3}\phi}-\frac23 e^{\sqrt{6}\phi}\ , \quad (C=0)\\
V(\phi)&=&\frac{4}{9}\mathrm{sech}(\sqrt{2/3}\phi)\ , \quad (C=C_*=1/4)\\
V(\phi)&=&\frac{2}{3}e^{\sqrt{2/3}\phi}\ , \quad (C=1/2)\\
V(\phi)&=&\frac{8}{9}e^{\sqrt{2/3}\phi}\ , \quad (C=3/4)\ .
\eea
These solutions are plotted, together with $V_t$, in figure~\ref{fig:VCs}.
We see that the $C=0$ case corresponds to the asymptotic behaviour of a type $-^*$ case, while the rest of cases are of type $0$. The last one has been already presented in subsection \ref{sect:simple0}.

For generic choices of $C$ outside the interval $(0,2C_*)$ the solutions above develop singularities at some $\phi_s$ and one should choose $\phi_m>\phi_s$.
The BoN action for the range $(\phi_m,\infty)$ can be obtained analytically but we do not give it here as it is not particularly illuminating.

To end this appendix, we use input from the higher dimensional theory to gain information on the range of parameters of the previous solutions which is of physical interest. From the discussion in section \ref{sec:topdown}, it follows that the $C\neq0$  instanton describes a BoN mediating the decay of a KK circle compactification. In this simple model, the parameters $A$ and $C$ can be related directly to the relevant scales of the BoN solution. First, the regularity conditions on the BoN surface ($\phi \to \infty$),  link $A$ to the KK radius, $R_{KK}$, as \eqref{UVresults} implies that $V_t(\phi \to \infty) \simeq - 3/(4R_{KK}^2) e^{\sqrt{6} \phi}$, and therefore  we must set
\be
A = \frac{3}{4 R_{KK}^2}\ .
\ee
Moreover, the constant $C$ can be related to the BoN nucleation radius, $\mathcal{R}$, as \eqref{phirhotop} implies that near the BoN core the metric function $\rho$ behaves as
\be
\rho(\phi \to \infty) \simeq \mathcal{R} e^{-\sqrt{\kappa/6} \phi}\ .
\label{eq:RhoInf}
\ee
On the other hand, the asymptotic behaviour of the function $\rho$ can be obtained from $V(\phi)$ and $V_t(\phi)$ taking the limit $\phi \to \infty$ in \eqref{rho}. Comparing the resulting expression with  \eqref{eq:RhoInf} we find
\be
C  = \frac{\mathcal{R}^2}{9}>0 \ .
\ee

\subsection{
\texorpdfstring{\bma{V_t=A e^\phi-e^{b\phi}}}{Vt=A Exp(phi)-Exp(b phi)}
\label{App:Family2}}

Consider now a more general family of examples with  
\be
V_t = A e^\phi-e^{b\phi}\ ,
\ee 
taking $A,b>0$ and $\kappa$ free.\footnote{$\kappa$ can eventually be set to 1 by rescaling the field. The $a$ parameter used in the general classification of section \ref{genasy}, with $\kappa$ rescaled to 1, is then given by $a=b/\sqrt{6\kappa}$}. This family of tunneling potentials was already discussed in \cite{EFH} to find CdL solutions, but we use it now to find BoN solutions instead. 
For this $V_t$, (\ref{Feq}) is solved by
\be 
F(\phi)=\frac{r}{q(b-1)}\left[\left(\frac{b}{A}\right)^q\left(e^{-\phi}-
\frac{A}{b}e^{-b\phi}\right)^rC+ e^{-b\phi}{}_2F_1\left(1,p;1+q;(A/b)e^{(1-b)\phi}\right)
\right]\ ,
\ee 
where $C$ is an integration constant and
\be
r\equiv 2\kappa/b\ ,\quad q\equiv 1+p/b\ , \quad p\equiv (b-2\kappa)/(b-1)\ .
\ee
The tunneling potential $V_t$ has a maximum at $\phi_{tT}=\log(A/b)/(b-1)$
and, imposing $V>V_t$ at $\phi_{tT}$ fixes the integration constant to be \cite{EFH}
\be
C=\frac{\pi\csc(\pi r)}{B(p,1+r)}\ ,
\ee
where $B(x,y)$ is the Euler beta function. 

The different possible asymptotic behaviours of $F$ and thus of $V$ when $\phi\to\infty$ can be studied generically depending on the parameter ranges. Instead of exploring that, we show how a judicious choice of the constants appearing in this general solution leads to a more or less simple expression for $V$, illustrating all the different types of solutions. In all the cases, we choose the integration constant that appears in 
$V$ so that $V_t(\phi_{tT})<V(\phi_{tT})$. This allows to give $V$ in 
a complete interval from its intersection with $V_t$ all the way to $\phi=\infty$ so that we can calculate the CdL part of the BoN tunneling action.

Many of the particular examples given in Subsection~\ref{sec:anex} belong to this family.
For instance, the example in subsection \ref{sect:CdLPlateau} is a case with $b=2$ and $\kappa=3/2$. The expressions given in that subsection  correspond to $\kappa$ being rescaled to 1, with $a=2/3$. Different choices and field rescalings generate the examples of Subsections \ref{sect:simple0} and \ref{sect:Counter}. Other examples are the following.

A type $0$ example can be obtained for $b=3/2$ and $\kappa=3/8$ (or $a=1$, $\kappa=1$) for which we get
\be
V(\phi)=A e^\phi-e^{3\phi/2}+\frac{(2A/3-e^{\phi/2})^2e^\phi}{e^{\phi/2}-A+A^2/(9e^{\phi/2}-3A)}\ .
\ee
It can be checked that $V(\phi\to\infty)\sim (2A/3)e^\phi$, which corresponds to a type $0$ example.

A type $-$ example is obtained for $b=2$ and $\kappa=3/4$ (or $a=2\sqrt{2}/3$ and $\kappa=1$) for which
\be
V=Ae^\phi-e^{2\phi}+\frac{2(A-2e^\phi)^2e^\phi/3}{3e^\phi-3A+\frac{5A^{5/4}e^\phi}{2(2e^\phi-A)^{3/4}e^{\phi/2}\sqrt{\pi}\Gamma(9/4)/\Gamma(7/4)+A^{5/4}{}_2F_1(1/2,1;9/4;Ae^{-\phi}/2)}}\ .
\ee
It can be checked that $V(\phi\to\infty)\sim -e^{2\phi}/9$, which indeed corresponds to a type $-$ example. In this case there is a Hawking-Moss instanton that can mediate vacuum decay. A numerical comparison shows that the BoN action is lower than the HM one, so that vacuum decay proceeds preferentially via a BoN.

We end up giving one further example of type $-^*$. We take $b=3/2$ and $\kappa=1$ (so that $a=\sqrt{3/8}<1$). The potential is quite simple:
\be
V=\frac{5A}{6}e^\phi-\frac23 e^{3\phi/2}\ .
\ee
The HM and BoN actions\footnote{As discussed before, this part of the action corresponds to the field interval from the contact point $\phi_{0\pp}$ of $V$ and $V_t$ to $\phi=\infty$. There is an additional contribution from the false vacuum $\phi_\pp$ to $\phi_{0\pp}$ that is common to both decays.} can be computed analytically as
\be
S_{\rm BoN} =\frac{64\pi^2}{A^3}\ ,\quad S_{\rm HM}= \frac{132}{125}S_{\rm BoN} \ ,
\ee
so that $ S_{\rm HM}>S_{\rm BoN} $ and BoN again dominates decay over HM.

\section{Parametric Dependences of the Tunneling Action\label{app:dSdA}}
In the text we have often found families of tunneling potentials, parametrized by some free parameter, $p$, related to the behaviour of $V_t(\phi)$ near the false vacuum. This parameter was $p=A$ for AdS/Minkowski vacua, with $A$ appearing in the low-field expansion of $V_t$ as in (\ref{VtexpMink}) and (\ref{VtexpAdS}),  and $p=\phi_i$ for dS vacua, with $\phi_i$ being the field value at which $V_t$ starts to depart from $V$. In this appendix we compute how the tunneling action depends on such parameters, that is, we calculate $dS/dA$ or $dS/d\phi_i$, although the formulas (\ref{dSdA}) and (\ref{dSdphii}) are of more general validity and apply for any parameter appearing in $V_t$ (and not in $V$).

Let us write the tunneling action in general as
\be
S[V_t]=\frac{6\pi^2}{\kappa^2}\int_{\phi_\pp}^{\phi_e}\frac{(D+V_t')^2}{D V_t^2}d\phi=\int_{\phi_\pp}^{\phi_e}s(V,V_t,V_t')\ d\phi .
\label{sVt}
\ee
As we have seen, for dS decay $\phi_i\neq \phi_\pp$ and the action integral has a HM-like contribution from the interval $(\phi_\pp,\phi_i)$ (in which $V_t=V$) and a CdL-like contribution from the interval $(\phi_i,\phi_e)$ (in which $V_t\neq V$). Concerning $\phi_e$, for pseudo-bounces or CdL bounces, $\phi_e$ takes some finite value while
for BoNs we have $\phi_e=\infty$. 

Consider  a family of tunneling potential solutions, $V_t(p;\phi)$, that depend on some parameter $p$  which does not appear in $V(\phi)$. Let us discuss first AdS/Minkowski vacua. We find
\be
\frac{dS}{dp}=s\left.\frac{d\phi}{dp}\right|_{\phi_\pp}^{\phi_e}+\int_{\phi_\pp}^{\phi_e}\left(\frac{\partial s}{\partial V_t}\frac{dV_t}{dp}+\frac{\partial s}{\partial V'_t}\frac{dV'_t}{dp}\right)d\phi\ .
\ee
Using $dV'_t/dp=d(dV_t/dp)/d\phi$ and integrating by parts, we get
\be
\frac{dS}{dp}=\left(s\left.\frac{d\phi}{dp}+\frac{\partial s}{\partial V'_t}\frac{dV_t}{dp}\right)\right|_{\phi_\pp}^{\phi_e}+\int_{\phi_\pp}^{\phi_e}\left[\frac{\partial s}{\partial V_t}-\frac{d}{d\phi}\left(\frac{\partial s}{\partial V'_t}\right)\right]\frac{dV_t}{dp}d\phi\ .
\ee
The integral above vanishes due to the EoM for $V_t$ and, using the explicit form of the action density $s$ from \eqref{sVt}, we get
\be
\frac{dS}{dp}=B(p;\phi_e)-B(p;\phi_\pp)
\label{dSdA}
\ee
with
\be
B(p;\phi)\equiv \frac{6\pi^2}{\kappa^2}\frac{(D+V_t')^2}{D^3 V_t^2}\left[D^2\frac{d\phi}{dp}+(2D-V_t')\frac{dV_t}{dp}\right]\ .
\ee
Although this formula is more general, in the text we apply it to the parameter $p=A$ describing families of $V_t$ solutions.

Let us discuss next dS vacua. We start from
\be
S[V_t]=\frac{6\pi^2}{\kappa^2}\int_{\phi_\pp}^{\phi_e}\frac{(D+V_t')^2}{D V_t^2}d\phi=-\left.\frac{24\pi^2}{\kappa^2 V}\right|_{\phi_\pp}^{\phi_i}+\int_{\phi_i}^{\phi_e}s(V,V_t,V_t')\ d\phi ,
\ee
where the first term comes from the HM-like part of the dS tunneling action. Following the same procedure as before we get
\be
\frac{dS}{dp}=B(p;\phi_e)-B(p;\phi_i)+\frac{24\pi^2}{\kappa^2}\frac{V'}{V^2}\left.\frac{d\phi}{dp}\right|_{\phi_\pp}^{\phi_i}\ .
\label{dSdphii}
\ee
As before, this formula is general but in the text we apply it to the parameter $p=\phi_i$ describing families of $V_t$ solutions.
In the expressions above,  $d\phi_\pp/dp=0$ and, for $\phi_e=\infty$, $d\phi_e/dp=0$.

The terms $B(A,\phi_\pp)$ and $B(\phi_i,\phi_i)$ only depend on the expansion of $V_t$ near $\phi_\pp$ and $\phi_i$ respectively. Therefore, they only depend on the type of false vacua considered and not on the type of instanton  driving the decay (pseudo-bounce, CdL or BoN). Before discussing these different types of decay let us then examine first these ``false vacuum'' terms, using the low field expansions derived in section~\ref{sec:BCs} to get the following results. 

For Minkowski false vacua, as $\phi\to 0$,
$V\sim \phi^2$, $V_t\sim \phi^2/\log\phi$, $V_t'\sim \phi/\log\phi$, and $dV_t/dA\sim \phi^2/\log^2\phi$ while
$D\sim \phi/\log\phi$ and $D+V_t'\sim \phi^3$. Using these asymptotics we find $B(A,\phi_\pp)=0$ in this case.

For AdS false vacua, we have, for $\phi\to 0$,
$V,V_t\sim V_\pp$, $V_t'\sim \phi$, and $dV_t/dA\sim \phi^{2+\alpha}$, with $\alpha=2\kappa V_\pp/m_t^2$, with $m_t^2=(3\kappa V_\pp/2)\sqrt{1-4m^2/(3\kappa V_\pp)}<0$, so that $0<\alpha<4/3$.  We also have $D\sim \phi^{1+\alpha/2}$. Using these asymptotics we find that $B(A,\phi)\sim\phi^{2-\alpha/2}$ and therefore $B(A,\phi_\pp)=0$  also in this case.

Finally, for dS decays, $p=\phi_i\neq \phi_\pp$, so that $d\phi_i/dp=1$. We also have, for $\phi\to \phi_i$, $V=V_t= V_i\equiv V(\phi_i)$, $V_t'=D= 3V_i'/4\equiv 3V'(\phi_i)/4$, and, taking a derivative of the boundary condition $V_t(p=\phi_i;\phi_i)=V(\phi_i)$ with respect to $\phi_i$ we find $ dV_t/dp|_i=V_i'/4$. Using these results we find 
\be
B(\phi_i;\phi_i)= \frac{24\pi^2 V_i'}{\kappa^2 V_i^2}\ ,
\ee
which cancels with the last term in (\ref{dSdphii}).

We conclude that 
\be
\frac{dS}{dp}=B(p;\phi_e)\ ,
\label{dSdp}
\ee
for any type of vacua. As $B(p;\phi_e)$ depends on the type of instanton considered, we discuss separately the different cases below.

\subsection{Pseudo-Bounces}
For pseudo-bounces, the boundary conditions at $\phi_e$ are  $V_t(\phi_e)=V(\phi_e)\equiv V_e$, $V_t'(\phi_e)=0$, which imply $D(\phi_e)=0$. In particular, pseudo-bounces are characterized \cite{PS} by $\rho(\phi\to\phi_e)\equiv \rho_e\neq 0$  and, using $\rho=3\sqrt{2(V-V_t)}/D$, we
deduce that $D$ and $V_t'$ go to zero as $\sqrt{V-V_t}$ when $\phi\to\phi_e$. More precisely,
\be
D\sim \frac{3\sqrt{2(V-V_t)}}{\rho_e}\ ,\quad
V_t'\sim -\frac{3}{\rho_e}\sqrt{2(V-V_t)}\sqrt{1-\frac{\kappa V_e\rho_e^2}{3}}\ .
\ee
Using the results above in (\ref{dSdp}) we get
\be
\frac{dS_{PS}}{dp}=\frac{6\pi^2}{\kappa^2V_e^2}\left(1-\sqrt{1-\frac{\kappa V_e\rho_e^2}{3}}\right)^2\left(2+\sqrt{1-\frac{\kappa V_e\rho_e^2}{3}}\right)\left.\frac{dV_t}{dp}\right|_{\phi_e}\ .
\label{dSPSdp}
\ee
Consider first the Minkowski/AdS case, with $p=A$.
As we have seen in the text, as $A$ increases, $\phi_e$ gets larger and $V_t(\phi_e)$ more negative. We thus expect $dV_t/dA|_{\phi_e}<0$ and therefore $dS/dA<0$. This corresponds
to the decreases of the pseudo-bounce action towards the CdL minimum as $A$ increases toward $A_{CdL}$.

For dS vacua ($p=\phi_i$), as $\phi_i$ increases towards $\phi_B$ (the field value of the top of the potential barrier), $\phi_e$ decreases (also toward $\phi_B$), with $V_t(\phi_e)$ increasing. We thus expect $dV_t/d\phi_i|_{\phi_e}>0$ and thus $dS/d\phi_i>0$. This corresponds to the increase of the pseudo-bounce action towards the HM maximum as $\phi_i\to\phi_B$.

Finally, we can also calculate the $\kappa\to 0$ limit of (\ref{dSPSdp}).
As in the $\kappa=0$ limit the value of $V_\pp$ is irrelevant, instead of $A$ or $\phi_i$ we can use $\phi_e$ as parameter (as was done in \cite{PS}).
With this trivial modification, the $\kappa\to 0$ expansion of (\ref{dSPSdp})
gives
\be
\frac{dS}{d\phi_e}=\frac{\pi^2}{2}\rho_e^4\left.\frac{dV_t}{d\phi_e}\right|_{\phi_e}\ ,
\ee
which reproduces the result in \cite{PS}.

\subsection{Coleman-De Luccia and Hawking-Moss Instantons}

We can also use (\ref{dSPSdp}) to show that both the CdL and the HM limits (of pseudo-bounce solutions) are stationary. The CdL limit corresponds to $\rho_e\to 0$. Expanding (\ref{dSPSdp}) at small $\rho_e$ does give $dS/dp=0$. The HM limit instead corresponds to $\phi_i\to\phi_B$ and $\phi_e\to\phi_B$. At the field value $\phi_T$ where $V_t$ has its maximum (not to be confused with $\phi_B$), we have $\rho^2=3/(\kappa V_t)$ and therefore, in the HM limit $\kappa V_e\rho_e^2/3\to 1$ and again we get $dS/dp=0$ from (\ref{dSPSdp}). 

\subsection{Bubbles of Nothing}

Consider first type-0 BoNs, for which $V_t\sim V_{tA}e^{\sqrt{6\kappa}\phi}$
and $D\sim D_\infty e^{\sqrt{8\kappa/3}\phi}$ for $\phi\to\infty$. Plugging these asymptotics in (\ref{dSdp}), with $d\phi_e/dp=0$, we get
\be
\frac{dS_{BoN,0}}{dp}
=-36\pi^2\sqrt{\frac{6}{\kappa}}\frac{V_{tA}}{D_\infty^3}\frac{dV_{tA}}{dp}\ .
\ee
As $V_{tA}<0$, we find that the sign of $dS_{BoN,0}/dp$ is the same as the sign of $dV_{tA}/dp$. We show this relation at work in subsection~\ref{sec:BoNvsOther}.
As a trivial but nice example, we can check that this formula reproduces the correct result for Witten's action, see section~\ref{sec:WBoN}, as follows. Using $V_{tA}=-3/(4\kappa R_{KK}^2)$ and $D_\infty^2=27/(2\kappa R_{KK}^4)$, and taking $p=R_{KK}$ we get $dS/dR_{KK}=2\pi^2 R_{KK}/\kappa$, integrating which one reproduces Witten's BoN action, (\ref{SWBoN}).

For type $+$ BoNs, with $V\sim V_A e^{a\sqrt{6\kappa}\phi}$ we have
\be
V_t= \frac{V_A}{1-a^2}e^{a\sqrt{6\kappa}\phi}+...+X e^{(a+1/a)\sqrt{6\kappa}\phi/2}+...,
\quad
D=D_\infty e^{(3a+1/a)\sqrt{\kappa/6}\phi}+...,
\ee
where $X$ and $D_\infty$ are free parameters, and $D$ is subleading with respect to $V_t'$. Using these expressions in the general formula (\ref{dSdp})
we get the action slope
\be
\frac{dS_{BoN,\pp}}{dX}=\sqrt{\frac{6}{\kappa}}\frac{36\pi^2 a^3 V_A}{(a^2-1)D_\infty^3}\ .
\ee

As shown in the text, BoNs of types $-$ and $-^*$ appear as limiting cases of families of other BoNs (of types $0$ or $+$). Therefore, it does not make sense to calculate the slope of the action of such BoNs except as the limit of actions of type $0$ or $+$, for which one can use the formulas given above.




\begin{thebibliography}{99}

\bibitem{Coleman:1977py}
S.~R.~Coleman,
``The Fate of the False Vacuum. 1. Semiclassical Theory,''
Phys. Rev. D \textbf{15} (1977) 2929
[erratum: Phys. Rev. D \textbf{16} (1977), 1248]

\bibitem{Callan:1977pt}
C.~G.~Callan, Jr. and S.~R.~Coleman,
``The Fate of the False Vacuum. 2. First Quantum Corrections,''
Phys. Rev. D \textbf{16} (1977) 1762.

\bibitem{CdL}
S.~R.~Coleman and F.~De Luccia,
``Gravitational Effects on and of Vacuum Decay,''
Phys. Rev. D \textbf{21} (1980) 3305.

\bibitem{SMdecay}
G.~Degrassi, S.~Di Vita, J.~Elias-Miro, J.~R.~Espinosa, G.~F.~Giudice, G.~Isidori and A.~Strumia,
``Higgs mass and vacuum stability in the Standard Model at NNLO,''
JHEP \textbf{08} (2012) 098,
\arXiv{1205.6497}{ph};
D.~Buttazzo, G.~Degrassi, P.~P.~Giardino, G.~F.~Giudice, F.~Sala, A.~Salvio and A.~Strumia,
``Investigating the near-criticality of the Higgs boson,''
JHEP \textbf{12} (2013) 089,
\arXiv{1307.3536}{ph}.

\bibitem{Denef:2007pq}
F.~Denef, M.~R.~Douglas and S.~Kachru,
``Physics of String Flux Compactifications,''
Ann. Rev. Nucl. Part. Sci. \textbf{57} (2007) 119.
\arXivold{th/0701050}.

\bibitem{Brown:1987dd}
J.~D.~Brown and C.~Teitelboim,
``Dynamical Neutralization of the Cosmological Constant,''
Phys. Lett. B \textbf{195} (1987) 177;
``Neutralization of the Cosmological Constant by Membrane Creation,''
Nucl. Phys. B \textbf{297} (1988) 787.

\bibitem{Bousso:2000xa}
R.~Bousso and J.~Polchinski,
``Quantization of four form fluxes and dynamical neutralization of the cosmological constant,''
JHEP \textbf{06} (2000) 006
\arXivold{th/0004134}.

\bibitem{Blanco-Pillado:2009lan}
J.~J.~Blanco-Pillado, D.~Schwartz-Perlov and A.~Vilenkin,
``Quantum Tunneling in Flux Compactifications,''
JCAP \textbf{12} (2009) 006
\arXiv{0904.3106}{th}.

\bibitem{BoN}
E.~Witten,
``Instability of the Kaluza-Klein Vacuum,''
Nucl. Phys. B \textbf{195} (1982) 481.

\bibitem{ColemanNegative}
S.~R.~Coleman,
``Quantum Tunneling and Negative Eigenvalues,''
Nucl. Phys. B \textbf{298} (1988) 178.

\bibitem{Young:1984jv}
R.~E.~Young,
``Some stability questions for higher dimensional theories,''
Phys. Lett. B \textbf{142} (1984) 149.

\bibitem{BS}
J.~J.~Blanco-Pillado and B.~Shlaer,
``Bubbles of Nothing in Flux Compactifications,''
Phys. Rev. D \textbf{82} (2010) 086015
\arXiv{1002.4408}{th}.

\bibitem{YBD}
I.~S.~Yang,
``Stretched extra dimensions and bubbles of nothing in a toy model landscape,''
Phys. Rev. D \textbf{81} (2010) 125020
\arXiv{0910.1397}{th}.

\bibitem{Brown:2011gt}
A.~R.~Brown and A.~Dahlen,
``On 'nothing' as an infinitely negatively curved spacetime,''
Phys. Rev. D \textbf{85} (2012) 104026
\arXiv{1111.0301}{th}.

\bibitem{Blanco-Pillado:2010vdp}
J.~J.~Blanco-Pillado, H.~S.~Ramadhan and B.~Shlaer,
``Decay of flux vacua to nothing,''
JCAP \textbf{10} (2010) 029
\arXiv{1009.0753}{th}.

\bibitem{Blanco-Pillado:2016xvf}
J.~J.~Blanco-Pillado, B.~Shlaer, K.~Sousa and J.~Urrestilla,
``Bubbles of Nothing and Supersymmetric Compactifications,''
JCAP \textbf{10} (2016) 002
\arXiv{1606.03095}{th}.

\bibitem{Ooguri:2017njy}
H.~Ooguri and L.~Spodyneiko,
``New Kaluza-Klein instantons and the decay of AdS vacua,''
Phys. Rev. D \textbf{96} (2017) 026016
\arXiv{1703.03105}{th}.

\bibitem{Dibitetto:2020csn}
G.~Dibitetto, N.~Petri and M.~Schillo,
``Nothing really matters,''
JHEP \textbf{08} (2020) 040
\arXiv{2002.01764}{th}.

\bibitem{Swamp}
C.~Vafa,
``The String landscape and the swampland,''
\arXivold{th/0509212}.

\bibitem{SwmpRvw}
M.~van Beest, J.~Calder\'on-Infante, D.~Mirfendereski and I.~Valenzuela,
``Lectures on the Swampland Program in String Compactifications,''
\arXiv{2102.01111}{th}.

\bibitem{CobConj}
J.~McNamara and C.~Vafa,
``Cobordism Classes and the Swampland,''
\arXiv{1909.10355}{th}.  



\bibitem{Garriga:1998ri}
J.~Garriga,
``Smooth 'creation' of an open universe in five-dimensions,''
\arXivold{th/9804106}.

\bibitem{BRS}
J.~J.~Blanco-Pillado, H.~S.~Ramadhan and B.~Shlaer,
``Bubbles from Nothing,''
JCAP \textbf{01} (2012) 045
\arXiv{1104.5229}{gr-qc}.

\bibitem{Gibbons:1994vm}
G.~W.~Gibbons, G.~T.~Horowitz and P.~K.~Townsend,
``Higher dimensional resolution of dilatonic black hole singularities,''
Class. Quant. Grav. \textbf{12} (1995) 297
\arXivold{th/9410073}.

\bibitem{DFG}
M.~Dine, P.~J.~Fox and E.~Gorbatov,
``Catastrophic decays of compactified space-times,''
JHEP \textbf{09} (2004) 037
\arXivold{th/0405190}.

\bibitem{DGL}
P.~Draper, I.~G.~Garcia and B.~Lillard,
``Bubble of nothing decays of unstable theories,''
Phys. Rev. D \textbf{104} (2021) 12
\arXiv{2105.08068}{th};
``De Sitter decays to infinity,''
JHEP \textbf{12} (2021) 154
\arXiv{2105.10507}{th}.

\bibitem{E}
  J.R.~Espinosa,
 ``A Fresh Look at the Calculation of Tunneling Actions,''
 JCAP\,{\bf 07} (2018) 36, \arXiv{1805.03680}{th}.
  
\bibitem{Eg}
  J.R.~Espinosa,
  ``Fresh look at the calculation of tunneling actions including gravitational effects,''
  Phys.\ Rev.\ D {\bf 100} (2019)  104007
  \arXiv{1808.00420}{th}.  

\bibitem{EFH}
J.R.~Espinosa, J.F.~Fortin and J.~Huertas,
``Exactly solvable vacuum decays with gravity,''
Phys. Rev. D \textbf{104} (2021)  065007
\arXiv{2106.15505}{th}.   

\bibitem{EK}
J.~R.~Espinosa and T.~Konstandin,
``A Fresh Look at the Calculation of Tunneling Actions in Multi-Field Potentials,''
JCAP \textbf{01} (2019) 051
\arXiv{1811.09185}{th}.

\bibitem{ESM}
J.~R.~Espinosa,
``Vacuum Decay in the Standard Model: Analytical Results with Running and Gravity,''
JCAP \textbf{06} (2020) 052
\arXiv{2003.06219}{ph}.

\bibitem{EGrav}
J.~R.~Espinosa,
``The Stabilizing Effect of Gravity Made Simple,''
JCAP \textbf{07} (2020) 061
\arXiv{2005.09548}{th}.
 
\bibitem{HM}
S.~W.~Hawking and I.~G.~Moss,
``Supercooled Phase Transitions in the Very Early Universe,''
Phys. Lett. B \textbf{110} (1982) 35.

\bibitem{PS}
J.~R.~Espinosa,
``Tunneling without Bounce,''
Phys. Rev. D \textbf{100} (2019) 105002
\arXiv{1908.01730}{th};
J.~R.~Espinosa and J.~Huertas,
``Pseudo-bounces vs. new instantons,''
JCAP \textbf{12} (2021) 029
\arXiv{2106.04541}{th}.

\bibitem{short}
J.J. Blanco-Pillado, J.R. Espinosa, J. Huertas and K. Sousa, [th/2312.xxxxx]

\bibitem{EF}
J.~R.~Espinosa and J.~F.~Fortin,
``Vacuum decay actions from tunneling potentials for general spacetime dimension,''
JCAP \textbf{02} (2023) 023
\arXiv{2211.13667}{th}.

\bibitem{Cvetic:1992st}
M.~Cvetic, S.~Griffies and S.~J.~Rey,
``Nonperturbative stability of supergravity and superstring vacua,''
Nucl. Phys. B \textbf{389} (1993) 3
\arXivold{th/9206004}.

\bibitem{Cvetic}
M.~Cvetic, S.~Griffies and S.~J.~Rey,
``Static domain walls in N=1 supergravity,''
Nucl. Phys. B \textbf{381} (1992) 301,
\arXivold{th/9201007};
M.~Cvetic, S.~Griffies and H.~H.~Soleng,
``Local and global gravitational aspects of domain wall space-times,''
Phys. Rev. D \textbf{48} (1993) 2613,
\arXivold{gr-qc/9306005}.

\bibitem{Freedman:2003ax}
D.~Z.~Freedman, C.~Nunez, M.~Schnabl and K.~Skenderis,
``Fake supergravity and domain wall stability,''
Phys. Rev. D \textbf{69} (2004) 104027
\arXiv{0312055}{th}.
  
\bibitem{CI}
I.~Affleck,
``On Constrained Instantons,''
Nucl.\ Phys.\ B {\bf 191} (1981) 429.
Y.~Frishman and S.~Yankielowicz,
``Large Order Behavior of Perturbation Theory and Mass Terms,''
Phys.\ Rev.\ D {\bf 19} (1979) 540.

\bibitem{DynCo}
R.~Angius, J.~Calder\'on-Infante, M.~Delgado, J.~Huertas and A.~M.~Uranga,
``At the end of the world: Local Dynamical Cobordism,''
JHEP \textbf{06} (2022) 142
\arXiv{2203.11240}{th}.   

\bibitem{Draper:2023ulp}
P.~Draper, B.~Lillard and C.~Skye,
``Neutralizing topological obstructions to bubbles of nothing,''
JHEP \textbf{10} (2023) 049
\arXiv{2305.17838}{th}.

\bibitem{Hebecker}
B.~Friedrich, A.~Hebecker and J.~Walcher,
``Cobordism and Bubbles of Anything in the String Landscape,''
\arXiv{2310.06021}{th}.

\bibitem{GMSV}
I.~Garc\'{\i}a Etxebarria, M.~Montero, K.~Sousa and I.~Valenzuela,
``Nothing is certain in string compactifications,''
\arXiv{2005.06494}{th}.



\bibitem{HW}
J.~C.~Hackworth and E.~J.~Weinberg,
``Oscillating bounce solutions and vacuum tunneling in de Sitter spacetime,''
Phys. Rev. D \textbf{71} (2005) 044014
\arXivold{th/0410142}.

\bibitem{Horowitz:2007pr}
G.~T.~Horowitz, J.~Orgera and J.~Polchinski,
``Nonperturbative Instability of AdS(5) x S**5/Z(k),''
Phys. Rev. D \textbf{77} (2008) 024004
\arXiv{0709.4262}{th}.

\bibitem{Bomans:2021ara}
P.~Bomans, D.~Cassani, G.~Dibitetto and N.~Petri,
``Bubble instability of mIIA on $\mathrm{AdS}_4\times S^6$,''
SciPost Phys. \textbf{12} (2022) 3, 099
\arXiv{2110.08276}{th}.

\bibitem{Breitenlohner:1982jf}
P.~Breitenlohner and D.~Z.~Freedman,
``Stability in Gauged Extended Supergravity,''
Annals Phys. \textbf{144} (1982) 249

\bibitem{Klebanov:1999tb}
I.~R.~Klebanov and E.~Witten,
``AdS / CFT correspondence and symmetry breaking,''
Nucl. Phys. B \textbf{556} (1999) 89
\arXiv{9905104}{th}.

\bibitem{Henneaux:2002wm}
M.~Henneaux, C.~Martinez, R.~Troncoso and J.~Zanelli,
``Black holes and asymptotics of 2+1 gravity coupled to a scalar field,''
Phys. Rev. D \textbf{65} (2002) 104007
\arXiv{0201170}{th}

\bibitem{Hertog:2004ns}
T.~Hertog and G.~T.~Horowitz,
``Designer gravity and field theory effective potentials,''
Phys. Rev. Lett. \textbf{94} (2005) 221301
\arXiv{0412169}{th}

\bibitem{Hertog:2004dr}
T.~Hertog and K.~Maeda,
``Black holes with scalar hair and asymptotics in N = 8 supergravity,''
JHEP \textbf{07} (2004) 051
\arXiv{0404261}{th}.

\bibitem{Brown:2010mf}
A.~R.~Brown and A.~Dahlen,
``Bubbles of Nothing and the Fastest Decay in the Landscape,''
Phys. Rev. D \textbf{84} (2011) 043518
\arXiv{1010.5240}{th}.

\bibitem{Bandos:2023yyo}
I.~Bandos, J.~J.~Blanco-Pillado, K.~Sousa and M.~A.~Urkiola,
``Brane nucleation in supersymmetric models,''
JHEP \textbf{10} 061 (2023)
\arXiv{2306.09412}{th}.

\bibitem{Brown:2014rka}
A.~R.~Brown,
``Decay of hot Kaluza-Klein space,''
Phys. Rev. D \textbf{90} (2014) 104017
\arXiv{1408.5903}{th}.

\bibitem{MPW}
A.~Masoumi, S.~Paban and E.~J.~Weinberg,
``Tunneling from a Minkowski vacuum to an AdS vacuum: A new thin-wall regime,''
Phys. Rev. D \textbf{94} (2016) 025023
\arXiv{1603.07679}{th}.

\bibitem{GHY}
  G.W.~Gibbons and S.W.~Hawking,
  ``Action Integrals and Partition Functions in Quantum Gravity,''
  Phys.\ Rev.\ D {\bf 15} (1977) 2752;
 J.~W.~York, Jr.,
  ``Role of conformal three geometry in the dynamics of gravitation,''
  Phys.\ Rev.\ Lett.\  {\bf 28} (1972) 1082.

\end{thebibliography}
\end{document}